\documentclass[%
 reprint,
 amsmath,amssymb,
 aps,
]{revtex4-2}

\usepackage{graphicx}% Include figure files
\usepackage{dcolumn}% Align table columns on decimal point
\usepackage{svg}
\usepackage{natbib}
\usepackage{subcaption}
\usepackage{caption}
\usepackage{float}
\usepackage{bm}% bold math
\usepackage{hyperref}
\hypersetup{
    colorlinks=true,
        linktoc=all,
    linkcolor=blue,
    citecolor=blue,
    urlcolor=blue
}
\begin{document}

\preprint{APS/123-QED}

\title{Detection and Mitigation of Glitches in LISA Data: A Machine Learning Approach}% Force line breaks with \\

\author{Niklas Houba}
\email{niklas.houba@erdw.ethz.ch}
 % \altaffiliation[Also at ]{Institute of Geophysics, ETH Zurich}%Lines break automatically or can be forced with \\
\author{Luigi Ferraioli}%
\author{Domenico Giardini}%
\affiliation{%
 Institute of Geophysics \\  Department of Earth and Planetary Sciences at ETH Zurich \\
 Sonneggstrasse 5, 8092 Zurich, Switzerland % \textbackslash\textbackslash
}%


%\affiliation{
%\affiliation{
%\affiliation{%


\date{\today}% It is always \today, today,

\begin{abstract}
The proposed Laser Interferometer Space Antenna (LISA) mission is tasked with the detection and characterization of gravitational waves from various sources in the universe. This endeavor is challenged by transient displacement and acceleration noise artifacts, commonly called glitches. Uncalibrated glitches impact the interferometric measurements and decrease the signal quality of LISA's time-delay interferometry (TDI) data used for astrophysical data analysis. The paper introduces a novel calibration pipeline that employs a neural network ensemble to detect, characterize, and mitigate transient glitches of diverse morphologies. A convolutional neural network is designed for anomaly detection, accurately identifying and temporally pinpointing anomalies within the TDI time series. Then, a hybrid neural network is developed to differentiate between gravitational wave bursts and glitches, while a long short-term memory (LSTM) network architecture is deployed for glitch estimation. The LSTM network acts as a TDI inverter by processing noisy TDI data to obtain the underlying glitch dynamics. Finally, the inferred noise transient is subtracted from the interferometric measurements, enhancing data integrity and reducing biases in the parameter estimation of astronomical targets. We propose a low-latency solution featuring generalized LSTM networks primed for rapid response data processing and alert service in high-demand scenarios like predicting binary black hole mergers. The research highlights the critical role of machine learning in advancing methodologies for data calibration and astrophysical analysis in LISA.
%\begin{description}
%use the optional argument of the \verb+\item+ command to give the category of %each item. 
%\end{description}
\end{abstract}

\maketitle

%\tableofcontents

\section{\label{sec:level1}Introduction}
%* Introduction
%    * Historical Context of Glitches in LISA Pathfinder
%    * Extrapolation to LISA Mission Data
%    * Overview of the Paper
The Laser Interferometer Space Antenna (LISA) is an ambitious space mission scheduled for launch in the late 2030s, with the objective of detecting gravitational waves in the Millihertz band. LISA aims to extend gravitational wave astronomy beyond the capabilities of ground-based detectors like the Laser Interferometer Gravitational-Wave Observatory (LIGO) and Virgo \cite{amaroseoane2017laser}. Its measurement strategy involves monitoring the relative motion of free-falling test masses across the spacecraft constellation caused by the time-varying curvature of spacetime  \cite{Thorpe2006, Shaddock2008}.

LISA is expected to encounter several intrinsic noise sources that challenge its scientific efficacy. One significant disturbance in LISA's measurement system is laser noise, primarily arising from the unequal lengths of the interferometer arms and intrinsic laser frequency instabilities \cite{Dhurandhar2010}. Ideally, the arms of LISA's inter-spacecraft interferometers would be equal in length, allowing for the inherent cancellation of laser noise. However, these arms are unequal in practice due to the optical configuration and the orbital dynamics of the constellation \cite{Martens2021}. This leads to residual laser noise spoiling the gravitational wave measurements by seven to eight orders of magnitude \cite{Bayle2018a, Bayle2018b}. To mitigate the effect, a post-processing technique called time-delay interferometry (TDI) is applied \cite{Tinto2002TDI1stGen, TDITINTO2005}. TDI works by combining time-shifted measurements from multiple spacecraft in a manner that effectively cancels out noise, thereby synthesizing virtual interferometers with equal arm lengths.

Another prevalent disturbance in LISA denotes tilt-to-length (TTL) coupling, where spacecraft angular jitter translates into longitudinal displacement measurements \cite{hartig2021nongeometric, hartig2022geometric}. The coupling occurs when the angular motion of the spacecraft and optomechanical components induce path length variations due to instrument imperfections, misalignments, and laser wavefront errors, adding another layer of complexity to the noise profile.

Amidst these quasi-stationary perturbations, LISA contends with transient artifacts, known as glitches. These glitches, first observed in the LISA Pathfinder (LPF) mission, are expected to appear in LISA's TDI data, as well \cite{Armano2018, Robson2019}. Glitches can vary in form, intensity, and duration, and their presence will interfere with accurately characterizing gravitational wave sources. Given a good understanding of suppressing and removing perpetual noise sources like laser noise \cite{PhysRevLett.104.211103, Mitryk2010} and TTL coupling \cite{PhysRevD.106.042005, HoubaPRD, houbaJoGEngNote}, understanding and mitigating glitch artifacts becomes the next challenge in data calibration. The transition to handling unpredictable transient glitches during LISA data analysis requires novel methodologies outlined in the paper.
\vspace{-14pt}
\subsection{\label{ssec1:level1} Glitches in LISA Pathfinder}
The LPF mission, launched in December 2015, marked a milestone in the demonstration of technology critical for LISA \cite{armano2019lisaPF}. LPF consisted of one spacecraft, and its primary goal was to measure the relative acceleration between two test masses in free fall \cite{Bortoluzzi2009}. The mission surpassed expectations by achieving remarkably high measurement sensitivity, marking a major technological breakthrough and providing validation for the foundational concepts of LISA's design \cite{PhysRevLett.116.231101,Wanner2019Space-based}.\newpage %\enlargethispage{5\baselineskip}
\noindent
Alongside these successes, LPF faced challenges in the form of transient signals, i.e., glitches, in its interferometric measurements \cite{PhysRevD.106.062001}. The glitches, varying significantly in intensity and duration, presented themselves as nonperiodic events uncorrelated with the regular operations of the mission. The glitches' enigmatic nature poses a considerable challenge in data analysis and system understanding, as their origins remain unknown \cite{BaghiPhysRevD.105.042002}. Extensive research involving simulation studies and on-ground testing has been undertaken, focusing on comprehending the physical processes behind these transient events. LPF's data analysis revealed two predominant types of glitches \cite{BaghiPhysRevD.105.042002}:
\begin{enumerate}
    \item Fast Rise and Exponential Decay (FRED) events: These glitches were characterized by a sudden, sharp increase in acceleration, followed by a gradual exponential decline back to the baseline. This pattern suggests a rapid energy release followed by a slower dissipation process.
    \item Sine-Gaussian glitches: These glitches were identified by their distinctive pattern of sequential movements in opposite directions, resembling the shape of a sine wave modulated by a Gaussian envelope. Sine-Gaussian glitches were less frequent than FRED events and appeared to be stochastic in nature, with a notably higher incidence during phases when the system was cooled down. This observation hinted at a potential correlation between environmental conditions and the emergence of glitches, suggesting that external factors might influence their manifestation.
\end{enumerate}

\subsection{\label{ssec2:level1}Projection to LISA}

In the context of the LISA mission, addressing glitches is paramount. Unmitigated glitches in TDI data pose a risk to the mission's primary objective of accurate gravitational wave detection and source parameter estimation \cite{spadaro2023glitch}. The interference caused by these glitches could potentially lead to estimation biases, misinterpretations of gravitational wave signals, or even mask transient astrophysical signals entirely.

The experience gained from handling glitches in the LPF mission plays a crucial role in shaping the strategies for dealing with similar issues in LISA. A significant part of this strategy involves utilizing shapelet models to characterize glitches mathematically \cite{SpritzLDF}. These models are adept at capturing a range of glitch behaviors, including sharp rises, exponential decay, and oscillatory patterns, which align with the characteristics observed in LPF's glitch events.

To effectively identify these glitches, the authors of \cite{BaghiPhysRevD.105.042002} implemented a matching pursuit algorithm. This iterative method employs matched filtering alongside a parametric model, allowing for a systematic approach to glitch detection. The algorithm's effectiveness was demonstrated through its application to LPF's $\Delta g$ noise-only measurements recorded from March 2016 to July 2017. The analysis of this data uncovered a variety of glitch events, thereby validating the utility of the approach in deriving statistics about the physical features of the glitch population. 

The complexity and unpredictability of glitch morphologies suggest that standard models like first-order shapelets might not be sufficient to capture the entire spectrum of potential glitches \cite{PhysRevD.106.042006}. The limitations of first-order models become apparent when confronted with more complex or atypical glitch patterns that may not conform to the shapes seen in LPF measurements. Therefore, while shapelet models and current calibration algorithms offer a solid foundation, they might need to be supplemented with more flexible modeling and mitigation techniques to cover a broader range of glitch morphologies. 
\subsection{\label{ssec3:level1}Overview of the Paper}
The paper introduces a machine-learning approach to address artifacts in LISA data, proposing a three-step pipeline of connected neural networks for detecting, characterizing, and mitigating transient glitches.  
Section \ref{sec:level2} deals with the mathematical modeling of glitches and their propagation through TDI. This section forms the theoretical backbone for understanding the nature and impact of glitches in the LISA context.
Section \ref{sec:level3} outlines the methodology for detecting anomalies in LISA data. It starts with the fundamentals of Convolutional Neural Networks (CNNs), which form the first step of the proposed pipeline. The section further discusses data pre-processing for robust anomaly detection, simulation procedures, and evaluates the detection performance. 
Section \ref{sec:level4} focuses on the characterization of anomalies detected in the previous step. It discusses template matching for glitch characterization and hybrid neural networks to differentiate between glitches and gravitational wave bursts while locating glitch injection points. 
Section \ref{sec:level5} explores a novel technique for mitigating the effects of glitches. It highlights the application of a long short-term memory (LSTM) network for glitch estimation, which forms the final step of the proposed pipeline before glitch removal. Additionally, the section discusses a low-latency solution.
Section \ref{sec:level6} summarizes the findings of the study and highlights future directions of research.

\section{\label{sec:level2} Mathematical Modeling}
%* Mathematical Modeling
%    * Time-Domain Glitch Model: A Random-Walk Approach
%    * Models of Displacement and Acceleration Transients
%    * TDI Response to Glitches and Glitch Cross-Coupling
%    * TDI Response to Gravitational Wave Bursts
The section lays the groundwork for comprehending the glitches LISA's interferometric measurements and TDI data are expected to encounter. We start with a parametric model that encapsulates the transient nature of FRED events, using insights drawn from the LPF mission. The model is further augmented with a stochastic component designed to account for more diverse and unforeseen morphologies of glitches in the context of LISA.
\subsection{\label{ssec1:level2} Time-Domain FRED Glitch Model}
\begin{figure}
	\centering
  \includegraphics[trim=0 0 0 0, clip,width=0.52\textwidth]{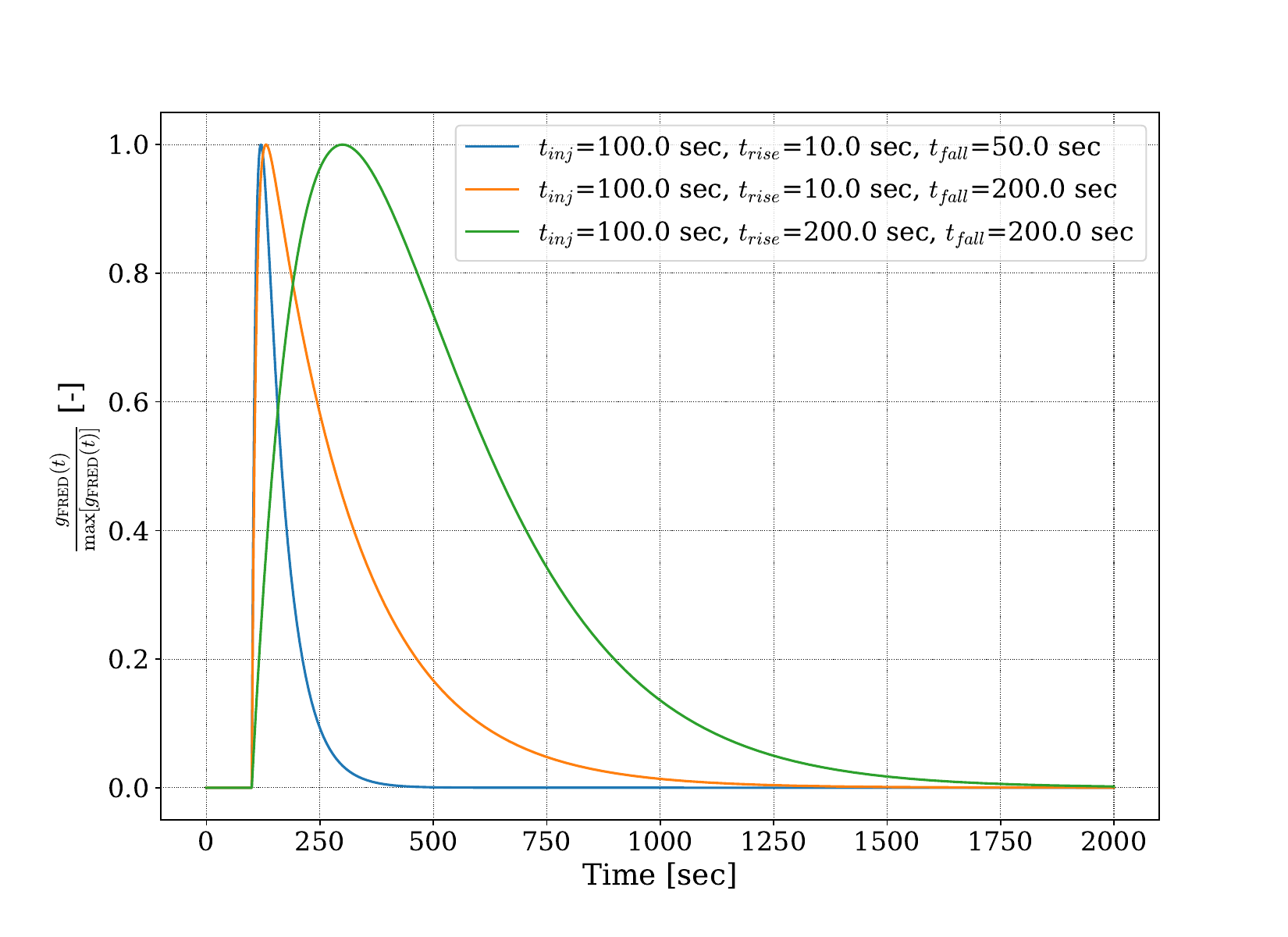}
	\caption{Normalized illustration of  $g_{\mathrm{FRED}}(t)$ for three different sets of parameters. The model will be modified by integrating stochastic fluctuations to accommodate more complex glitch shapes.}
	\label{im:LPGFlitch}
\end{figure}

The observations of glitches in LPF measurements have catalyzed the advancement of multiple glitch models. These models have been systematically integrated into the Python package \texttt{lisaglitch} \cite{lisaglitch21} tailored for application within the context of LISA. The paper primarily concentrates on a FRED-like model embodied in \texttt{lisaglitch}, distinguished by a rapid onset and subsequent gradual decay. Mathematically, the model is defined as a one-sided double-exponential function that starts at a specified injection time $t_{inj}\geq 0$ and ramps up to a maximum of $A/(t_{rise}-t_{fall})$ during $t_{rise}> 0$ before flattening off to zero during $t_{fall} > 0$. The glitch signal for $t\geq t_{inj}$, and $t_{rise} \neq t_{fall}$ is:
\begin{equation}
g_{\mathrm{FRED}}(t) = A\cdot \frac{ e^{-\frac{t - t_{inj}}{t_{rise}}} - e^{-\frac{t - t_{inj}}{ t_{fall}}}}{t_{rise}- t_{fall}}. \label{eq:FREDGl}
\end{equation}
The model captures the essential features of a FRED-like glitch, including its rise, peak, and decay phases. When $t_{rise}=t_{fall}$, the model uses a continuous extension to avoid singularity:
\begin{equation}
g_{\mathrm{FRED}}(t) = A \cdot \frac{t - t_{inj}}{t_{fall}^2} \cdot e^{-\frac{t - t_{inj} }{ t_{fall}}}.
\end{equation}
This extension is valid for $t\geq t_{inj}$ and ensures a smooth transition between the rise and fall phases of the glitch. 
Figure \ref{im:LPGFlitch} presents the model evaluated for three sets of parameters. As the glitch maximum is a function of the amplitude $A$ as well as the rising and falling time parameters, the curves are normalized to facilitate visual comparison.

To account for more complex glitch shapes the FRED model will be modified by stochastic fluctuations. 
\subsection{\label{ssec2.0:level2} Stochastic Model Extension}
To comprehensively capture the diverse morphological characteristics of glitches, we have expanded the FRED model by incorporating a stochastic component. This extension reflects the random-walk nature of noise transients, characterized by random fluctuations occurring on time scales shorter than $t_{rise}+t_{fall}$. The stochastic element within the model is described as a standard Wiener process $W(t)$, which is a continuous-time stochastic process representing cumulative random fluctuations \cite{Levy2019}:
\begin{align}
n(t) = W(t), \label{eq:Wiener}
\end{align}
To model temporal correlations and prevent unrealistic high-frequency variations, we apply a moving average smoothing to Eq. \eqref{eq:Wiener}, represented as an integration over a continuous window:
\begin{equation}
n_{filtered}(t) = \frac{1}{w} \int_{t-w}^{t} n(\tau)d\tau.
\end{equation}
Here, $w$ is the window size of the moving average, determining the extent of smoothing.

\begin{figure}[t!]
  \includegraphics[trim=0 0 0 0, clip,width=0.52\textwidth]{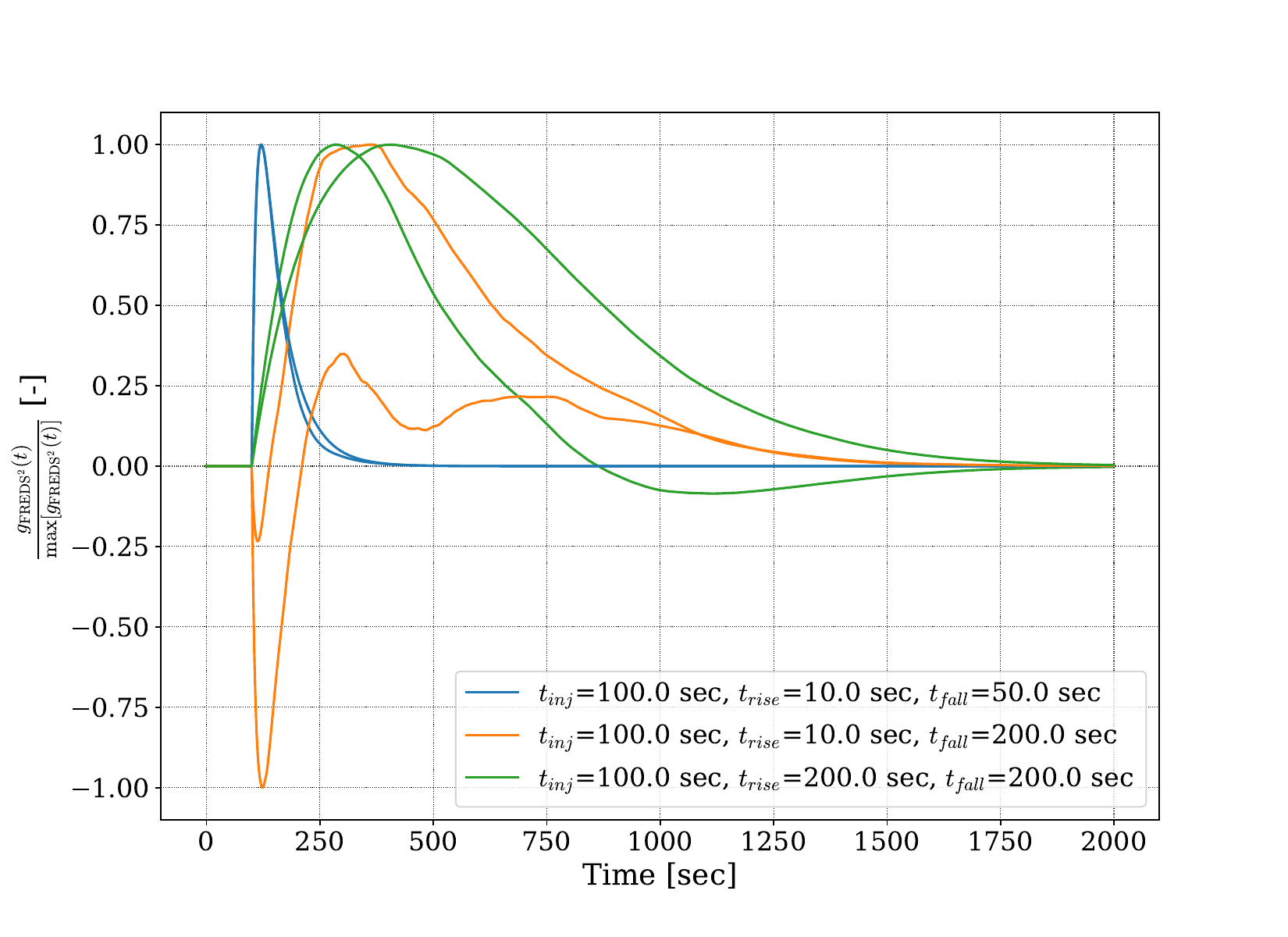}
	\caption{Normalized illustration of  $g_{\mathrm{FREDS}^2}(t)$ for the \hspace{2pt} three parameter sets of Fig. \ref{im:LPGFlitch} and $w = 5000$ at 4 Hz. The same color denotes glitches with identical FRED model parameters, highlighting that while their broader behavior aligns, short-term variations can differ significantly due to the stochastic extension.}
	\label{im:StochLPGFlitch}
\end{figure}

The FRED model with stochastic shaping denoted as FREDS$^2$ combines both the deterministic rise and fall pattern of the FRED model with the smoothed stochastic extension, resulting in a glitch signal that represents both LPF-inspired and unforeseen glitch characteristics:
\begin{equation}
g_{\mathrm{FREDS}^2}(t) = g_{\mathrm{FRED}}(t) \cdot n_{filtered}(t).\label{eq:FREDSSGl}
\end{equation}
Figure \ref{im:StochLPGFlitch} provides a normalized illustration of the extended glitch model for the three sets of parameters from Fig. \ref{im:LPGFlitch}. The figure demonstrates the impact of stochastic variations on the glitch morphology. Equal colors represent glitch pairs with identical FRED model parameters, illustrating that while their general behavior aligns, the introduction of stochastic elements can lead to significant short-term variations. 

In this paper, we treat the moving average smoothing window size $w$ as independent of the glitch duration. If it is desired in future studies to align the stochastic smoothing process with the inherent glitch timescale, it may be beneficial to adjust the window size proportionately to the sum of $t_{rise}$ and $t_{fall}$.

The stochastic model extension includes erratic and complex exponential decay distortions and is employed in the paper to simulate glitches in interferometric LISA measurements. It is important to note that introducing the stochastic extension complicates glitch mitigation during TDI data calibration. The stochastic nature of the extended model implies that mere estimation and subtraction of a glitch signal obtained from FRED model matching will be insufficient. The stochastic extension introduces random variations that significantly modify the glitch morphology. Consequently, even when applying FRED model-based glitch correction with the true FRED model parameters, residual glitch noise remains in the TDI data.  This residual impact can obscure or distort the gravitational wave signals that LISA aims to detect, thereby reducing the accuracy and reliability of the mission's scientific return. 

To address this challenge, a more sophisticated approach is required to predict the stochastic variations in glitch time series. Here, the use of neural networks for glitch time series prediction emerges as a promising solution. Neural networks, with their ability to learn complex patterns and dependencies in data, are well-suited for recognizing the intricate and stochastic nature of these glitches. By training a network on a diverse set of glitch time series encompassing a range of stochastic behaviors, a powerful tool can be developed capable of accurately estimating and mitigating the impact of glitches in TDI data.

Before, we will explore how the derived glitch model can be integrated into LISA's interferometer measurement model. This investigation is essential to understand their effect on the various TDI channels that need to be handled by the neural network-driven glitch detection and mitigation pipeline.
\subsection{\label{ssec2:level2} Acceleration and Displacement Transients}
\begin{figure}[b!]
%\captionsetup{justification=centering}
	\centering
  \includegraphics[trim=45 10 20 0, clip,width=0.50\textwidth]{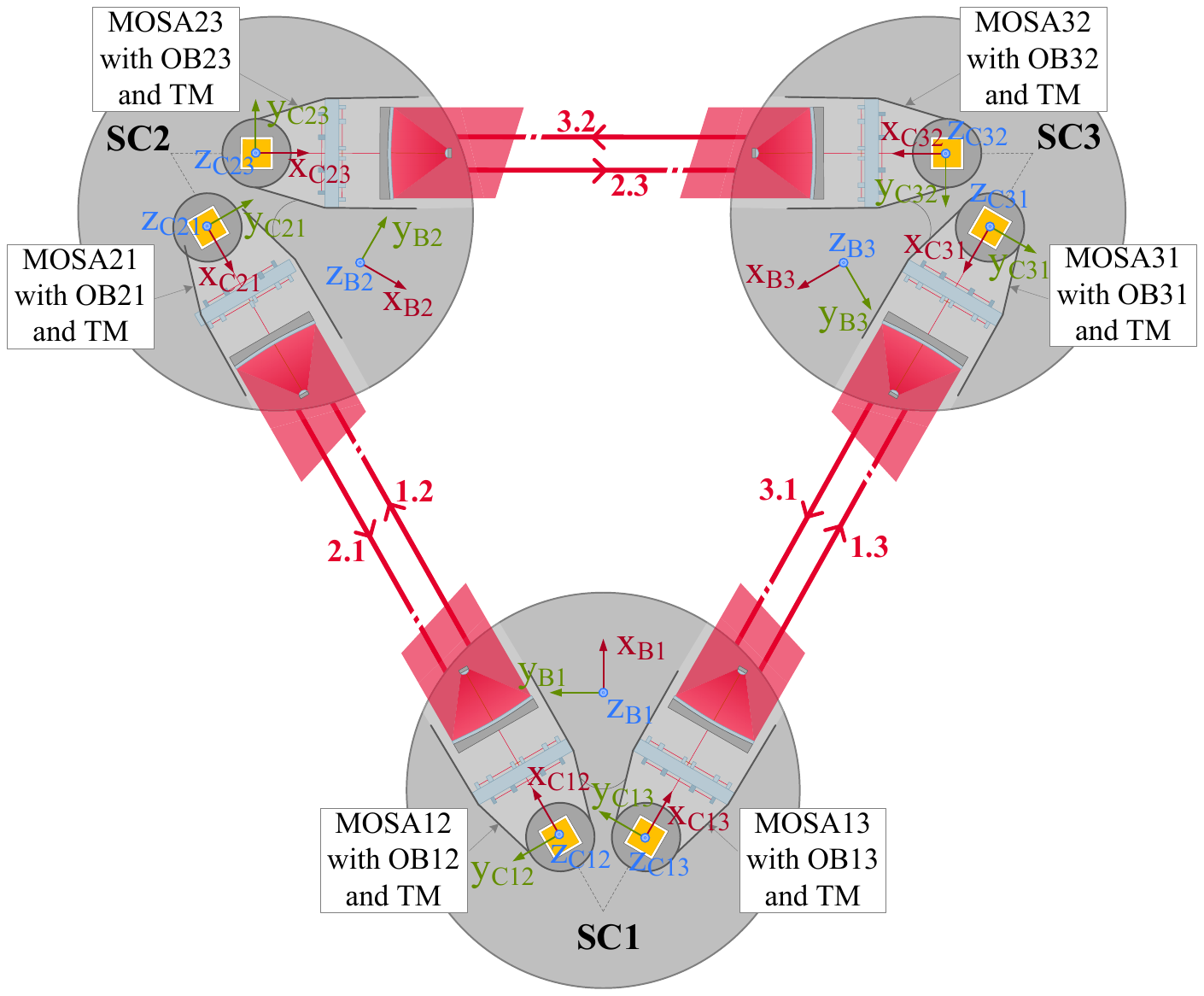}
	\caption{Configuration of the LISA constellation, including the index convention for this paper analog to \cite{Houba2023}. MOSA, OB, and TM acronyms stand for moving optical subassembly, optical bench, and test mass. A MOSA mainly accommodates an optical bench and a test mass.}
	\label{im:LISAConstellation}
\end{figure}

According to LISA's optical design, there are three distinct interferometers per optical bench: the inter-spacecraft interferometer, the test mass interferometer, and the reference interferometer. The configuration of the six optical benches of the LISA constellation is depicted in Fig \ref{im:LISAConstellation}. 
Focusing on optical bench 12 as a representative case, the corresponding measurement models are as follows:
\newline\newline
\noindent
1. Inter-spacecraft interferometer
\begin{equation}
\label{eq:inter_spacecraft_interferometer}
\mathrm{isi}_{12}(t) = p_{21;2.1}(t) - p_{12}(t) + H_{12}(t) + N^\mathrm{isi}_{12}(t).
\end{equation}
\newline
2. Test mass interferometer
\begin{equation}
\label{eq:test_mass_interferometer}
\mathrm{tmi}_{12}(t) = p_{13}(t) - p_{12}(t) + \frac{2}{\lambda}\boldsymbol{n}_{1.2}\cdot \boldsymbol{v}_{12}(t)+ N^\mathrm{tmi}_{12}(t).
\end{equation}
\newline
3. Reference interferometer
\begin{equation}
\label{eq:reference_interferometer}
\mathrm{rfi}_{12}(t) = p_{13}(t) - p_{12}(t) + N^\mathrm{rfi}_{12}(t).
\end{equation}
These measurements are quantified in units of frequency. In Eq. \eqref{eq:inter_spacecraft_interferometer}, the term $p_{12}(t)$ represents the laser noise from optical bench 12, which interferes with the laser beam of optical bench 21. The laser noise $p_{21;2.1}(t)$ from optical bench 21 undergoes a time delay, $\tau_{2.1}(t)$, due to the photon travel between optical bench 21 and 12. This delay is incorporated in the model as 
$p_{21;2.1}(t) \equiv p_{21}(t-\tau_{2.1}(t))$. For the test mass and reference interferometers, outlined in Eqs. \eqref{eq:test_mass_interferometer} and \eqref{eq:reference_interferometer}, the laser beams from optical bench 12 and the adjacent optical bench 13 on the same spacecraft are combined. As a result, these equations include $p_{13}(t)$  instead of  $p_{21;2.1}(t)$. No inter-spacecraft time delay is required.

In Eq. \eqref{eq:inter_spacecraft_interferometer}, $H_{12}(t)$ represents the gravitational wave contribution and the terms $N_{12}^{\mathrm{isi}}(t)$, $N_{12}^{\mathrm{tmi}}(t)$, and $N_{12}^{\mathrm{rfi}}(t)$ represent optical path length noise in the respective interferometers. 

The test mass interferometer is affected by displacements of the test mass. When considering units of frequency, the test mass velocity $\boldsymbol{v}_{12}(t)$ impacts the measurement, as detailed in Eq. \eqref{eq:test_mass_interferometer}. Here,  $\boldsymbol{v}_{12}(t)$ is projected onto the line of sight using the laser link unit vector  $\boldsymbol{n}_{1.2}$. Analog to \cite{phdthesisJB}, it is presumed that the direction of the link remains constant while laser light travels along the arm. The factor $2$ accounts for the round-trip journey of the laser beam as it hits and reflects off the test mass, and $\lambda = 1064.5$ nm denotes the nominal wavelength of the laser source.

A complete model of LISA's  measurement setup would incorporate additional disturbances, such as clock noise, TTL noise, and backlink fiber noise \cite{https://doi.org/10.15488/11372, HoubaJoG, Fleddermann2018}. These components need not be included in the modeling scope of this paper. Nonetheless, they will be regarded as disturbances in the simulation, affecting the performance of the glitch mitigation pipeline.  Note that the measurement equations for the remaining five optical benches can be obtained from Eqs. \eqref{eq:inter_spacecraft_interferometer} to \eqref{eq:reference_interferometer} through a cyclic permutation of the indices.

By establishing the measurement system containing the three interferometers per optical bench in LISA, we can now outline the specific glitch injection points considered in this framework. Optical displacement glitches are taken into account via $N_{12}^{\mathrm{isi}}(t)$, $N_{12}^{\mathrm{tmi}}(t)$, and $N_{12}^{\mathrm{rfi}}(t)$ while test mass acceleration glitches are addressed through the test mass velocity vector $\boldsymbol{v}_{12}(t)$ in Eq. \eqref{eq:test_mass_interferometer}. To represent these acceleration glitches, the extended glitch model $g_{\mathrm{FREDS}^2}(t)$ is integrated over time, yielding:
\begin{align}
\begin{aligned}
         \boldsymbol{n}_{1.2}\cdot \boldsymbol{v}_{\mathrm{FREDS}^2,12}(t)  &\equiv v_{\mathrm{FREDS}^2,12}(t) \\ &= \int_{0}^{t} g_{\mathrm{FREDS}^2,12}(\tau)d\tau.\label{eq:fredsinteg}
\end{aligned}        
\end{align}
For optical displacement glitches within the various interferometers, we consider differentiating $g_{\mathrm{FREDS}^2}(t)$. The relation between frequency shift and displacement is given by:
\begin{align}
    \begin{aligned}
        N_{\mathrm{FREDS}^2,12}^{\mathrm{x}}(t) = \frac{1}{2\pi}\frac{2\pi}{\lambda}\frac{dg_{\mathrm{FREDS}^2,12}^\mathrm{x}(t)}{dt},\label{eq:NxFREDS2}
    \end{aligned}
\end{align}
with $\mathrm{x} \in \{\mathrm{isi,tmi,rfi}\}$. The paper categorizes instrumental glitches into two primary types: test mass acceleration glitches and optical displacement glitches, also known as phasemeter glitches. This classification aligns with the glitches outlined in \cite{spadaro2023glitch}. Future studies may consider test mass velocity glitch profiles resulting from closed-loop interactions of the drag-free and attitude control system. The investigation is ongoing and represents an important aspect of research.

\subsection{\label{ssec3:level2}TDI Response to Acceleration and Displacement Transients}
Test mass acceleration and displacement artifacts are obscured in the raw interferometer measurements of LISA due to the dominant presence of laser noise. These transient anomalies become visible after TDI is applied. Therefore, understanding the response of the TDI post-processing algorithm to glitches is crucial for mitigating their impact during astrophysical data analysis. 

The initial noise reduction pipeline processes raw measurements of the LISA constellation to obtain noise-calibrated TDI data streams. Before executing TDI, intermediary variables $\xi_{ij}(t)$, $Q_{ij}(t)$, and $\eta_{ij}(t)$ with $i,j \in \{1,2,3\}$ and $i\neq j$ are computed in this process. These variables mitigate translational optical displacement noise as well as clock noise and reduce the number of lasers from six to three. The algorithms for calculating $\xi_{ij}(t)$, $Q_{ij}(t)$, and $\eta_{ij}(t)$ are expounded upon in \cite{https://doi.org/10.15488/8545}. The $\eta_{ij}(t)$ variables are used as input for TDI.

Note that there are several ways for linearly combining $\eta_{ij}(t)$ variables to suppress laser noise. These linear combinations, known as TDI channels, offer distinct benefits in the context of gravitational wave analysis and instrument diagnosis \cite{muratore2020revisitation}. The second-generation TDI Michelson variable, which is a four-link combination, is focused on in this paper. TDI second generation (also known as TDI 2.0) suppresses laser noise to the required level for a rotating spacecraft configuration with time-variant arm lengths. For the TDI Michelson channel  $X(t)$ associated with spacecraft 1, measurements from the optical benches 12, 13, 21, and 31 are utilized. TDI Michelson $X(t)$ is formulated as follows:
\begin{align}
\begin{aligned}
    X(t) &= {}  \eta_{13} - \eta_{12} - \eta_{21;2.1} + \eta_{31;3.1}  + \eta_{12;g} - \eta_{13;i} \\ &  + \eta_{21;Y} - \eta_{31;Z} - \eta_{13;W}   + \eta_{12;U}  + \eta_{21;M}\\ & - \eta_{31;N}  - \eta_{12;I} + \eta_{13;K} - \eta_{21;A} + \eta_{31;B}.
\end{aligned}\label{eq:TDI20X}
\end{align}
The time argument is omitted in Eq. \eqref{eq:TDI20X} for readability. The equation incorporates multiple nested delays, essential for laser noise cancellation, with delay arrangements summarized using letter notations for clarity as given in Table \ref{tab:translationtable}. Similar combinations for $Y(t)$ and $Z(t)$ of spacecraft 2 and 3 are obtained by cyclic permutation of measurement and delay indices.
\begin{table}
\begin{center}
\captionsetup{justification=centering}
\caption{Translation of delay index letters \cite{HoubaPRD}.}\label{tab:translationtable} 
\vspace{5pt}
\begin{tabular}{cccc}
\hline
\hline
Index & Delays & Index & Delays \\
\hline

$A$ & 2.1,1.3,3.1,1.3,3.1,1.2,2.1 & $W$ & 1.3,3.1,1.2,2.1 \\ %\hline
$B$ & 3.1,1.2,2.1,1.2,2.1,1.3,3.1 & $X$ & 2.3,3.2,2.1,1.2 \\ %\hline
$I$ & 1.3,3.1,1.3,3.1,1.2,2.1     & $Y$ & 2.1,1.3,3.1 \\ 
%\hline
$K$ & 1.2,2.1,1.2,2.1,1.3,3.1     & $Z$ & 3.1,1.2,2.1       \\ %\hline
$M$ & 2.1,1.2,2.1,1.3,3.1         & $f$ & 3.2,2.3    \\
$N$ & 3.1,1.3,3.1,1.2,2.1         & $g$ & 1.3,3.1  \\ 
$U$ & 1.2,2.1,1.3,3.1        & $i$ & 1.2,2.1  \\

\hline
\hline
\end{tabular}%\vspace{20pt}
\end{center}
\end{table}

We consider optical bench 12 as an example to assess the impact of glitches on $X(t)$, $Y(t)$, and $Z(t)$. The different types of glitches introduced in the previous section will reveal distinctive patterns in those TDI channels.  
\subsubsection{Test Mass Acceleration Glitches}
For a test mass acceleration glitch in the test mass interferometer of optical bench 12, all terms in Eqs. \eqref{eq:inter_spacecraft_interferometer} to \eqref{eq:reference_interferometer} are set to zero except for $\frac{2}{\lambda}\boldsymbol{n}_{1.2}\cdot \boldsymbol{v}_{12}(t)$. The scenario is represented as:
\begin{align}
\begin{aligned}
\mathrm{tmi}_{12}(t) &= \frac{2}{\lambda}\boldsymbol{n}_{1.2}\cdot \boldsymbol{v}_{\mathrm{FREDS}^2,12}(t) \\ &=\frac{2}{\lambda}v_{\mathrm{FREDS}^2,12}(t).
\end{aligned}
\end{align}
with $v_{\mathrm{FREDS}^2,12}(t)$ according to Eq. \eqref{eq:fredsinteg}. Following the noise reduction procedure of \cite{https://doi.org/10.15488/8545}, this leads to the following expressions for $\eta_{ij}(t)$:
\begin{align}
\eta_{12}(t) &= -\frac{1}{\lambda}v_{\mathrm{FREDS}^2,12}(t), \\
\eta_{21}(t) &= -\frac{1}{\lambda}\mathcal{D}_{1.2} v_{\mathrm{FREDS}^2,12}(t),
\end{align}
with all other terms equal to zero and $\mathcal{D}_{i.j}f(t)\equiv f(t-\tau_{i.j})$. Using Eq. \eqref{eq:TDI20X}, we derive the impact of a test mass acceleration glitch of optical bench 12 on the TDI Michelson channel $X(t)$:
\begin{align}
    X(t) = (\mathcal{D}_{g}-1)(\mathcal{D}_{i}+1)(\mathcal{D}_{W}-1)\frac{v_{\mathrm{FREDS}^2,12}(t)}{\lambda}. \label{eq:XTMAccGlitch}
\end{align}
The TDI Michelson channels $Y(t)$ and $Z(t)$ are obtained through a cyclic permutation of indices:
\begin{align}
Y(t) &= -2 \mathcal{D}_{1.2} (\mathcal{D}_{f} - 1)  (\mathcal{D}_{X} - 1) \frac{v_{\mathrm{FREDS}^2,12}(t)}{\lambda},\label{eq:YTMAccGlitch}
\end{align}
\begin{align}
Z(t) &= 0.\label{eq:ZTMAccGlitch}
\end{align}
For the analytical investigation presented in this study, we adopt the commonly used assumption that the delay operator is commutative. This assumption is equivalent to treating the arm lengths in the LISA constellation as time-invariant.
\subsubsection{Optical Displacement Glitches}
Next, we will proceed to derive the TDI response to optical displacement glitches. For the inter-spacecraft interferometer of optical bench 12, all terms in Eqs. \eqref{eq:inter_spacecraft_interferometer} to \eqref{eq:reference_interferometer} are set to zero except for $N_{12}^{\mathrm{isi}}(t)$. The scenario is represented as:
\begin{align}
\mathrm{isi}_{12}(t) &= N_{\mathrm{FREDS}^2,12}^{\mathrm{isi}}(t),
\end{align}
with $N_{\mathrm{FREDS}^2,12}^{\mathrm{isi}}(t)$ from Eq. \eqref{eq:NxFREDS2}. Applying the initial noise reduction pipeline of \cite{https://doi.org/10.15488/8545}, this translates to:
\begin{align}
\eta_{12}(t) &= N_{\mathrm{FREDS}^2,12}^{\mathrm{isi}}(t),
\end{align}
with all other terms equal to zero.
Using Eq. \eqref{eq:TDI20X}, the impact on $X(t)$ becomes:
\begin{align}
X(t) &= -(\mathcal{D}_{g} - 1)(\mathcal{D}_{W} - 1)N_{\mathrm{FREDS}^2,12}^{\mathrm{isi}}(t).
\end{align}
For $Y(t)$ and $Z(t)$, we find:
\begin{align}
Y(t) &= \mathcal{D}_{1.2}(\mathcal{D}_{f} - 1)(\mathcal{D}_{X} - 1)N_{\mathrm{FREDS}^2,12}^{\mathrm{isi}}(t),
\end{align}
\begin{align}
Z(t) &= 0.
\end{align}
In the case of an optical displacement glitch in the test mass interferometer of optical bench 12, the measurement model of Eqs. \eqref{eq:inter_spacecraft_interferometer} to \eqref{eq:reference_interferometer} reduces to:
\begin{align}
\mathrm{tmi}_{12}(t) &= N_{\mathrm{FREDS}^2,12}^{\mathrm{tmi}}(t).
\end{align}
This translates to changes in the intermediary variables:
\begin{align}
\eta_{12}(t) &= -\frac{1}{2}N_{\mathrm{FREDS}^2,12}^{\mathrm{tmi}}(t), \\
\eta_{21}(t) &= -\frac{1}{2} \mathcal{D}_{1.2} N_{\mathrm{FREDS}^2,12}^{\mathrm{tmi}}(t).
\end{align}
With Eq. \eqref{eq:TDI20X}, the impact on $X(t)$ can be obtained:
\begin{align}
\begin{aligned}
X(t) &= \frac{1}{2} (\mathcal{D}_{g} - 1)(D_{i} + 1)(\mathcal{D}_{W} - 1)\\ &\hspace{32pt} \cdot N_{\mathrm{FREDS}^2,12}^{\mathrm{tmi}}(t). \label{eq:XTMOptGlitch}
\end{aligned}
\end{align}
Similarly, for $Y(t)$ and $Z(t)$:
\begin{align}
Y(t) &= - \mathcal{D}_{1.2} (\mathcal{D}_{f} - 1)(\mathcal{D}_{X} - 1)N_{\mathrm{FREDS}^2,12}^{\mathrm{tmi}}(t),\label{eq:YTMOptGlitch}
\end{align}
\begin{align}
Z(t) &= 0. \label{eq:ZTMOptGlitch}
\end{align}
\begin{table*}[]
\begin{center}
\captionsetup{justification=centering}
\caption{TDI responses to different types of glitches injected on the optical bench 12.}
\vspace{5pt}
\label{tab:tdi_responses}
\begin{tabular}{ccccc}
\hline
\hline
& Test mass glitch & \multicolumn{3}{c}{Optical glitch} \\
& & Inter-spacecraft interferometer & Test mass interferometer & Reference interferometer \\
\hline
$X(t)$ \hspace{4pt} & $(\mathcal{D}_{g}-1)(\mathcal{D}_{i}+1)(\mathcal{D}_{W}-1)$ & $-(\mathcal{D}_{g} - 1)(\mathcal{D}_{W} - 1)$ & $\frac{1}{2} (\mathcal{D}_{g} - 1)(D_{i} + 1)(\mathcal{D}_{W} - 1)$ & $-\mathcal{D}_{i} (\mathcal{D}_{g} - 1) (\mathcal{D}_{W} - 1)$ \\
$Y(t)$ \hspace{4pt} & $-2 \mathcal{D}_{1.2} (\mathcal{D}_{f} - 1)  (\mathcal{D}_{X} - 1)$ & $\mathcal{D}_{1.2}(\mathcal{D}_{f} - 1)(\mathcal{D}_{X} - 1)$ & $- \mathcal{D}_{1.2} (\mathcal{D}_{f} - 1)(\mathcal{D}_{X} - 1)$ & $\mathcal{D}_{1.2}(\mathcal{D}_{f} - 1)(\mathcal{D}_{X} - 1)$ \\
$Z(t)$ \hspace{4pt} & $0$ & $0$ & $0$ & $0$ \\
\hline
\hline
\end{tabular}\label{tab:TDIGlitchModelSC1}
\end{center}
\end{table*}
\noindent\,\hspace{-10.7pt} For the reference interferometer of optical bench 12, it holds:
\begin{align}
\mathrm{rfi}_{12}(t) &= N_{\mathrm{FREDS}^2,12}^{\mathrm{rfi}}(t).
\end{align}
Following the approach as before, we obtain:
\begin{align}
\eta_{12}(t) &= \frac{1}{2}N_{\mathrm{FREDS}^2,12}^{\mathrm{rfi}}(t), \label{eq:RefODG12}\\
\eta_{31}(t) &= -\frac{1}{2}\mathcal{D}_{1.3}N_{\mathrm{FREDS}^2,12}^{\mathrm{rfi}}(t), \\
\eta_{13}(t) &= \frac{1}{2}N_{\mathrm{FREDS}^2,12}^{\mathrm{rfi}}(t), \label{eq:RefODG13}\\
\eta_{21}(t) &=  \frac{1}{2}\mathcal{D}_{1.2}N_{\mathrm{FREDS}^2,12}^{\mathrm{rfi}}(t).
\end{align}
Substituting into Eq. \eqref{eq:TDI20X} leads to:
\begin{align}
X(t) &= -\mathcal{D}_{i} (\mathcal{D}_{g} - 1) (\mathcal{D}_{W} - 1)N_{\mathrm{FREDS}^2,12}^{\mathrm{rfi}}(t). 
\end{align}
Finally, $Y(t)$ and $Z(t)$ are:
\begin{align}
Y(t) &= \mathcal{D}_{1.2}(\mathcal{D}_{f} - 1)(\mathcal{D}_{X} - 1)N_{\mathrm{FREDS}^2,12}^{\mathrm{rfi}}(t),
\end{align}
\,
\vspace{-13pt}
\,
\begin{align}
Z(t) = 0.
\end{align}
The TDI glitch responses for the remaining five optical benches can be derived through a cyclic permutation of indices from those detailed for optical bench 12.

From Eqs. \eqref{eq:XTMAccGlitch} to  \eqref{eq:ZTMAccGlitch} and \eqref{eq:XTMOptGlitch} to \eqref{eq:ZTMOptGlitch}, it becomes apparent that distinguishing between test-mass velocity glitches and optical displacement glitches within the test mass interferometer of an optical bench is not possible when solely analyzing TDI data. Both glitch types exhibit identical echo patterns in the TDI Michelson set. This agrees with the findings of \cite{phdthesisJB}. Note that here, we assume glitches in the test mass acceleration profile. Under this assumption, a distinction will be attainable. 

An interesting observation regarding the occurrence of optical displacement glitches in the reference interferometer is evident when examining Eqs. \eqref{eq:RefODG12} and \eqref{eq:RefODG13}. The variables $\eta_{12}(t)$ and $\eta_{13}(t)$ are affected identically. Consequently, these two variables, being the only non-delayed quantities in the definition of $X(t)$, cancel in Eq. \eqref{eq:TDI20X}. Therefore, the effect of an optical displacement glitch in the reference interferometer of optical bench 12 is not immediately measurable in $X(t)$ at the moment of injection. This is also evident from Table \ref{tab:TDIGlitchModelSC1} summarizing the TDI responses for optical bench 12. For each glitch category, there is one TDI response, which, when expanded, includes a factor of $\pm 1$, except for the reference interferometer. Here, in the case of optical bench 12 in $X(t)$, there is the delay $\mathcal{D}_i=D_{1.2,2.1}$, and in $Y(t)$, the delay $D_{1.2}$ as a prefactor so that a corresponding glitch is detectable in the TDI set at the earliest by an amount $\tau_{1.2}$ after the actual injection time in the interferometer. This fact will complicate the precise determination of the glitch injection time in the first step of the glitch mitigation pipeline, where the glitch source and, thus, the associated TDI response model are unknown.

From Table \ref{tab:TDIGlitchModelSC1}, we see that one of the TDI Michelson combinations remains unaffected by glitches originating from a single optical bench, for example, $Z(t)$ for glitches caused within optical bench 12 (or 21). This characteristic will be valuable for identifying the origin of the glitch in the second step of the glitch mitigation pipeline. Additionally, it aids in differentiating between glitches and transient astrophysical phenomena. Unlike the selective effect observed for glitches, the latter tends to impact all three TDI combinations simultaneously. 

In scenarios where noise is present, advanced concepts, such as neural networks, become essential to accurately distinguish between transient astrophysical sources and glitches and determine the glitch source. This is particularly true for noise artifacts masked by quasi-stationary noise yet still impacting the accuracy of source parameter estimation. These faint glitches will be analyzed in the sections that follow.

\subsection{\label{ssec4:level2}TDI Response to Gravitational Wave Bursts}
Gravitational wave bursts represent key astrophysical signals in LISA \cite{SciReqDoc}. It is essential to differentiate and preserve these signals in the data during calibration rather than erroneously removing them as noise artifacts. Therefore, the glitch detection and mitigation pipeline aims to identify and discern gravitational wave bursts accurately. Achieving this requires modeling the influence of gravitational wave bursts on LISA's TDI data. 

The impact of gravitational wave bursts on the change in interferometric path length was presented in \cite{Robson2019} and is briefly summarized below. The observed path length change, denoted as $\delta l_{i.j}(t)$, in the link between spacecraft $i$ and $j$ is as follows:
\begin{equation}
\delta l_{i.j}(t) = D_{i.j} :  \int_{\xi_i}^{\xi_j} h(t) \, dt,
\end{equation}
where the colon indicates full contraction between tensors, with $t$ being the Solar system barycenter time. The wave variable $\xi_i = t_i - \boldsymbol{k}\cdot \boldsymbol{x}_i$ defines surfaces of constant phase for the gravitational wave. The detector tensor $D_{i.j}$ is given by:
\begin{equation}
D_{i.j} = \frac{1}{2} \left( \boldsymbol{r}_{i.j} \otimes \boldsymbol{r}_{i.j} \right) / \left( 1- \boldsymbol{k} \cdot \boldsymbol{r}_{i.j} \right),
\end{equation}
where $\boldsymbol{k}$ is the gravitational wave propagation direction, and $\boldsymbol{r}_{i.j}$ is the unit-separation vector between the spacecraft pointing from $i$ to $j$. The gravitational wave tensor $h(t)$ comprises two polarization states, $h_{+}(t)$ and $h_{\times}(t)$:
\begin{equation}
h(t) = h_+(t) e{'}_{+}(\psi, \theta, \phi) + h_\times(t) e{'}_{\times}(\psi, \theta, \phi),
\end{equation}
where $e{'}_{+}$ and $e{'}_{\times}$ are polarization tensors rotated by the polarization angle $\psi \in [0, \pi ]$, while $\theta$ and $\phi$ denote the source position in spherical polar coordinates stated in the Solar system barycentric frame \cite{Robson2019}.

The FREDS$^2$ model is employed for modeling the gravitational wave polarizations. To simulate the instrument measurement, we make use of the Python packages \texttt{lisagwresponse} and \texttt{lisainstrument} \cite{lisagwresponse21, lisainstrument22}. The packages enable simulating the response of the six inter-spacecraft interferometers to the gravitational wave signal of interest. The resulting data is then processed by the \texttt{pyTDI} software \cite{pytdi23} to obtain the manifestation of the gravitational wave burst in the TDI Michelson channels.
\section{\label{sec:level3} Anomaly Detection}
This section is dedicated to robust anomaly detection, representing the initial step of the glitch mitigation pipeline developed for LISA. Classical anomaly detection employs diverse methods, including statistical techniques, clustering-based approaches, and dimensionality reduction strategies, each with its own strengths tailored for specific scenarios \cite{Len2020, Imani2018Anomaly}.

Within the scope of this paper, we define an anomaly as a transient signal that can signify either a gravitational wave source or an instrument glitch. Such an anomaly may exhibit complicated features and be obscured by quasi-stationary noise. In this context, the capabilities of CNNs appear highly promising. The adeptness in managing anomalies with unforeseen morphologies and noisy environments positions CNNs as a compelling choice for the anomaly detection step of the glitch mitigation pipeline. The applicability for LISA is studied in the following.
\subsection{\label{ssec1:level3} Convolutional Neural Networks}
CNNs are specialized feed-forward neural networks that significantly impacted the evolution of machine learning with applications ranging from image and video recognition to natural language processing and financial time series analysis \cite{rawat2017deep}. A notable advantage of CNNs is their ability to counteract issues such as vanishing and exploding gradients, commonly encountered in traditional neural networks, by applying weight regularization across reduced connections. One of the hallmarks of CNNs is their reduced need for data pre-processing compared to other image analysis techniques. CNNs learn to optimize their filters autonomously, a significant advantage in fields where feature extraction is complex \cite{alzubaidi2021review}.

In traditional neural networks, particularly those fully connected, each neuron in a layer is linked to all neurons in the subsequent layer. This can lead to a tremendous number of weights, especially for larger inputs like high-resolution images. CNNs drastically reduce this number by employing convolutional layers. These layers use cascaded convolution kernels to process the input in small segments, significantly scaling down the number of required neurons and, consequently, the number of free parameters \cite{golinko2018learning}.
\begin{figure*}[t!]
%\captionsetup{justification=centering}
	\centering
  \includegraphics[trim=0 0 0 0, clip,width=0.99\textwidth]{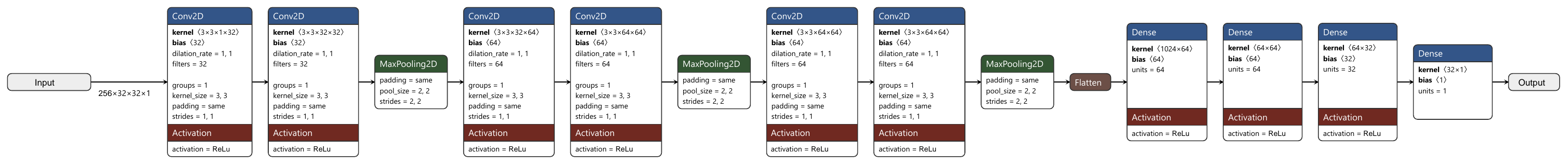}
	\caption{CNN architecture and attributes employed for anomaly detection and anomaly injection time estimation. The architecture and subsequent network architectures are visualized using the open-source tool \texttt{Netron} \cite{netron22}.}
	\label{im:CNNArchitecture}
\end{figure*}

Figure \ref{im:CNNArchitecture} illustrates the CNN architecture with \texttt{TensorFlow} attributes \cite{tensorflow2015whitepaper} applied in the research presented. The architecture is tailored to enable the precise detection of anomalies within LISA’s TDI data. The network processes TDI data streams as images, thus capitalizing on the pattern recognition capabilities of CNNs.

The CNN structure presented in Fig. \ref{im:CNNArchitecture} initiates with two convolutional layers, each containing 32 filters of size 3$\times$3. This is followed by four convolutional layers, each with 64 filters. All convolutional layers employ the rectified linear unit (ReLU) activation function defined as $f(x) = \mathrm{max}(0,x)$. Utilizing a non-linear activation function allows the network to learn and represent non-linear relations between the network's inputs and outputs. The non-linearity is important as the TDI data streams will contain complex patterns with non-linear characteristics.

Each pair of convolutional layers is followed by a 2$\times$2 maximum pooling layer. In CNNs, maximum pooling layers are used for reducing the spatial dimensions of the input data, which in turn diminishes the computational load and the number of parameters. Maximum pooling layers operate on the output of the convolutional layers, known as feature maps, with the primary goal of downsampling these maps while retaining relevant information. The process involves a specific-sized window that slides over the feature map. As it moves, typically in strides of two pixels, the pooling layer performs a maximum operation. This operation selects the maximum value from the pixels under the window, representing the entire region on the output feature map. As a result, the spatial dimensions of the feature map, both height and width, can be reduced. The reduction process is vital for preserving the most important features in the data. Features with higher numerical values in the feature maps, indicative of critical information like edges or specific textures, are prioritized. By focusing on these maximum values, the pooling layer ensures the retention of salient parts of the information. Note that pooling layers contribute to the model's generalizability and robustness. By abstracting the features learned and lowering the resolution of the feature maps, maximum pooling layers help in reducing the risk of overfitting. Consequently, the network is expected to perform better on unseen data \cite{Galanis2022Convolutional, Mahmoudi2020Learnable}.

The network of Fig. \ref{im:CNNArchitecture} then proceeds to a flattening step, where the two-dimensional feature maps are converted into a one-dimensional vector compatible with fully connected layers.  The flattened data is fed into three successive fully connected dense layers with 64, 64, and 32 nodes, respectively. The function of these layers is to refine the feature representation further, focusing on subtle aspects that might indicate anomalies.

\subsection{\label{ssec2:level3} Time Series to Image Conversion}
The preprocessing of TDI time series data into an image format is critical in enabling its analysis with CNNs. The inherent structure and patterns of instrumental artifacts expected in TDI lend themselves well to image-based analysis. 

The conversion process begins with segmenting TDI data streams into intervals of 7.5 min each. The duration strikes a pragmatic balance, being sufficiently brief to ensure that, ideally, each segment encapsulates no more than one anomaly while also being long enough to include a mix of anomalous and non-anomalous data. This approach enables the network to be trained on transient artifacts embedded in quasi-stationary noise. 

Considering a sampling rate of 4 Hz, the 7.5-min segment translates to an initial image width of 1800 pixels, calculated as 
7.5 min $\times$ 60 sec $\times$ 4 samples/sec. To accommodate a subsequent resizing of the image to a targeted value of 256 pixels, we opt for a slightly reduced initial image width of 1792 pixels. This adjustment ensures that the resizing process can be executed with an integer factor, thereby reducing interpolation errors.

Following the segmentation process, the data stream is subjected to a moving average filter with a window size of 20. The window size is chosen to smooth short-term fluctuations while preserving the data's longer-term trends, which are important for anomaly detection. Subsequently, the values are converted to their absolute forms and normalized, ensuring their range is confined within the interval $[0, 1]$.

For graphical representation, the normalized and filtered data is mapped onto an image matrix. We select an image height of 16 pixels and perform the time series to image (TTI) conversion for the quasi-orthogonal TDI combination:
\begin{align}
E(t)=\frac{X(t)- 2Y(t) + Z(t)}{\sqrt{6}}\equiv E_Y(t), 
\end{align}
as derived in \cite{LISAOptSens} and its variant:
\begin{align}
E_Z(t)=\frac{Y(t)- 2Z(t) + X(t)}{\sqrt{6}}. 
\end{align}
Each of these combinations is represented as an image of size 16$\times$1792 pixels. The final composite, measuring 32$\times$1792 pixels, is obtained by stacking the two images.

Note that $E_Y(t)$ and $E_Z(t)$ represent six link combinations containing all three TDI Michelson variables in parallel. This ensures that glitches are captured in both $E_Y(t)$ and $E_Z(t)$ regardless of their injection point throughout the LISA constellation. This applies to gravitational wave bursts as well and proves particularly valuable in the detection step, where no differentiation is made between bursts and glitches. 

Note that the quasi-orthogonal TDI variable $A(t)=(X(t) - Z(t))/\sqrt{2}$ \cite{LISAOptSens} also involves six laser links but only two TDI Michelson variables. It is, thus, less preferable in the process of obtaining consistent features for image generation. To illustrate, in the event of a test mass acceleration glitch of optical bench 21, $A(t)$ does not immediately reflect the impact since it lacks the inclusion of the associated TDI Michelson variable $Y(t)$. This leads to a delayed response in $A(t)$ due to the time it takes for photons to travel between spacecraft 2 and 3 or spacecraft 2 and 1. In contrast, $E_Y(t)$ and $E_Z(t)$ both immediately capture the effect of the glitch as they include $Y(t)$ in their formulation. Using $A(t)$ and $E_Y(t)$ or $A(t)$ and $E_Z(t)$ together can lead to glitches not consistently mapping to the same pixel position in the TTI conversion, depending on their source. This justifies the choice of $E_Y(t)$ and $E_Z(t)$. 

The quasi-orthogonal TDI variable $T(t)$, expressed as $(X(t) + Y(t) + Z(t))/\sqrt{3}$ \cite{LISAOptSens} was initially considered as an alternative to either $E_Y(t)$ or $E_Z(t)$. However, it is observed that test mass acceleration glitches are less pronounced in $T(t)$.  A more detailed examination of this observation based on analytical TDI response models is omitted here.

Following the TTI conversion, the image width undergoes resizing for reasons primarily related to computational efficiency. The width is reduced to 256 pixels. The reduction to 256 pixels facilitates faster training times for the CNN but also tends to boost the detection performance by making artifacts more evident in the resized images with unchanged height. The resizing step makes use of the Python imaging library.

\begin{figure}[t!]
%\captionsetup{justification=justified,singlelinecheck=false}
\captionsetup[subfigure]{justification=centering,singlelinecheck=false}
\begin{subfigure}{\columnwidth}
  \includegraphics[width=\linewidth]{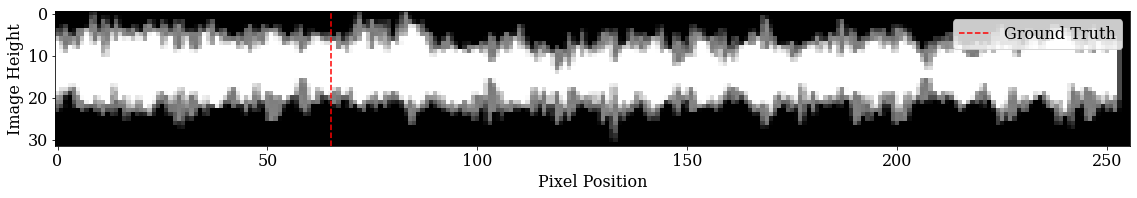}
  \caption{}
  \label{fig:sub1}
\end{subfigure}
\begin{subfigure}{\columnwidth}
  \includegraphics[width=\linewidth]{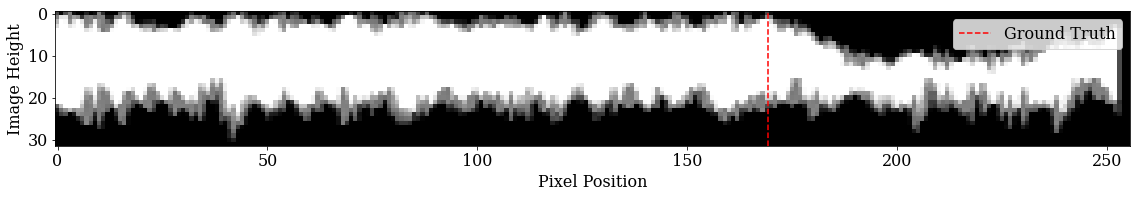}
  \caption{}
  \label{fig:sub2}
\end{subfigure}
\begin{subfigure}{\columnwidth}
  \includegraphics[width=\linewidth]{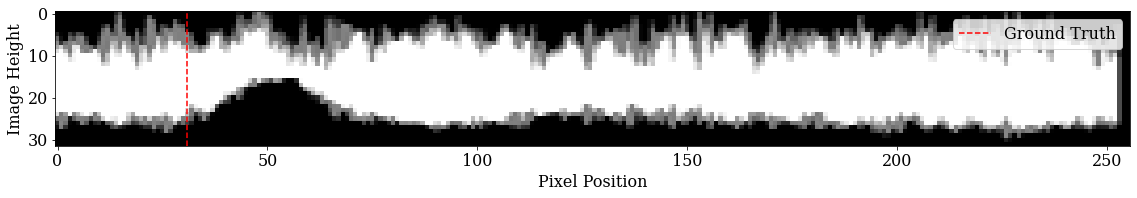}
  \caption{}
  \label{fig:sub3}
\end{subfigure}
\begin{subfigure}{\columnwidth}
  \includegraphics[width=\linewidth]{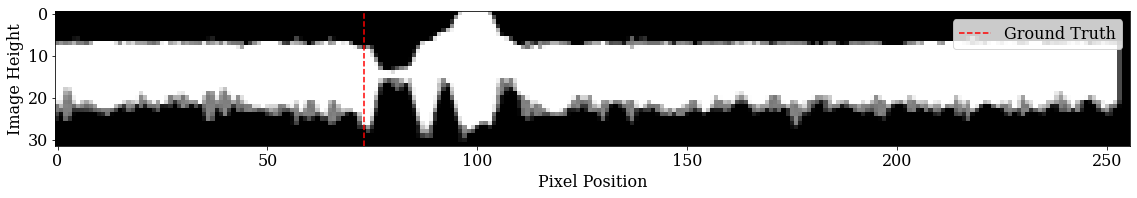}
  \caption{}
  \label{fig:sub4}
\end{subfigure}
%\end{figure}
%\begin{figure}
\caption{Example outputs of the dual-channel TTI conversion process, including ground truth labels that highlight the anomaly injection time. The subfigures illustrate the diverse ways in which an anomaly may manifest.  Note that in (a), the anomaly is scarcely discernible to the human observer. This prompts the exploration of CNN capabilities in detecting such subtle artifacts in LISA's TDI data.}
\label{fig:TTIConversionDemonstration}
\end{figure}

Examples of the stacked and resized images of TDI $E_Y(t)$ and $E_Z(t)$ time segments are given in Fig \ref{fig:TTIConversionDemonstration}, where the ground truth label highlights the anomaly injection time across varying anomaly dynamics.
Each subfigure displays dual-channel information from $E_Y(t)$ and $E_Z(t)$, as explained earlier, with the upper and lower half corresponding to $E_Y(t)$ and $E_Z(t)$, respectively. Figure \ref{fig:sub1} shows an anomaly that eludes visual detection in the presence of noise despite its several-minute span. In Fig. \ref{fig:sub2}, $E_Y(t)$ reveals a slow-dynamic transient that stands out against the background. The same holds for $E_Z(t)$ in Fig. \ref{fig:sub3}. Figure \ref{fig:sub4} marks an example of an anomaly being readily apparent against the background noise in both the $E_Y(t)$ and $E_Z(t)$ combinations. The CNN will be trained with these images across a spectrum of different anomaly dynamics and anomaly-to-noise ratios, a process whose efficacy will be evaluated in the following section.
\subsection{\label{ssec3:level3} Detection Performance}
The training and testing TDI data sets of the CNN model are generated using a suite of Python packages developed by the LISA consortium, including \texttt{lisaglitch}, \texttt{lisainstrument}, \texttt{lisagwresponse}, and \texttt{pyTDI} \cite{lisaglitch21, lisainstrument22, lisagwresponse21, pytdi23}. Here, the FREDS$^2$ model is locally integrated into \texttt{lisaglitch} to simulate glitches with stochastic fluctuations, as described in Section \ref{ssec2.0:level2}. The instrument response is acquired with \texttt{lisainstrument}, which is then fed into \texttt{pyTDI} to obtain the TDI variables at 4 Hz in units of frequency. The simulation assumes static yet unequal distances between the spacecraft to streamline computation, an important aspect to be considered when generating large data sets. The noise profile in the TDI variables adheres to the SciRDv1 model outlined in \cite{SciReqDoc}. Glitch characteristics are varied, with durations extending from a few seconds to 15 min and strengths ranging from 5 \% to 200 \% of the maximum quasi-stationary noise amplitude. The rising time of these glitches is set to fluctuate between 1 sec and 200 sec. Note that the window size of the moving average filter equals the one used in Fig. \ref{im:StochLPGFlitch}.

The CNN of Fig. \ref{im:CNNArchitecture} is applied in two stages. The initial stage classifies time segments as anomalous or non-anomalous, while the second stage focuses on predicting the injection times of anomalies. This dual-stage process improves the accuracy in predicting glitch injection times but requires training the network on two distinct data sets: one with a mixture of anomalous and non-anomalous segments and another exclusively containing anomalous data. Non-anomalous data are uniquely labeled with negative injection times. Consequently, a predicted negative injection time is evaluated as a non-anomalous data segment.

For the classification task, the training set comprises 250,000 samples, containing a randomly generated mix of quasi-stationary and transient noise time series. This data set includes both anomalous data (such as gravitational wave bursts and instrumental glitches from all test masses and interferometers) and non-anomalous data. The testing set, containing 20,000 samples, follows a similar composition. In the second stage, focusing on anomaly injection time prediction, the training and testing sets contain 250,000 and 20,000 samples, as well, consisting solely of anomalous events, as the first stage is tasked to filter out non-anomalous TDI segments.

During both training phases, we employ the Adam optimizer, setting the learning rate at a constant 0.0005. The loss function is the mean squared error, and the network undergoes training over 50 epochs with a batch size of 16. To ensure optimal performance, we select the model weights corresponding to the epoch that yields the lowest loss value on an independent validation data set.

\subsubsection{True Negative and True Positive Detections}
Considering the testing data set of 20,000 TTI-converted TDI samples, the CNN exhibits a 100 \% rate for correctly detecting non-anomalous data as non-anomalous;
99.99915  \% of anomalous events are detected as such, while 0.00085 \% of anomalous data are missclassified as non-anomalous.

\subsubsection{Estimation of Anomaly Injection Times}
\begin{figure}[t!]
%captionsetup{justification=justified,singlelinecheck=false}
\captionsetup[subfigure]{justification=centering,singlelinecheck=false}
\begin{subfigure}{\columnwidth}
  \includegraphics[width=\linewidth]{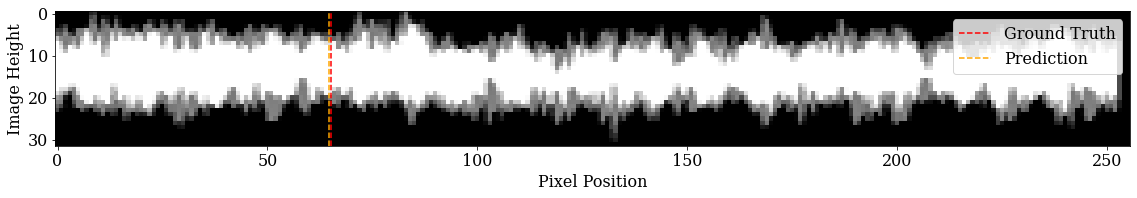}
  \caption{}
  \label{fig:sub1P}
\end{subfigure}
\begin{subfigure}{\columnwidth}
  \includegraphics[width=\linewidth]{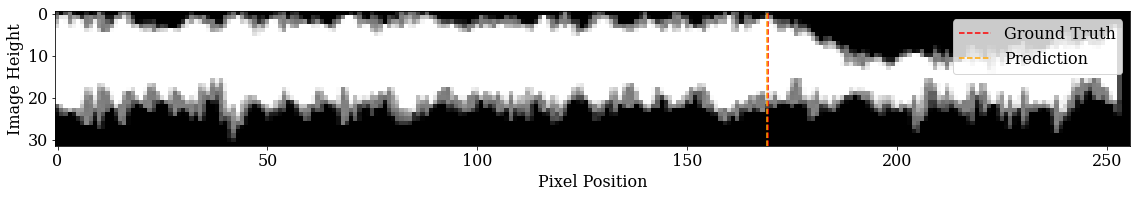}
  \caption{}
  \label{fig:sub2P}
\end{subfigure}
\begin{subfigure}{\columnwidth}
  \includegraphics[width=\linewidth]{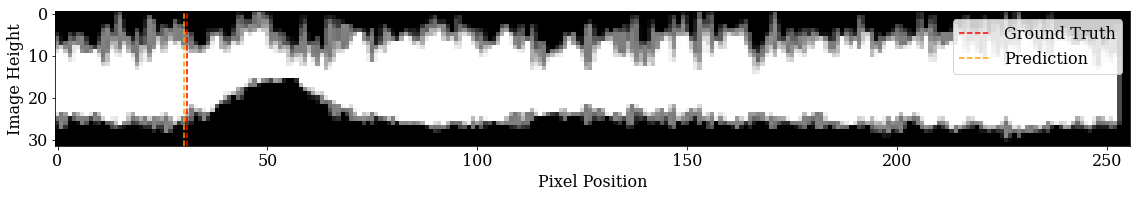}
  \caption{}
  \label{fig:sub3P}
\end{subfigure}
\begin{subfigure}{\columnwidth}
  \includegraphics[width=\linewidth]{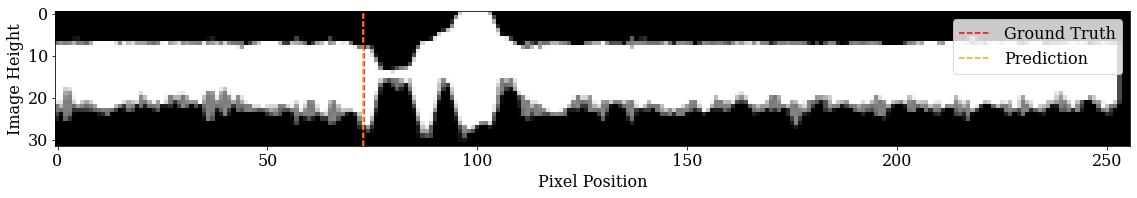}
  \caption{}
  \label{fig:sub4P}
\end{subfigure}
\caption{Estimated anomaly injection times for the four scenarios presented in Fig. \ref{fig:TTIConversionDemonstration}, contrasted with the corresponding ground truth data.}

\label{fig:TTIConversionDemonstrationPred}
\end{figure}

The proficiency of the CNN in estimating anomaly injection times is presented in Fig. \ref{fig:TTIConversionDemonstrationPred} for the examples of Fig. \ref{fig:TTIConversionDemonstration}.
The estimation accuracy for Fig. \ref{fig:sub1P} is remarkable. To validate this performance, comparable scenarios were tested. Three of them are depicted in the appendix. Overall, the estimation error maintains a zero mean with a standard deviation of 3.995 sec, considering the testing data set of the second stage, which comprises 20,000 samples.

\begin{figure}[]
%\captionsetup{justification=justified,singlelinecheck=false}
\captionsetup[subfigure]{justification=centering,singlelinecheck=false}
\begin{subfigure}{\columnwidth}
  \includegraphics[width=\linewidth]{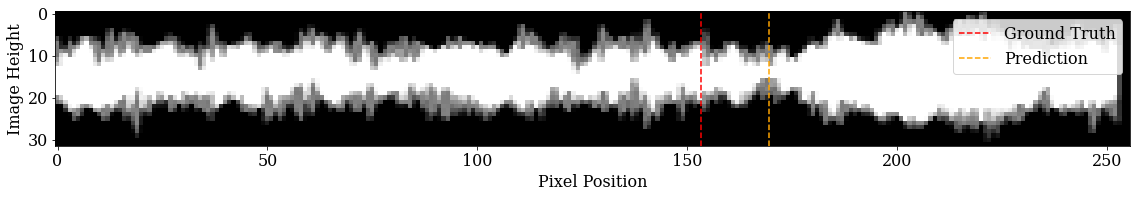}
  \caption{}
  \label{fig:sub1PLE}
\end{subfigure}
\begin{subfigure}{\columnwidth}
  \includegraphics[width=\linewidth]{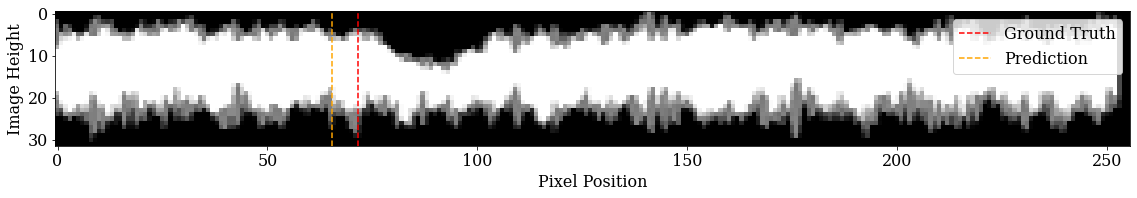}
  \caption{}
  \label{fig:sub2PLE}
\end{subfigure}
\caption{Estimated anomaly injection times compared to ground truth data in scenarios with larger estimation errors.}
\label{fig:TTIConversionDemonstrationPredWorse}
\end{figure}

\noindent
Figure \ref{fig:TTIConversionDemonstrationPredWorse} presents two scenarios for cases with larger estimation errors. The primary challenge in injection time estimation arises with slow-rising anomalies, which are less distinct in their initial phases due to the differentiating behavior of TDI 2.0. It turns out that pinpointing the onset of such anomalies becomes difficult if the glitch peak is reached more than 20 sec post-injection. Conversely,  anomalies with less than 20 sec rising time show estimation errors of less than 1 sec owing to more substantial changes in the statistical properties of $E_Y(t)$ and $E_Z(t)$ around the anomaly injection. This also holds for artifacts embedded in quasi-stationary noise that remain undetectable through visual inspection.

Besides slow-rising artifacts, the network's estimation accuracy is impacted by optical displacement glitches from LISA's reference interferometers. These glitches are labeled in the training data according to their injection time in the interferometer, but they manifest in the TDI data with an 8.3-sec delay, increasing network confusion when handling various types of artifacts simultaneously.

In conclusion, while the CNN effectively detects anomalies and estimates injection times,  determining the onset of faint and slow-rising artifacts with sub-second accuracy is challenging for the current CNN design. Note that the subsequent steps of the glitch mitigation pipeline will operate based on the estimation performance presented here.

\section{\label{sec:level4} Anomaly Characterization}
In the previous section, we detected anomalies in LISA's TDI data. Building upon this, the focus shifts to the task of anomaly characterization. This involves differentiating between astrophysical sources and instrumental glitches and accurately identifying the source of these glitches. Such a source could be either a test mass glitch (with the need to specify the particular test mass involved) or an optical displacement glitch (pinpointing the exact interferometer and optical bench). We integrate a neural network with a glitch template matching methodology to address this challenge, leveraging the glitch TDI response models derived in Section \ref{ssec3:level2}.
\begin{figure*}[t!]
\captionsetup{justification=centering}
	\centering
  \includegraphics[trim=0 0 0 0, clip,width=0.75\textwidth]{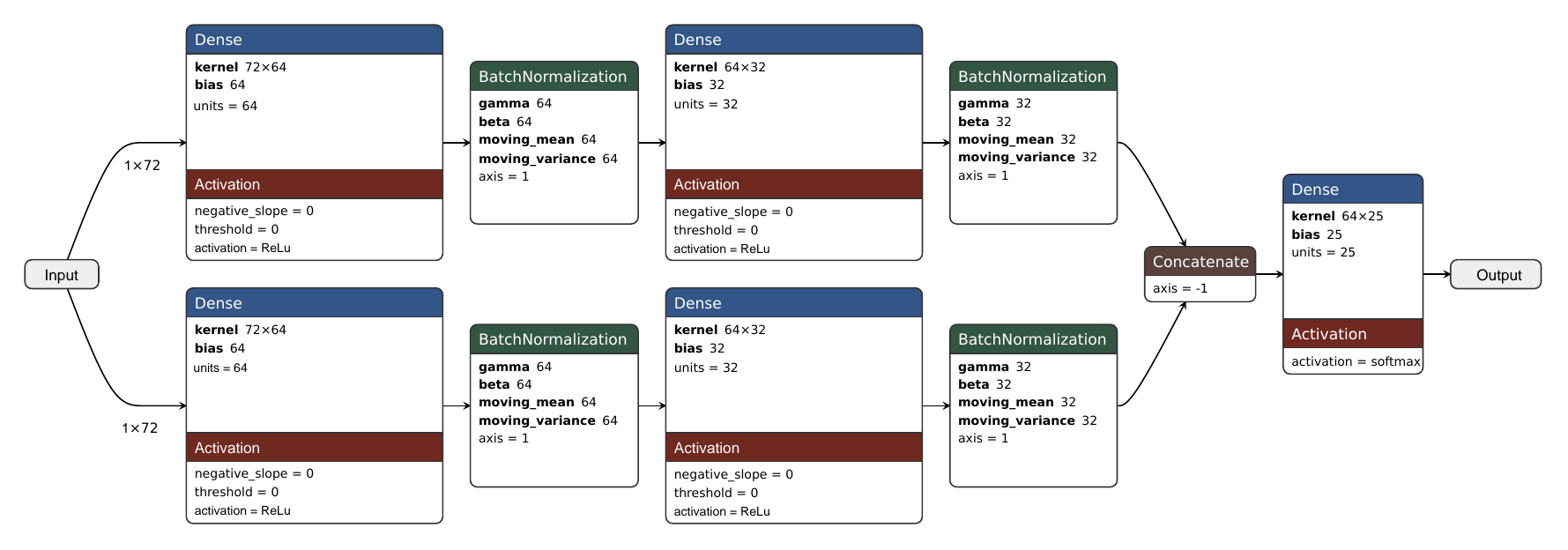}
	\caption{WDNN architecture and attributes employed for anomaly characterization.}
	\label{im:WDNNArchitecture}
\end{figure*}
\subsection{Glitch Template Matching}
In gravitational wave astrophysics, template matching is a systematic approach for contrasting observed data with a predefined set of waveforms, either theoretical or empirical \cite{Sampson2013Mismodeling, PhysRevD.77.104017}. This process is instrumental in recognizing and characterizing potential gravitational wave events. We adapt this method to the context of glitch localization within LISA's TDI data. Therefore, we introduce the glitch-to-noise ratio (GNR) $\rho$ in Eq. \eqref{eq:snrRho} analogously to the LISA signal-to-noise ratio (SNR) provided by \cite{StrubGPU}:
\begin{equation}
\rho=\frac{\left\langle d(t) \mid g\left(t,\theta\right)\right\rangle}{\sqrt{\left\langle g\left(t,\theta\right) \mid g\left(t,\theta\right)\right\rangle}}.\label{eq:snrRho}
\end{equation}
Here, the scalar product $\langle x(t) \mid y(t)\rangle$ between two time-domain signals 
$x(t)$ and $y(t)$ is defined by:
\begin{equation}
\langle x(t) \mid y(t)\rangle=4 \mathcal{R}\left(\int_0^{\infty} \frac{\tilde{x}(f) \tilde{y}^*(f)}{S(f)} d f\right),
\end{equation}
with $\tilde{x}(f)$ representing the Fourier transform of $x(t)$, $\tilde{x}^*(f)$ as the complex-conjugate of $\tilde{x}(f)$, and $S(f)$ signifying the one-sided noise power spectral density. 
In this context, the data $d(t)$ refers to the TDI variable of interest, while $g(t,\theta)$ denotes the glitch response in $d(t)$ characterized by a set of parameters $\theta$.

In the case of a deterministic FRED glitch, it would be possible to infer the intrinsic FRED model parameters $\theta = \{t_{inj},t_{rise},t_{fall}\}$ of Eq. \eqref{eq:FREDGl} from $d(t)$ when knowing the glitch injection point. For example, the parameter inference could be performed through maximum likelihood estimation utilizing Eq. \eqref{eq:snrRho} as proposed in \cite{PhysRevD.106.062003} for astrophysical targets. 

When dealing with the FREDS$^2$ model and an unknown glitch source, we confront a more intricate challenge. To determine the glitch source, we first approximate the FREDS$^2$ glitch of Eq. \eqref{eq:FREDSSGl} by the FRED model of Eq. \eqref{eq:FREDGl} adopting the average of the expected minimum and maximum values for the rising and falling time parameters, along with the injection time estimate derived from the preceding section. This results in a very rough approximation of the actual glitch transient. Then, the approach involves evaluating Eq. \eqref{eq:snrRho} for all 24 feasible glitch sources across the TDI variables $X(t)$, $Y(t)$, and $Z(t)$ knowing that the Fourier transform of $g(t,\theta)$ is given by the Fourier transform of the glitch approximation and the TDI response models derived in Section \ref{ssec3:level2}. This computation yields a GNR triplet $\{\rho_X, \rho_Y, \rho_Z\}$ for each glitch injection point. 

Selecting the triplet with the highest aggregate GNR could serve as a starting point for predicting the glitch origin. This method will prove insufficient given the stochastic nature of glitch shapes, substantial uncertainties in the model parameters and the model itself, as well as quasi-stationary noise. 
Therefore, we feed the 24 GNR triplets into a hybrid classification neural network, which maps the inputs to specific interferometers or test masses or identifies them as gravitational wave bursts.
\subsection{Hybrid Neural Network Model}
The neural network employed for anomaly characterization is hybrid in two key respects. 
Firstly, it incorporates model information by utilizing frequency-domain TDI glitch responses from Section \ref{ssec3:level2} while avoiding explicit waveform models for classifying gravitational wave bursts. This design enables the network to accommodate unexpected burst morphologies without relying on assumptions for waveform parameters. Secondly, the network combines wide and deep learning components, forming a so-called wide and deep neural network (WDNN). The WDNN architecture merges the strengths of both wide and deep components: the wide part consists of a linear model that directly connects input features to the output, excelling in memorizing feature interactions, while the deep component comprises multiple layers of non-linear transformations, adept at capturing complex feature interactions \cite{10.1145/2988450.2988454}. The architecture of the WDNN is illustrated in Fig. \ref{im:WDNNArchitecture}. 

The neural network model comprises a wide and deep component. The wide component has two layers with 64 and 32 neurons: ReLU activation, L2 regularization, and batch normalization. The deep component consists of three layers with 64, 64, and 32 neurons, using ReLU activation, L2 regularization, and batch normalization, as well. L2 regularization mitigates overfitting by adding a penalty term to the loss function, encouraging smaller weight magnitudes. Batch normalization improves training stability and accelerates convergence by normalizing the activations within each sub-batch, reducing internal covariate shift and enabling higher learning rates during training. The final layer employs the softmax activation function. The softmax activation function is given by:
\begin{equation}
f\left(x_i\right)=\frac{\exp \left(x_i\right)}{\sum_{j=1}^n \exp \left(x_j\right)},\label{eq:softmax}
\end{equation}
for $x_i \in X$, where $X \in \mathbb{R}^n$ denotes an input vector with $n\geq 1$.
Equation \eqref{eq:softmax} is commonly used in the context of classification tasks \cite{Goodfellow-et-al-2016,DUBEY202292}. It takes the raw output scores from the previous network layers and converts them into a probability distribution over multiple classes, justifying its application for the glitch characterization network.
%\newpage
\subsection{Characterization Performance}
To train and test the WDNN, TDI data sets are generated, simulating both glitch and gravitational wave events with properties described in the previous section. The GNR triplets calculated for each TDI glitch model of each simulation run are normalized and labeled using one-hot encoding according to Table \ref{tab:injection_points} in the appendix. One-hot encoding categorizes each sample into one of 25 classes, encompassing 24 possible glitch injection points and a class for gravitational wave bursts. We apply a data set comprising 150,000 training samples and 20,000 testing samples for the network's training and evaluation.

\begin{figure}[]
	\centering
  \includegraphics[trim=190pt 200pt 190pt 120pt, clip,width=0.47\textwidth]{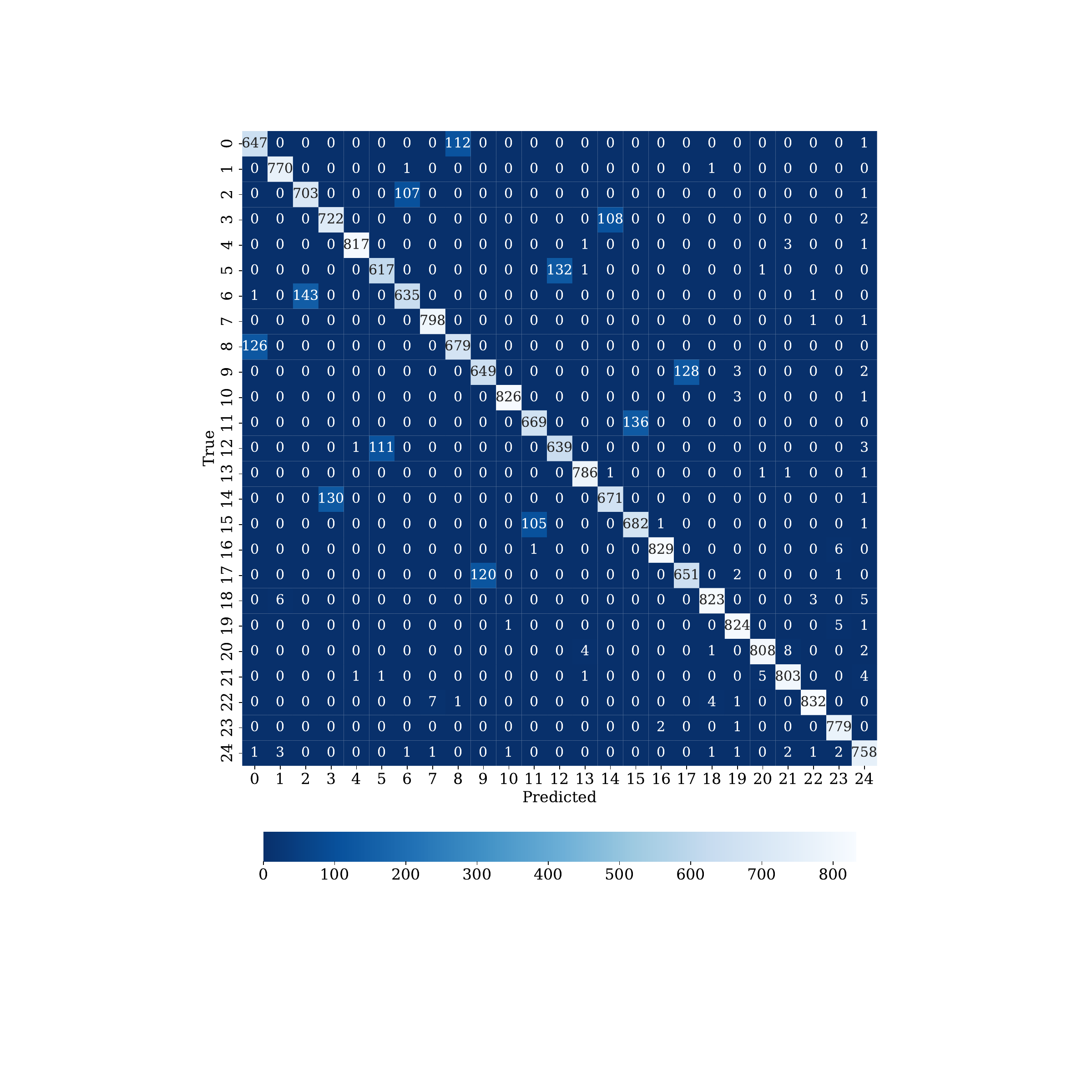}
	\caption{Confusion matrix demonstrating the classification accuracy of the hybrid network for the classes outlined in Table \ref{tab:injection_points}. The matrix differentiates between the true classes, represented by rows, and the predicted classes, indicated by columns, offering a detailed assessment of the model's ability to classify anomalous TDI data segments.}
	\label{im:Confusionmatrix}
%\end{figure}
%\begin{figure}[h!]
	\centering
  \includegraphics[trim=190pt 200pt 190pt 100pt, clip,width=0.47\textwidth]{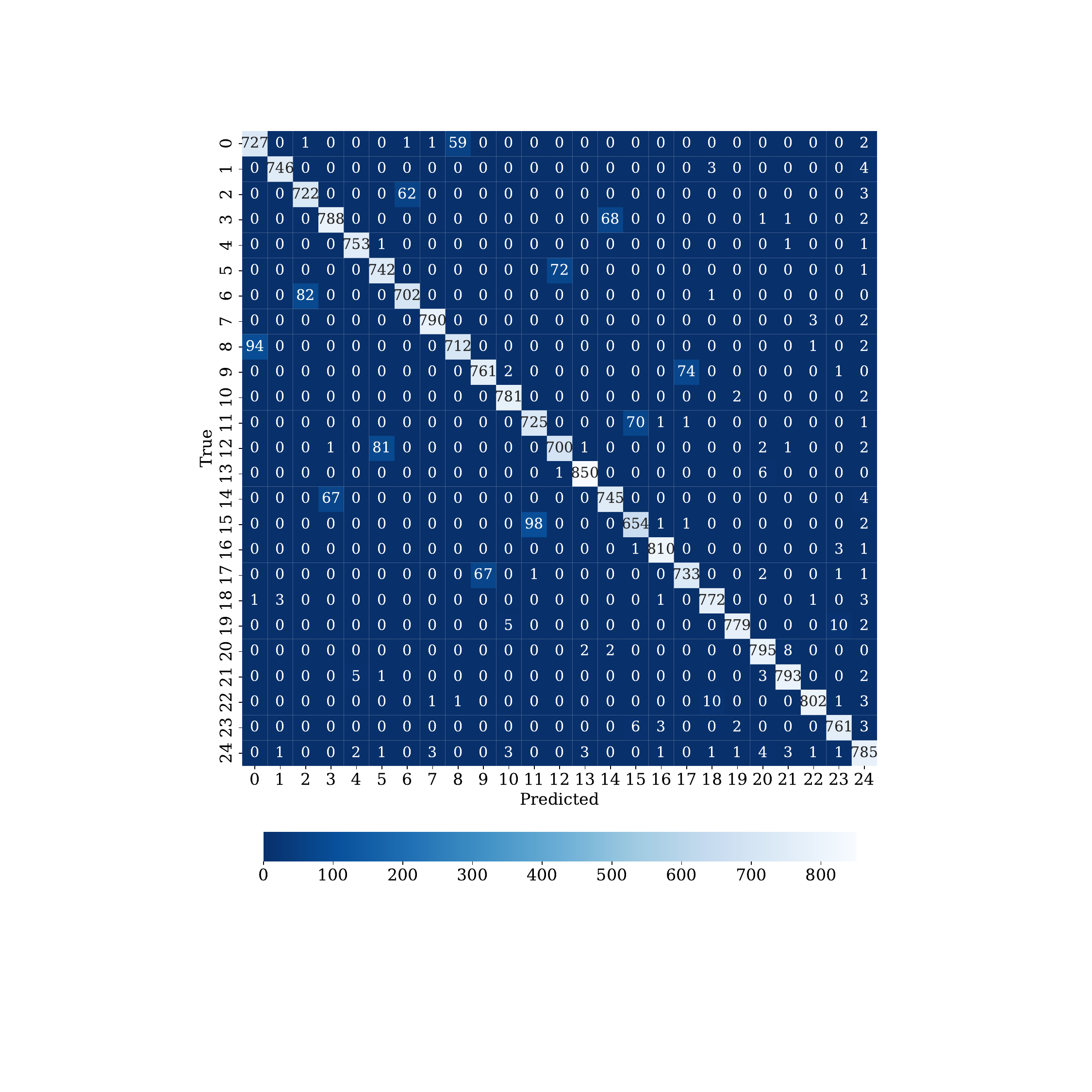}
	\caption{Confusion matrix for a standard deviation error in the injection times reduced from 3.995 sec of Fig. \ref{im:Confusionmatrix} to 3.0 sec. This reduction results in a decrease of misclassifications within the off-diagonals. }
	\label{im:Confusionmatrix30}
\end{figure}

The performance of the network is gauged using a confusion matrix, which offers a detailed view of the model's classification accuracy across the existing classes. The matrix is presented in Fig. \ref{im:Confusionmatrix}. We find gravitational wave bursts correctly classified in 98.187 \% of all cases, while 0.065 \% of FREDS$^2$ glitches are misclassified as gravitational wave bursts. Overall, the glitch injection points are correctly characterized with a 92.085 \% rate. The hybrid network model demonstrates a remarkable increase in classification correctness for stochastic glitch morphologies compared to when selecting the injection point without machine learning solely based on the highest GNR triplet sum. In this case, only 68.613 \% of the glitch origins would be correctly identified. 

The distinctive pattern observed in the three off-diagonals in Fig. \ref{im:Confusionmatrix} reveals misclassifications of optical displacement glitches between the inter-spacecraft interferometer $ij$ and the reference interferometer $ji$ of two opposite spacecraft. Optical benches of opposite spacecraft inject their glitches into the same bi-directional laser link. Consequently, they share the same glitch-unaffected TDI variable, such as TDI 2.0 Michelson $Z(t)$ for glitches originating from optical benches 12 and 21, see Table \ref{tab:TDIGlitchModelSC1}. Assuming equal arm lengths, the response to a glitch of the inter-spacecraft interferometer 12 in $X(t)$ is given by $-(\mathcal{D}^2 - 1)(\mathcal{D}^4 - 1)$, while the response to a glitch of the reference interferometer 21 in $X(t)$ would be $\mathcal{D}(\mathcal{D}^2 - 1)(\mathcal{D}^4 - 1)$. Here, the primary distinction lies in the additional delay $\mathcal{D}$ in the response related to the reference interferometer 21.
Considering that the standard deviation of the estimated injection times is 3.995 sec, and by that approximately half the order of the light travel times between two spacecraft, there are instances in which an optical glitch of the inter-spacecraft interferometer $ij$ may resemble an optical glitch of the reference interferometer $ji$ for the network. This occurs when the injection time has been overestimated. Similarly,  an optical glitch of the reference interferometer $ji$  can be mistaken for an optical glitch of the inter-spacecraft interferometer $ij$ when the injection time has been underestimated.

Note that confusion between, for instance,  optical glitches of the inter-spacecraft interferometer 12 and reference interferometer 12 is generally avoided due to their TDI response difference characterized by two delays $\mathcal{D}_{i}=\mathcal{D}_{1.2,2.1}=\mathcal{D}^2$ (where the latter notation applies for the equal arm length assumption), as outlined in Table \ref{tab:translationtable} and \ref{tab:TDIGlitchModelSC1}. Such confusion would only arise if the injection time estimation error were to increase to the order of two times the light travel time, which, fortunately, is not the case.
On the contrary, the misclassification of optical glitches between the inter-spacecraft and reference interferometer of opposite optical benches would diminish with further enhancements of the CNN's performance. For illustration, another confusion matrix is provided in Fig. \ref{im:Confusionmatrix30}, with the difference to Fig. \ref{im:Confusionmatrix} being a reduction in the standard deviation of the injection time estimation error from 3.995 sec (the performance of the previous section) to 3.0 sec.

\begin{figure}[]
	\centering
  \includegraphics[trim=100pt 0 0 0, clip,width=0.58\textwidth]{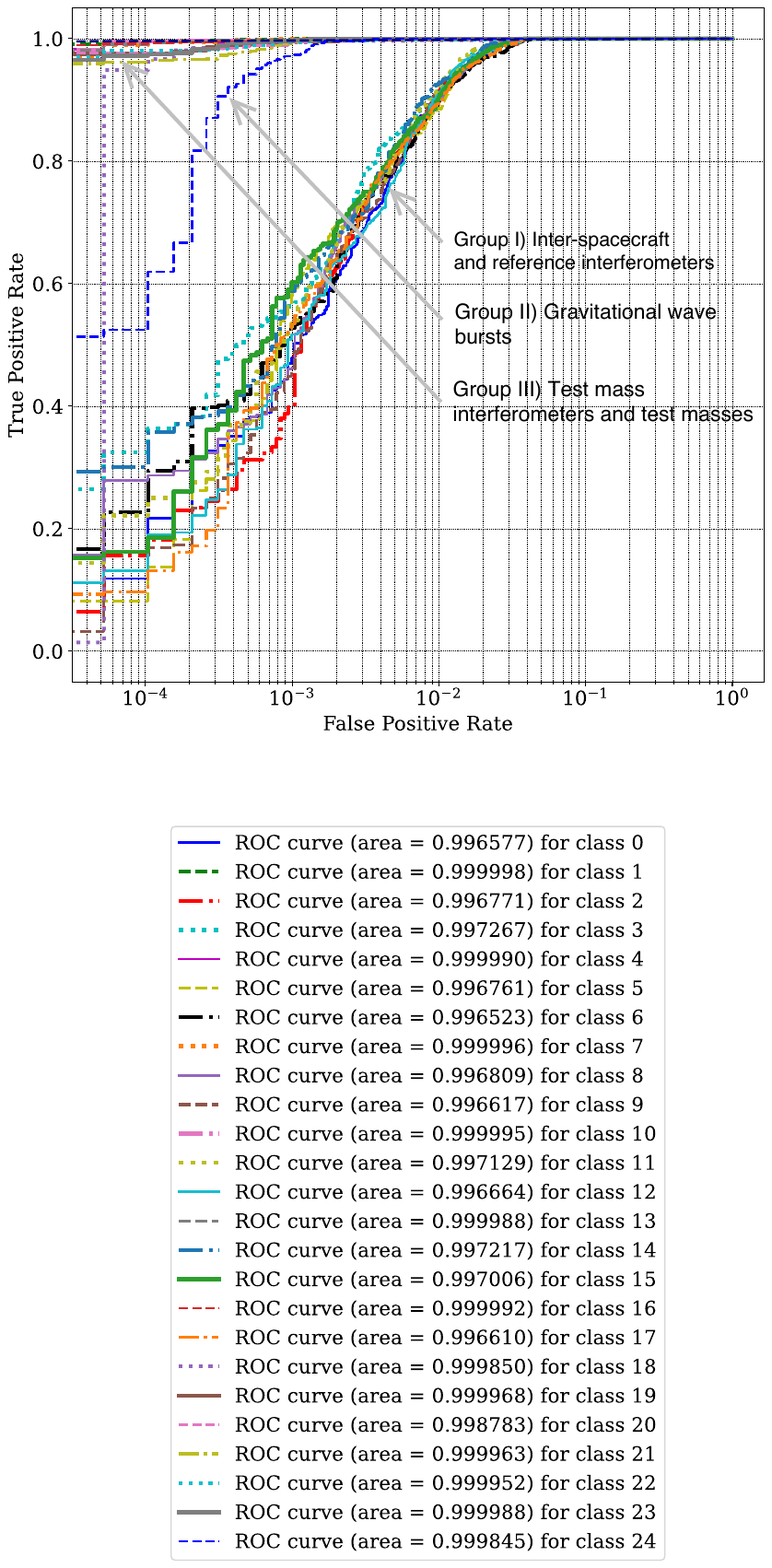}
	\caption{Semi-logarithmic ROC curves representing the model's proficiency for binary classification using the one-vs-rest approach. Translations of class labels are provided in Table \ref{tab:injection_points}. Group I corresponds to the off-diagonal elements in Fig. \ref{im:Confusionmatrix}.}
	\label{im:ROCCurve}
\end{figure}

Finally, in Fig. \ref{im:ROCCurve}, we utilize a semi-logarithmic representation of the receiver operating characteristic (ROC) curve, adapted for multi-class classification using the one-vs-rest approach. Similar to confusion matrices, ROC curves are a commonly employed metric for evaluating the performance of classification networks and are consequently presented here for reference. They visually represent the model's proficiency for binary classification in distinguishing between positive and negative classes across a spectrum of thresholds, plotting the true positive rate against the false positive rate. The area under the curve (AUC) measures the classifier's overall efficacy. An AUC value of 0.5 represents random guessing, while an AUC of 1.0 indicates perfect classification. In the one-vs-rest method, described, for example, in \cite{gimeno:hal-03447745}, each class is treated as positive, with the rest as negative, facilitating an assessment of the model's ability to differentiate between various anomaly types.

The ROC curves of Fig. \ref{im:ROCCurve} are categorized into three groups. Group I specifically relates to the misclassifications represented by the off-diagonal elements in Fig. \ref{im:Confusionmatrix}, occurring at a false positive rate of less than 0.04.
\section{Glitch Mitigation}\label{sec:level5}
After determining the glitch characteristics, including the identification of the injection point and timing through machine learning, we move to the final step, that is, glitch mitigation. The section focuses on estimating and removing the glitch from the TDI data. An LSTM network plays the key role in this process.
\subsection{Long Short-Term Memory Neural Networks}
LSTM networks are a subset of Recurrent Neural Networks (RNNs), distinguishing themselves from traditional feed-forward neural networks. In contrast to the one-directional flow of information from input to output in feed-forward networks, RNNs, including LSTMs, incorporate looping connections that preserve a hidden state or memory of prior inputs \cite{Hochreiter1997, Yu2019}. In contrast to CNNs, which are proficient in spatial recognition within images, LSTMs can make continuous predictions of time series. 
\begin{figure*}[t!]%\captionsetup{justification=centering}
	\centering
  \includegraphics[trim=0 0 0 0, clip,width=0.8\textwidth]{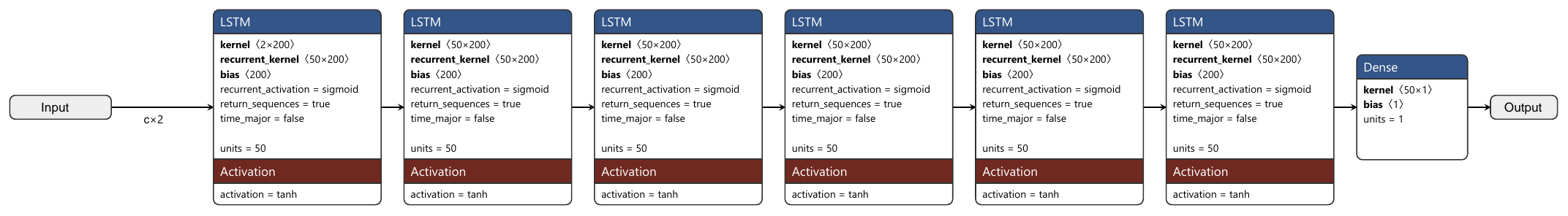}
	\caption{LSTM architecture and attributes employed for glitch time series prediction. The parameter c represents the length of the TDI inputs, which depends on the FRED model parameters acquired pre-training via differential evolution optimization.}
	\label{im:LSTMArchitecture}
\end{figure*}
LSTMs are distinct from standard RNNs in their ability to circumvent long-term dependency.
The core of their architecture lies in a system of gates – specifically, the input, forget, and output gates. These gates function as regulators of data flow within the network, enabling LSTMs to make informed decisions about what information to store, update, or discard at each step in the sequence \cite{Gers2000}. This selective memory process could make LSTMs adept at handling time-dependent and stochastic glitch patterns in TDI data.
The input gate controls the extent to which new information flows into the cell state, the internal memory of the network. The forget gate decides which information is no longer relevant and should be removed from the cell state. Finally, the output gate determines what part of the cell state should be used to compute the output at each time step. This layered approach to information processing allows LSTMs to preserve relevant historical data.

The LSTM network applied in this paper for glitch time series prediction is detailed in Fig. \ref{im:LSTMArchitecture}. It consists of multiple stacked LSTM layers, each with 50 neurons and tanh activation functions. The network culminates in a dense layer with a single neuron for continuous value prediction. It takes as input the two TDI variables affected by the glitch and produces the glitch dynamics responsible for the observed TDI data, essentially functioning as a TDI inverter from the perspective of glitches.
\subsection{Differential Evolution of FRED Model Parameters}
The FREDS$^2$ model of Eq. \eqref{eq:FREDSSGl}, blending deterministic FRED characteristics with stochastic fluctuations, presents considerable complexity. Therefore, the LSTM network shall primarily focus on the stochastic elements while incorporating as much a priori knowledge of the deterministic glitch properties as possible. Since the glitch origin is determined by the probability presented in the previous section, we can now refine our understanding of the deterministic glitch components. Therefore, differential evolution is employed for optimizing the intrinsic FRED model parameters $t_{rise}$ and $t_{fall}$ of Eq. \eqref{eq:FREDGl}. This step is vital for providing a more accurate description of the glitch's deterministic nature, which the LSTM network will incorporate into its training process.

The optimization is performed by maximizing the GNR, as defined in Eq. \eqref{eq:snrRho}. Remember, $g(t,\theta)$ denotes the TDI response to the glitch. Therefore, the glitch injection point is essential for this computation. By integrating all the information from the previous steps of the pipeline, we gain a comprehensive understanding of the glitch characteristics, which are to be made available to the LSTM. This includes selecting TDI variables based on the injection point and guiding the training with tailored data sets. The intrinsic FRED model parameters $t_{rise}$ and $t_{fall}$ are estimated similarly to the approach presented in \cite{StrubGPU} for estimating parameters of galactic binaries via differential evolution. For that, we make use of the implementation provided by \texttt{SciPy} \cite{2020SciPy-NMeth}.
\subsection{Mitigation Performance}
Due to the complexity of the glitch shapes focused on in this paper, the LSTM network of Fig. \ref{im:LSTMArchitecture} is retrained whenever there is a change in the glitch's injection point or FRED parameters. This process is computationally intensive but leads to accurate predictions. We suggest a secondary strategy to enable a quicker response for high-demand scenarios in a low-latency sense: training the LSTM with a data set encompassing a range of FRED model parameters. This approach offers ready-to-deploy generalized networks with a compromise in accuracy, as presented later.

The network is trained using the Adam optimizer with a learning rate of 0.001 and mean squared error as the loss function. Model checkpoints save the weights with the lowest loss on a validation set, ensuring peak performance. This approach is consistent with those followed for the first two neural networks in our pipeline.

\begin{figure}[]
	\centering
  \includegraphics[trim=0 0 0 0, clip,width=0.445\textwidth]{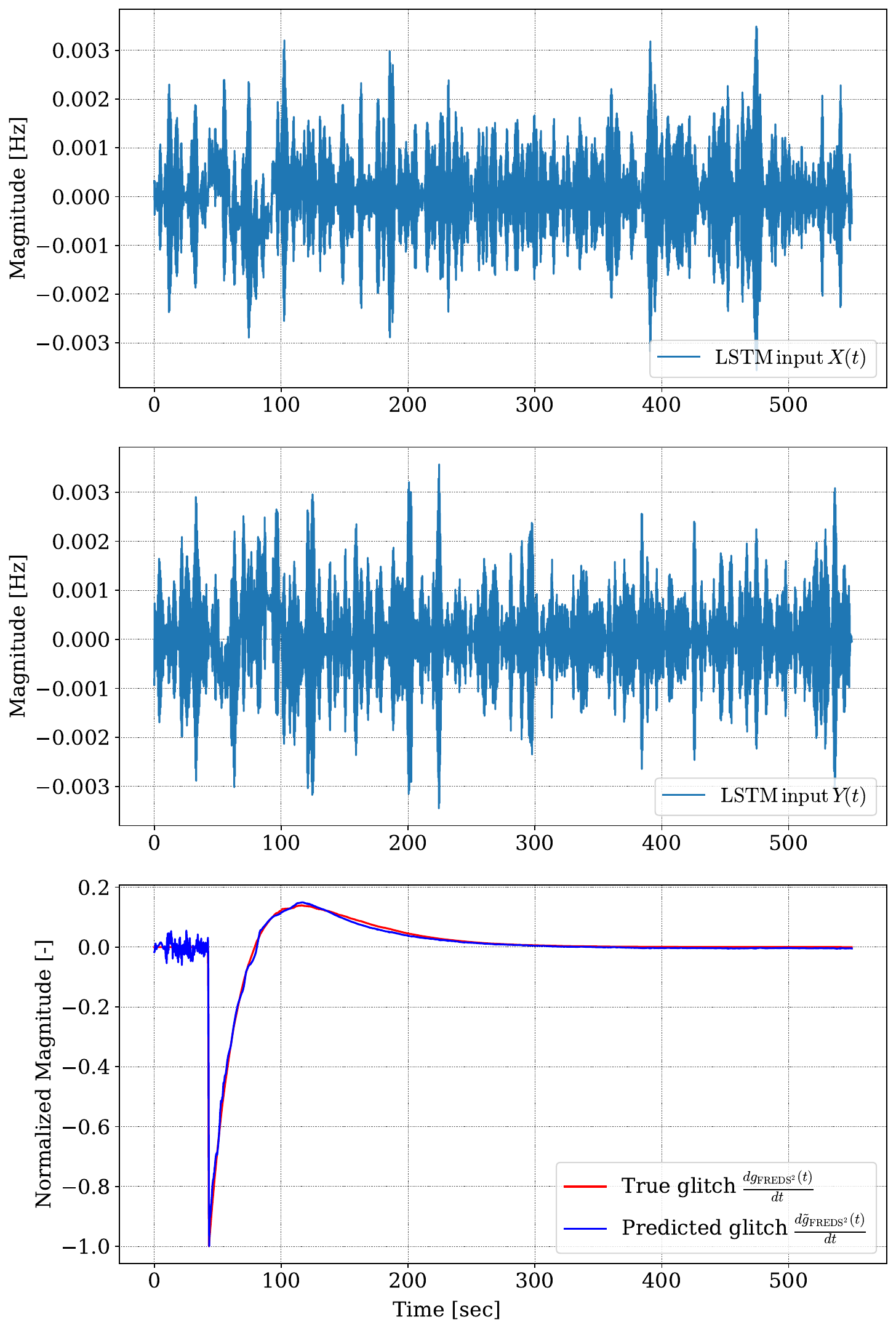}
	\caption{LSTM neural network prediction for an optical displacement glitch injected in the inter-spacecraft interferometer of optical bench 12, alongside the true glitch and the LSTM inputs. }
	\label{im:LSTMRes1}
\end{figure}

A first example showcasing the glitch prediction performance is presented in Fig. \ref{im:LSTMRes1}. The figure depicts the glitch time series prediction and the TDI inputs applying the high-accuracy LSTM model. The model was trained using a data set comprising 100,000 samples. The TDI training data set exclusively incorporates quasi-stationary noise and FREDS$^2$ glitches, with $t_{rise}$ and $t_{fall}$ estimates derived from the previously explained differential evolution process as well as the $t_{inj}$ estimate obtained from the CNN. The $t_{rise}$ and $t_{fall}$ model parameters correspond to the glitch depicted in Fig. \ref{fig:sub1P}. Note that the preceding step in the pipeline successfully identified this anomaly as a glitch within the inter-spacecraft interferometer of optical bench 12. 

\begin{figure}[]
	\centering
  \includegraphics[trim=0 0 0 0, clip,width=0.47\textwidth]{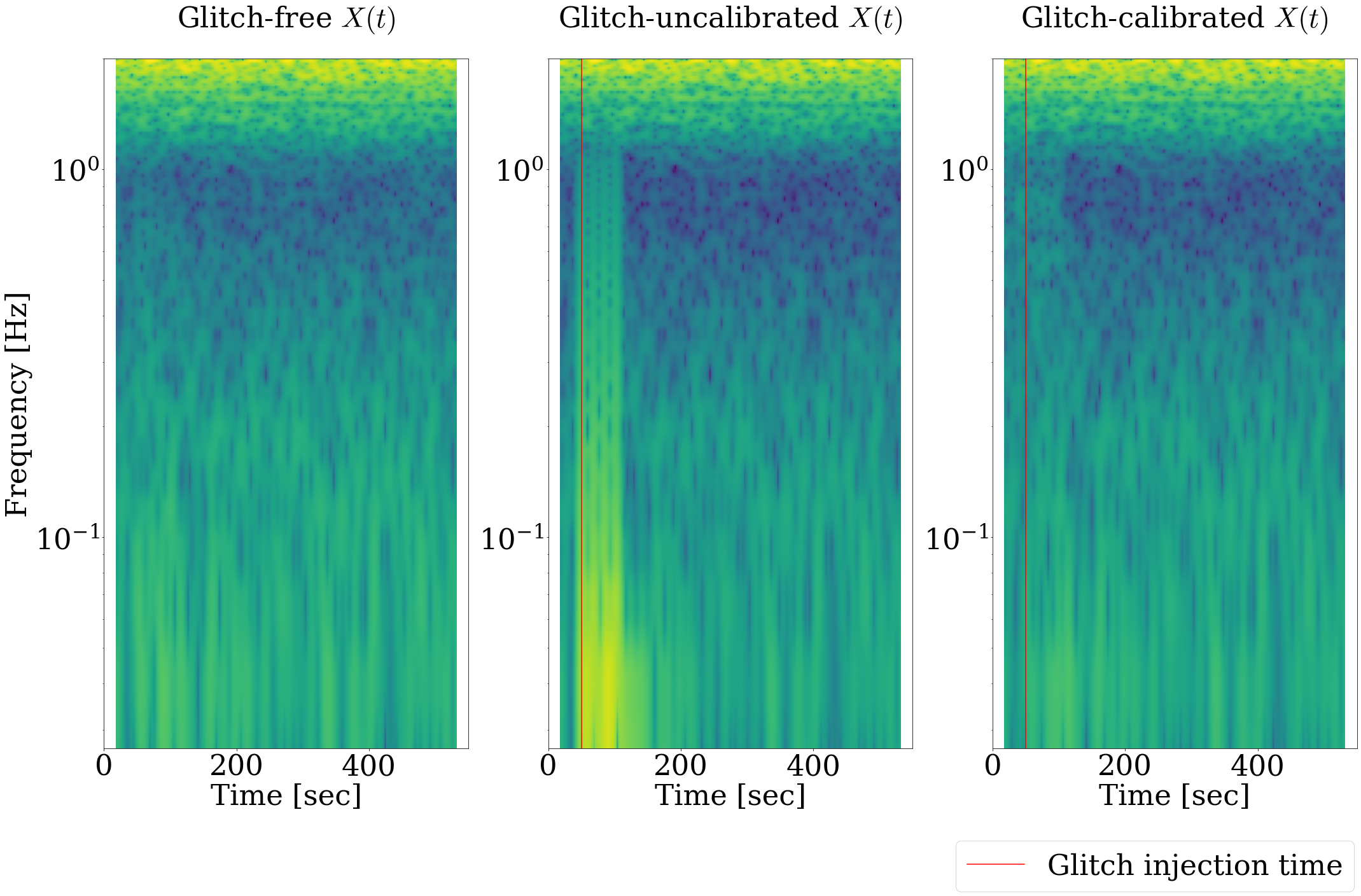}
	\caption{Semi-logarithmic time-frequency spectrograms for the scenario depicted in Fig. \ref{im:LSTMRes1}, showcasing the glitch-free, glitch-uncalibrated, and glitch-calibrated data using the glitch-affected TDI Michelson $X(t)$ channel as an illustrative example. Brighter colors indicate increased noise levels. The vertical red line denotes the moment of glitch injection.}
	\label{im:SingleHighAccuracy_Example}
\end{figure}

The spectrograms of $X(t)$ for the scenario of Fig. \ref{im:LSTMRes1} are presented in Fig. \ref{im:SingleHighAccuracy_Example}. Spectrograms serve as a visual representation of the temporal evolution of frequency content in a non-stationary signal. The figure illustrates the spectrogram before and after calibration for the achieved estimation performance. It also includes the spectrogram of the glitch-free case, representing an ideal scenario with perfect glitch mitigation. The glitch appears prominently as a bright yellow vertical stripe before calibration. It is worth noting that for frequencies greater than 1 Hz, the TDI 2.0 zero locations appear as faint horizontal blue lines. These zeros are not discernible for frequencies below 1 Hz due to the semi-logarithmic plotting and particular spectrogram calculation. To make these zero locations visible in the lower frequency range, one needs to increase the segment length of the Fourier calculation. Nonetheless, such an adjustment would further blur the contribution of the glitch. The selected segment length was carefully chosen to optimize the appearance of the glitch.

\begin{figure}[]
	\centering
  \includegraphics[trim=0 0 0 0, clip,width=0.46\textwidth]{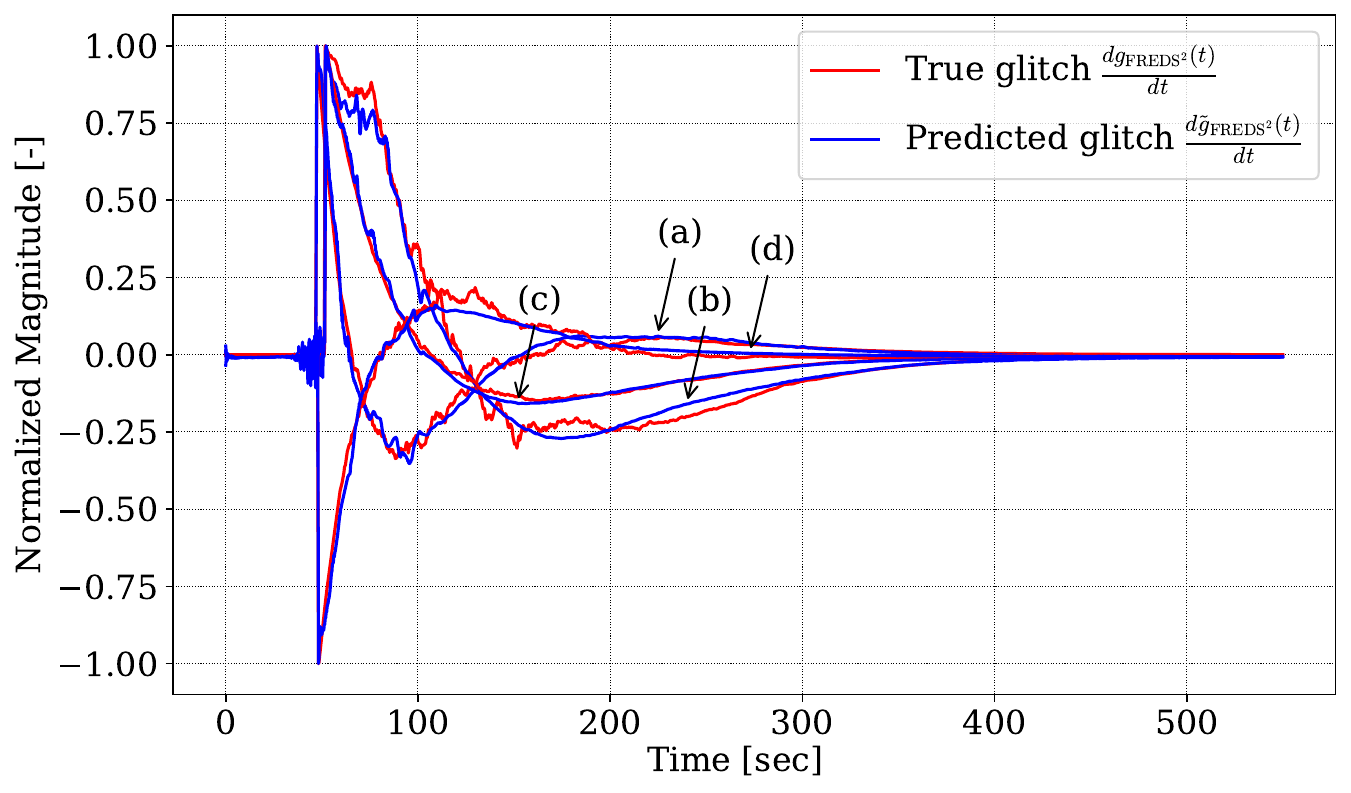}
	\caption{Performance comparison of the LSTM neural network in predicting the glitch of Fig. \ref{im:LSTMRes1} under conditions of amplified stochastic fluctuations. The TDI inputs for the LSTM predictions are not depicted. The spectrograms for (a) to (d) can be found in the appendix.}
	\label{im:LSTM_overviewHA}
\end{figure}

Figure \ref{im:LSTM_overviewHA} presents additional realizations of this glitch with amplified levels of stochastic fluctuations. The corresponding spectrograms can be found in the appendix.

Figure \ref{im:LSTM_overviewQR} provides a comparison with predictions obtained from the low-latency LSTM neural network. Although the quick-response variant may not attain the same level of precision as the high-accuracy alternative, it stands as a practical option for TDI glitch calibration in real-time situations requiring swift action and alert services. Additionally, it may offer benefits in Athena-LISA interactions, as suggested in \cite{Fabian2019AthenaLISAS}.

\newpage
In the final step of the calibration pipeline, the prediction for the specific glitch under consideration is subtracted from the affected interferometer measurement. Subsequently, the glitch-calibrated TDI data streams are calculated. To evaluate the effectiveness of the mitigation process, we subject the glitch-calibrated TDI data to a re-analysis by the anomaly detection step. In 93.93 \% of cases, the network classifies the calibrated data set as anomaly-free. One could explore the potential of iterating through the pipeline to increase this value further. In each iteration, a prediction of the (residual) glitch in the underlying TDI channels would be generated. The final step would involve subtracting the sum of these predictions from the original data.
\begin{figure}[]
	\centering
  \includegraphics[trim=0 0 0 0, clip,width=0.46\textwidth]{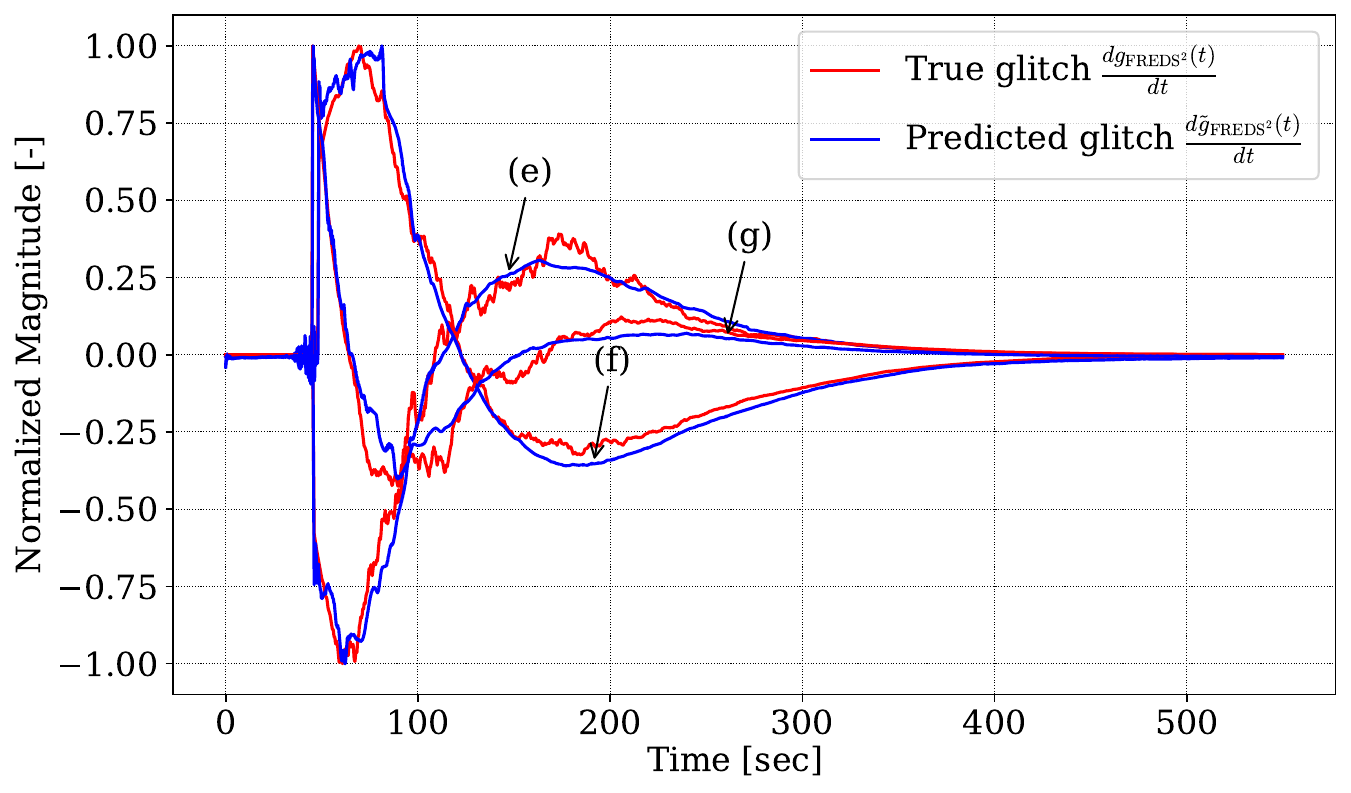}
	\caption{Performance evaluation of the LSTM neural network in the quick response mode considerable for low-latency pipeline applications with reduced emphasis on accuracy. The spectrograms for (e) to (g) can be found in the appendix.}
	\label{im:LSTM_overviewQR}
\end{figure}

\begin{figure*}[]
%\captionsetup{justification=centering}
	\centering
  \includegraphics[trim=240pt 650pt 315pt 20pt, clip,width=0.95\textwidth]{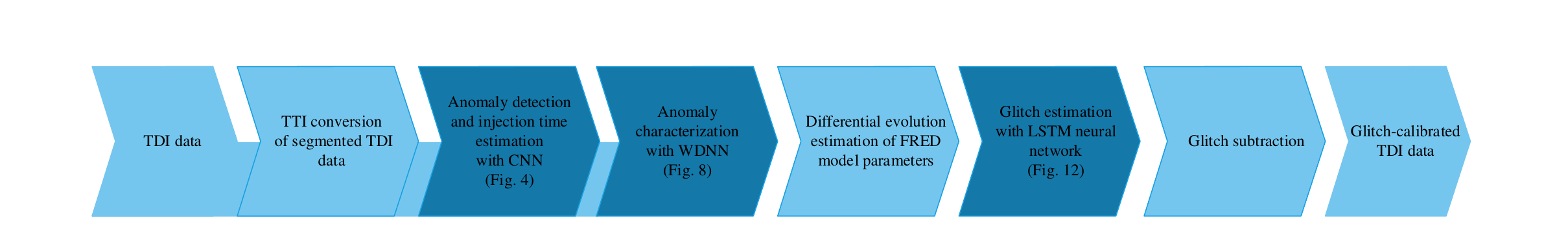}
	\caption{Outline of the glitch detection and mitigation process as detailed in this paper, with the steps involving machine learning highlighted in dark blue.}
	\label{im:pipeline}
\end{figure*}
\noindent
Training the LSTM network requires significantly more time than training the other two network architectures. As a result, we have yet to perform tests for all glitch injection points and the entire parameter space. So far, we have run two additional tests, one involving a test mass acceleration glitch associated with optical bench 32 and another focusing on an optical displacement glitch related to the reference interferometer of optical bench 13. We conducted a thorough analysis of all three scenarios using two different glitch parameter sets for each, one with a 20-sec duration and the other lasting 13 min. All three scenarios exhibited comparable glitch residuals, with the 13-min glitch predictions slightly less accurate. To assess the overall mitigation performance, further simulations must be conducted, encompassing various glitch types and a broader range of deterministic FRED model parameters. Additionally, it would be advantageous to compare the residual glitch noise against a specific requirement to ascertain the pipeline's global performance.%\enlargethispage{2\baselineskip}
\section{Conclusion}\label{sec:level6}
The study introduced a data calibration pipeline designed to detect, characterize, and mitigate glitches for the Laser  Interferometer Space Antenna (LISA) mission. The pipeline is structured around an ensemble of  three neural networks. An overview of the glitch detection and mitigation process is given in Fig. \ref{im:pipeline}.

The first component of our approach employs a convolutional neural network (CNN) with the capability to identify anomalies within LISA's time-delay interferometry (TDI) time series data. The CNN correctly identified non-anomalous data 100 \% of the time and achieved a 99.99915 \% accuracy in detecting anomalous events. It faced challenges in precisely pinpointing the onset of faint and slow-rising anomalies.

In the second phase, we employed a hybrid neural network, which distinguishes between gravitational wave bursts and glitches while accurately identifying the locations where glitches have been injected. The hybrid neural network correctly classified gravitational wave bursts in 98.187 \% of cases and characterized glitch injection points with a 92.085 \% success rate, surpassing a conventional reference method by roughly 20 \%.

Then, the calibration pipeline incorporates a long short-term memory (LSTM) neural network to estimate glitch time series based on noisy TDI input data and information obtained by the preceding networks. The LSTM network is designed to operate in a high accuracy or quick response mode, depending on the configuration of the training data sets.

The final step involves subtracting the estimated stochastic glitch transient from the affected interferometric measurement to obtain glitch-calibrated TDI variables. A re-analysis of the calibrated TDI data set was performed to verify the overall performance of our approach. The calibrated TDI data streams were classified as anomaly-free at 93.93 \%. Iteratively executing the pipeline could further increase this value and is under investigation.

Future efforts will focus on tackling the challenge of overlapping gravitational wave signals and glitches, as this is another area that demands careful attention.
%extending the machine-learning approach to autonomously determine the number of anomalies within a TDI segment when multiple anomalies are expected to be encountered. This functionality would be similar to that found in reversible jump Markov chain Monte Carlo methods oftentimes employed in LISA literature. 
Moreover, the exploration of Tensor Processing Units (TPUs) is underway. TPUs are specialized hardware accelerators designed for machine learning applications. Renowned for their processing power and efficiency in managing computational workloads, they are expected to further enhance the performance of the calibration pipeline, primarily by enabling the handling of larger training data sets and expediting the evaluation of diverse network architectures.
%Then, it is interesting to assess the residual glitch noise level that occurs for LSTM estimations based on incorrectly characterized glitch injection points. 
\begin{acknowledgments}
%\vspace{-5pt}
The authors thank Michele Vallisneri, Stefan Strub, Martina Muratore, and Jean-Baptiste Bayle for the valuable discussions. Moreover, Jean-Baptiste Bayle and Martin Staab for their practical support with \texttt{lisagwresponse} and \texttt{pyTDI}. The authors acknowledge support from GW-Learn, a project funded through a Sinergia grant from the Swiss National Science Foundation.
\end{acknowledgments}

\appendix
\section*{Appendix}
\begin{figure}[h!]
%captionsetup{justification=justified,singlelinecheck=false}
\captionsetup[subfigure]{justification=centering,singlelinecheck=false}
\begin{subfigure}{\columnwidth}
  \includegraphics[width=\linewidth]{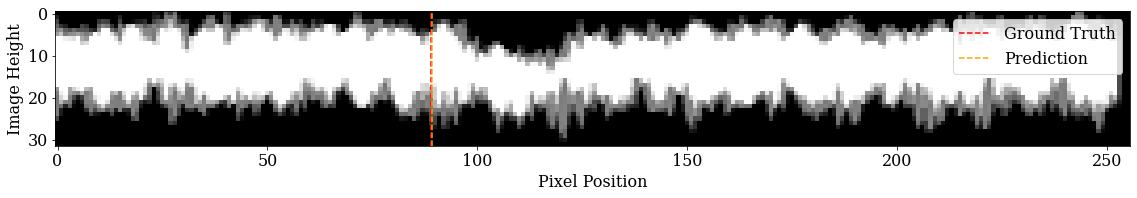}
  \caption{}
  \label{fig:sub5P}
\end{subfigure}
\begin{subfigure}{\columnwidth}
  \includegraphics[width=\linewidth]{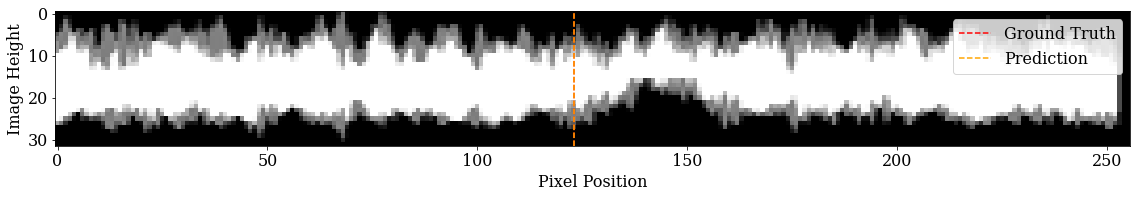}
  \caption{}
  \label{fig:sub6P}
\end{subfigure}
\begin{subfigure}{\columnwidth}
  \includegraphics[width=\linewidth]{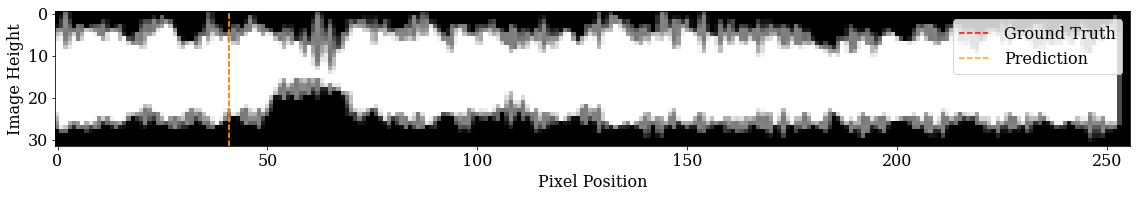}
  \caption{}
  \label{fig:sub7P}
\end{subfigure}
\caption{Examples of estimated anomaly injection times for glitch characteristics comparable to Fig. \ref{fig:sub1P}.}\label{fig:3ex}
\vspace{20pt}
%\end{figure}
%\newpage
%\begin{figure}[]
	\centering
  \includegraphics[trim=0 0 0 0, clip,width=0.47\textwidth]{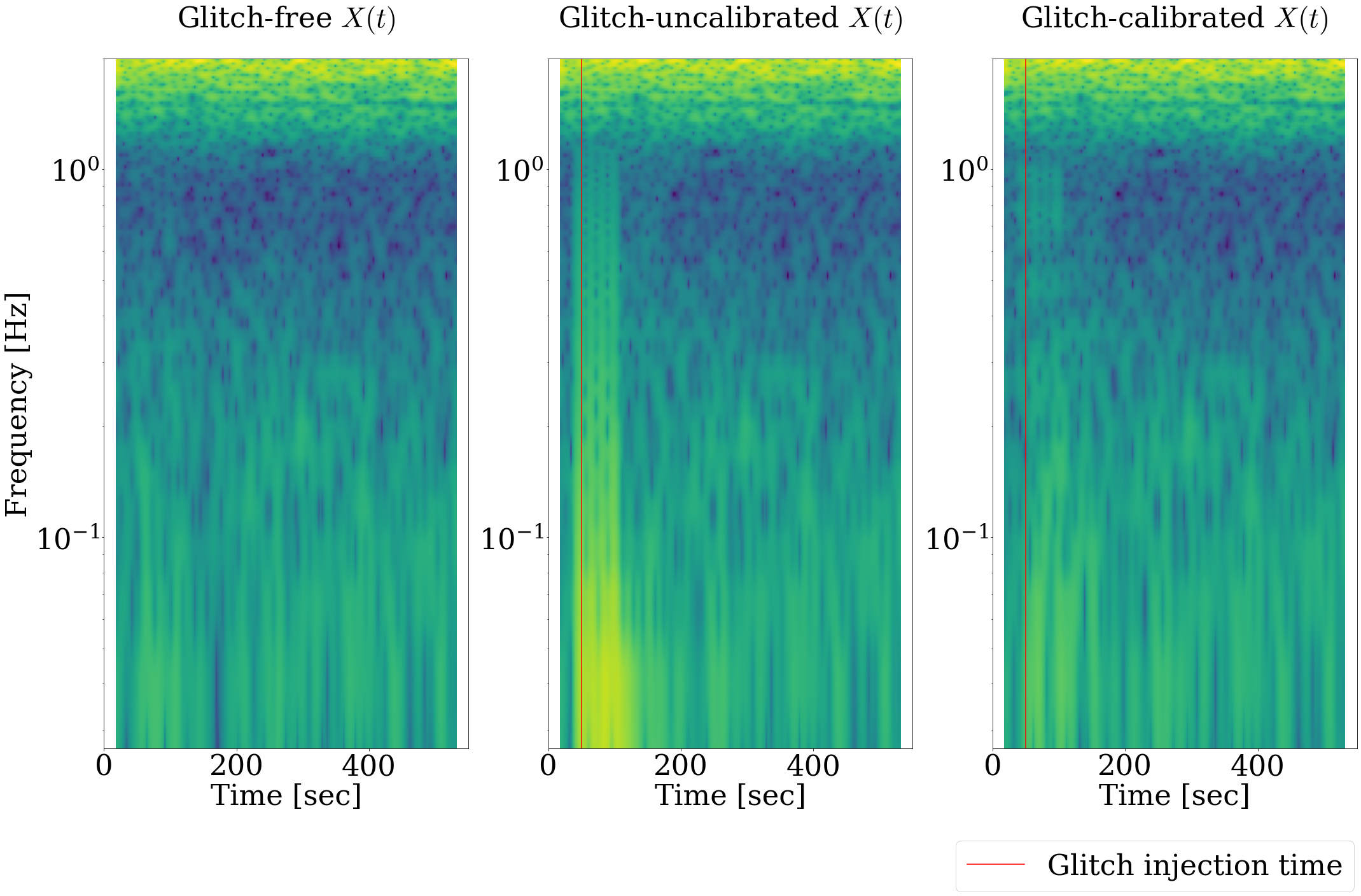}
	\caption{Spectrograms of $X(t)$ for case (a) of Fig. \ref{im:LSTM_overviewHA}.}
	\label{im:Spec_a}
 \vspace{20pt}
%\end{figure}
%\begin{figure}[]
	\centering
  \includegraphics[trim=0 0 0 0, clip,width=0.47\textwidth]{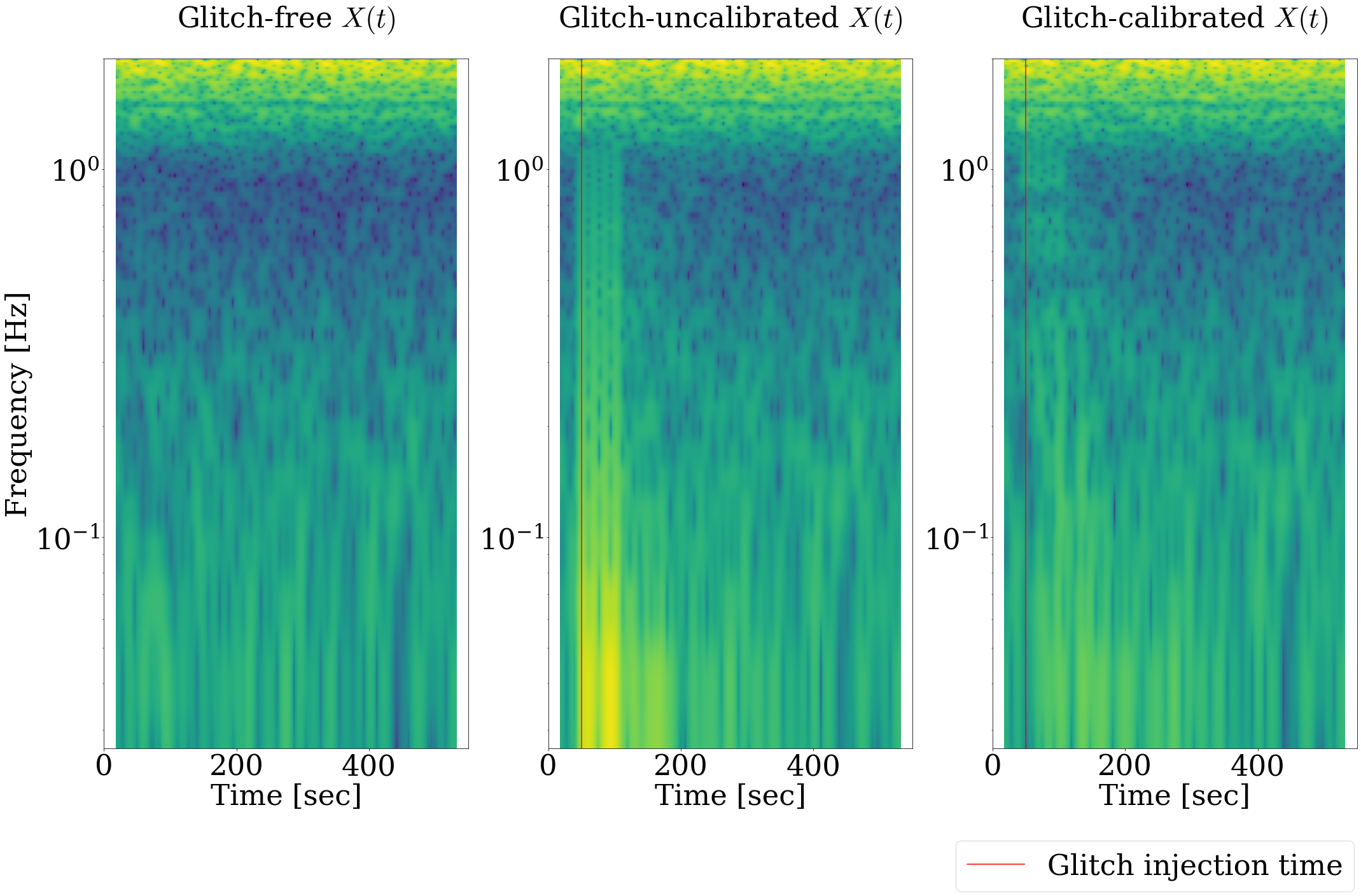}
	\caption{Spectrograms of $X(t)$ for case (b) of Fig. \ref{im:LSTM_overviewHA}.}
	\label{im:Spec_b}
  \vspace{20pt}
\end{figure}

\begin{figure}[]
	\centering
  \includegraphics[trim=0 0 0 0, clip,width=0.47\textwidth]{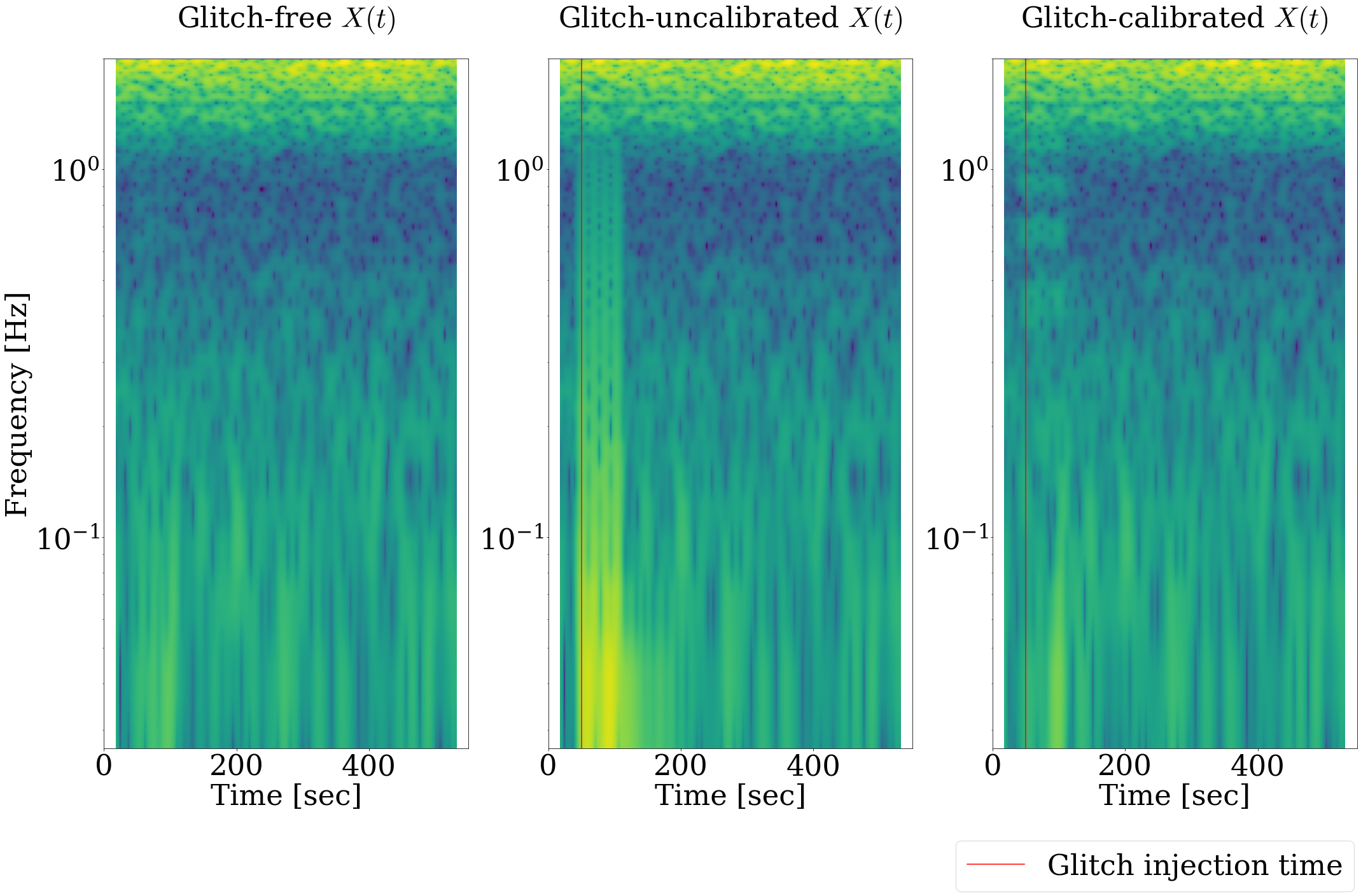}
	\caption{Spectrograms of $X(t)$ for case (c) of Fig. \ref{im:LSTM_overviewHA}.}
	\label{im:Spec_c}
  \vspace{20pt}
%\end{figure}
%\begin{figure}[]
	\centering
  \includegraphics[trim=0 0 0 0, clip,width=0.47\textwidth]{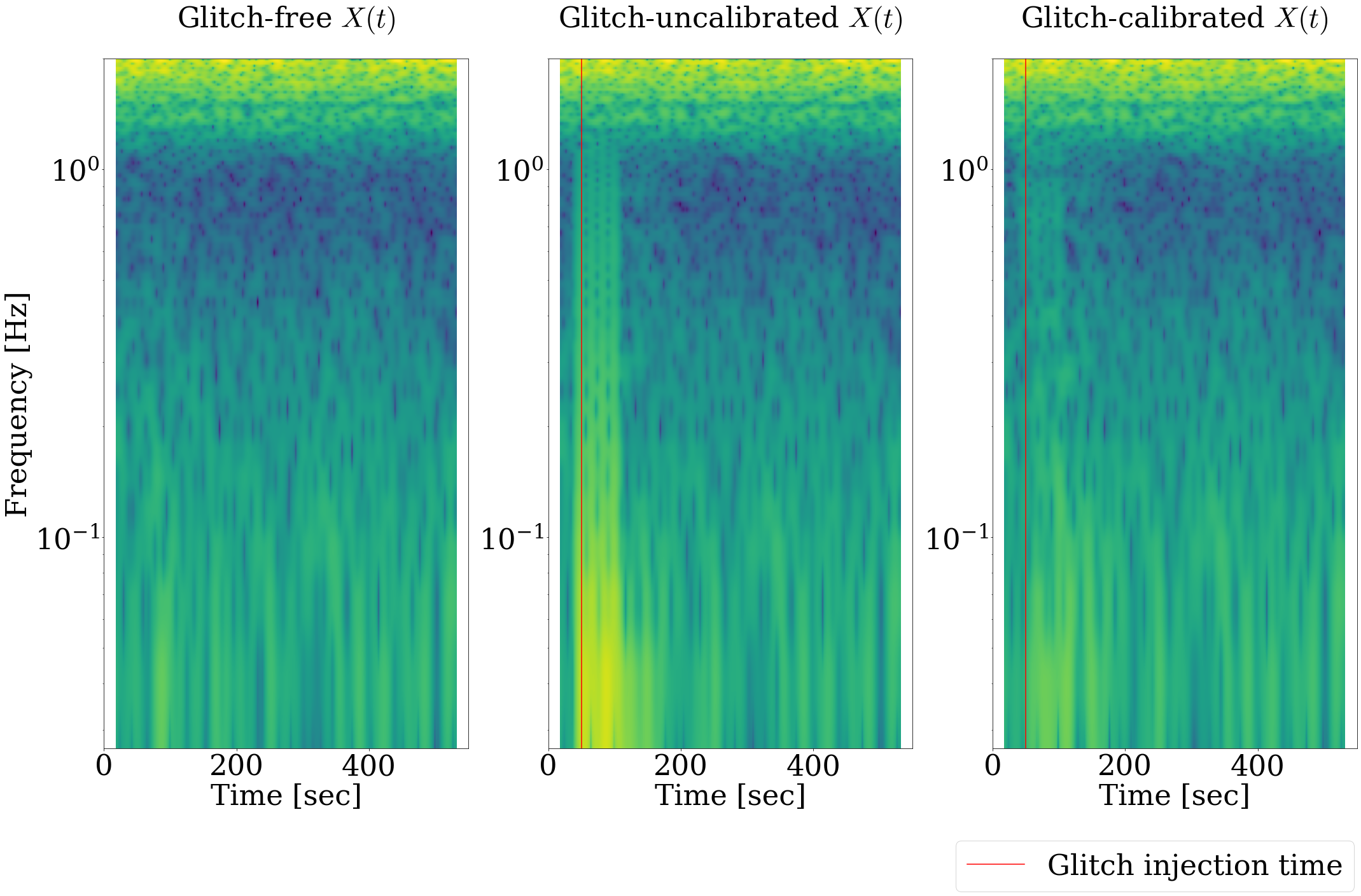}
	\caption{Spectrograms of $X(t)$ for case (d) of Fig. \ref{im:LSTM_overviewHA}.}
	\label{im:Spec_d}
  \vspace{20pt}
%\end{figure}
%\begin{figure}[]
	\centering
  \includegraphics[trim=0 0 0 0, clip,width=0.47\textwidth]{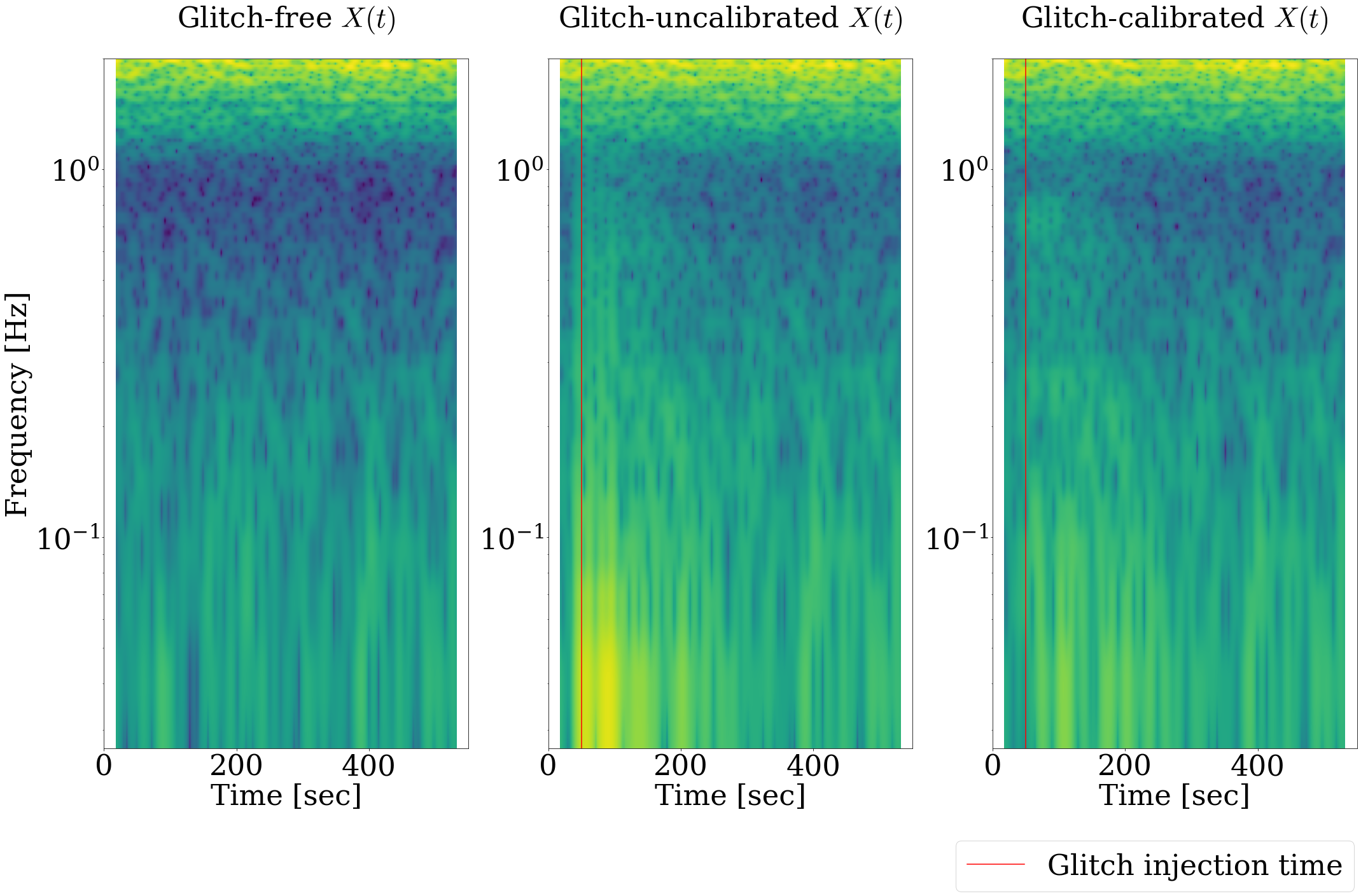}
	\caption{Spectrograms of $X(t)$ for case (e) of Fig. \ref{im:LSTM_overviewQR}.}
	\label{im:Spec_e}
  \vspace{20pt}
\end{figure}

\begin{figure}[]
	\centering
  \includegraphics[trim=0 0 0 0, clip,width=0.47\textwidth]{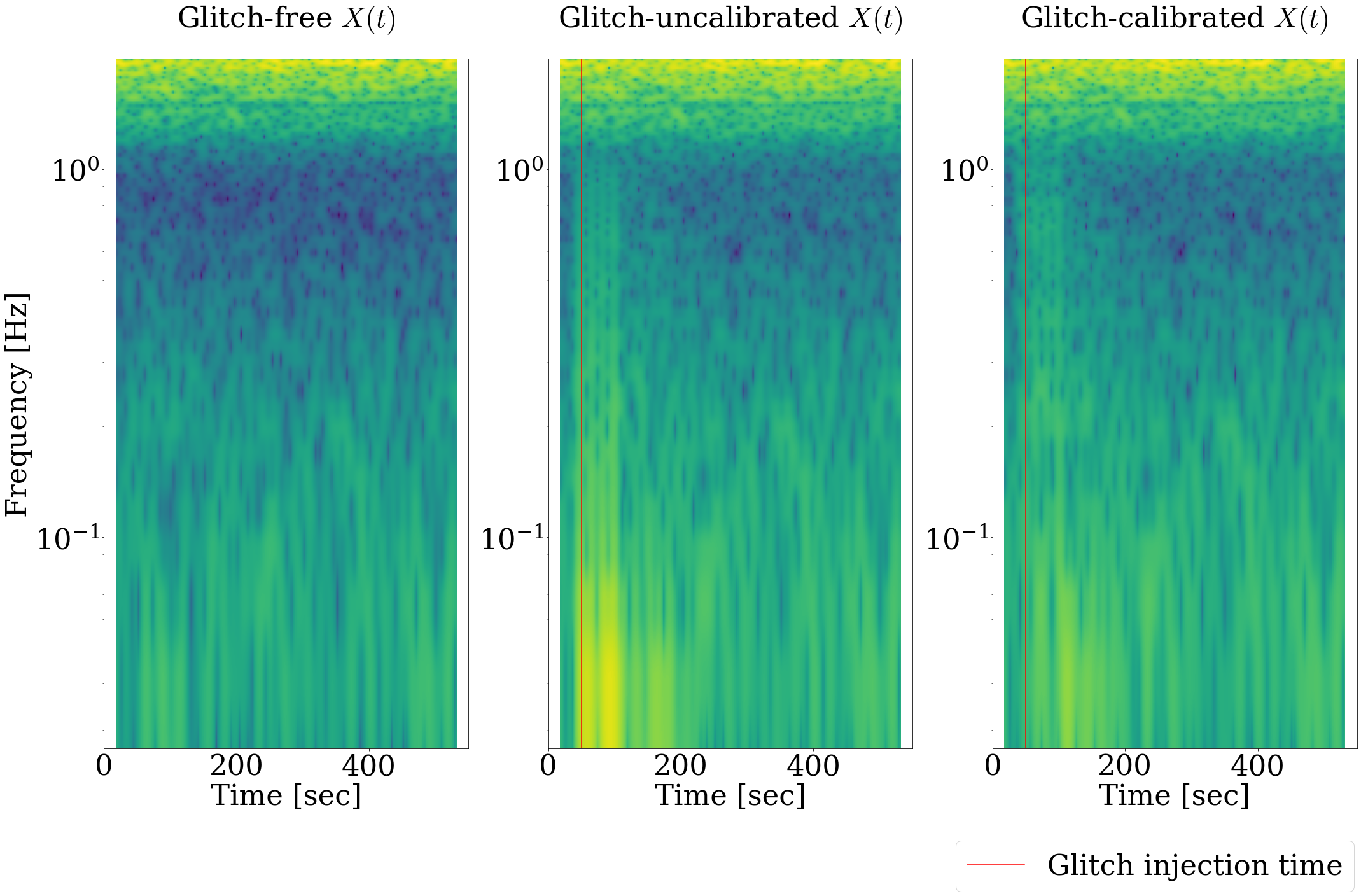}
	\caption{Spectrograms of $X(t)$ for case (f) of Fig. \ref{im:LSTM_overviewQR}.}
	\label{im:Spec_f}
\end{figure}

\begin{figure}[]
	\centering
  \includegraphics[trim=0 0 0 0, clip,width=0.47\textwidth]{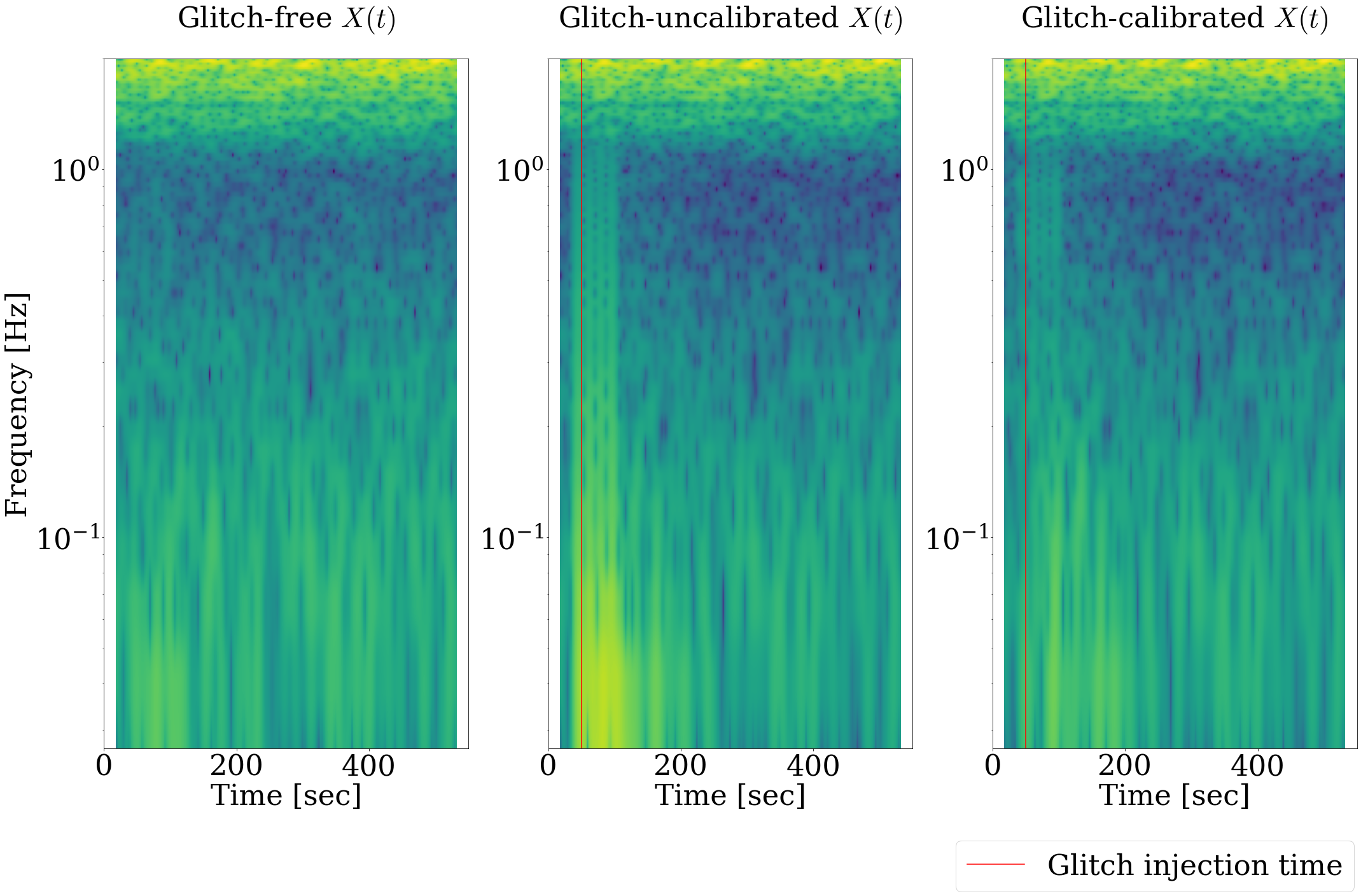}
	\caption{Spectrograms of $X(t)$ for case (g) of Fig. \ref{im:LSTM_overviewQR}.}
	\label{im:Spec_g}
\end{figure}

%\vspace{-5pt}
We conducted additional evaluations to validate the performance of anomaly injection time estimation under conditions similar to those in Fig. \ref{fig:sub1P}.  Figure \ref{fig:3ex} displays three such comparable scenarios where visual detection of the exact injection time is challenging but managed by the CNN of Fig. \ref{im:CNNArchitecture}. 
The spectrograms for the TDI 2.0 Michelson $X(t)$ data sets of Figs. \ref{im:LSTM_overviewHA} and \ref{im:LSTM_overviewQR} are given by Figs. \ref{im:Spec_a} to \ref{im:Spec_g}.
%\vspace{6pt}
\begin{table*}[t]
\captionsetup{justification=centering}
\caption{Translation of neural network labels.}
\begin{center}
\begin{tabular}{ccc}
\hline\hline
\textbf{Label} & \textbf{Optical Bench} & \textbf{Injection Point} \\
\hline
0 & 12 & Inter-spacecraft interferometer \\
1 & 12 & Test mass interferometer \\
2 & 12 & Reference interferometer \\
3 & 13 & Inter-spacecraft interferometer \\
4 & 13 & Test mass interferometer \\
5 & 13 & Reference interferometer \\
6 & 21 & Inter-spacecraft interferometer \\
7 & 21 & Test mass interferometer \\
8 & 21 & Reference interferometer \\
9 & 23 & Inter-spacecraft interferometer \\
10 & 23 & Test mass interferometer \\
11 & 23 & Reference interferometer \\
12 & 31 & Inter-spacecraft interferometer \\
13 & 31 & Test mass interferometer \\
14 & 31 & Reference interferometer \\
15 & 32 & Inter-spacecraft interferometer \\
16 & 32 & Test mass interferometer \\
17 & 32 & Reference interferometer \\
18 & 12 & Test mass \\
19 & 23 & Test mass \\
20 & 31 & Test mass \\
21 & 13 & Test mass \\
22 & 21 & Test mass \\
23 & 32 & Test mass \\
\cline{2-3}
24 & \multicolumn{2}{c}{Gravitational wave burst} \\ 
\hline\hline
\end{tabular}
\end{center}
\label{tab:injection_points}
\end{table*}

%Table \ref{tab:injection_points} provides the physical meaning of numerical class labels used by the glitch characterization network. 
%\bibliographystyle{plainnat}
\newpage
\bibliography{apssamp}% Produces the bibliography via BibTeX.

\begin{thebibliography}{63}%
\makeatletter
\providecommand \@ifxundefined [1]{%
 \@ifx{#1\undefined}
}%
\providecommand \@ifnum [1]{%
 \ifnum #1\expandafter \@firstoftwo
 \else \expandafter \@secondoftwo
 \fi
}%
\providecommand \@ifx [1]{%
 \ifx #1\expandafter \@firstoftwo
 \else \expandafter \@secondoftwo
 \fi
}%
\providecommand \natexlab [1]{#1}%
\providecommand \enquote  [1]{``#1''}%
\providecommand \bibnamefont  [1]{#1}%
\providecommand \bibfnamefont [1]{#1}%
\providecommand \citenamefont [1]{#1}%
\providecommand \href@noop [0]{\@secondoftwo}%
\providecommand \href [0]{\begingroup \@sanitize@url \@href}%
\providecommand \@href[1]{\@@startlink{#1}\@@href}%
\providecommand \@@href[1]{\endgroup#1\@@endlink}%
\providecommand \@sanitize@url [0]{\catcode `\\12\catcode `\$12\catcode `\&12\catcode `\#12\catcode `\^12\catcode `\_12\catcode `\%12\relax}%
\providecommand \@@startlink[1]{}%
\providecommand \@@endlink[0]{}%
\providecommand \url  [0]{\begingroup\@sanitize@url \@url }%
\providecommand \@url [1]{\endgroup\@href {#1}{\urlprefix }}%
\providecommand \urlprefix  [0]{URL }%
\providecommand \Eprint [0]{\href }%
\providecommand \doibase [0]{https://doi.org/}%
\providecommand \selectlanguage [0]{\@gobble}%
\providecommand \bibinfo  [0]{\@secondoftwo}%
\providecommand \bibfield  [0]{\@secondoftwo}%
\providecommand \translation [1]{[#1]}%
\providecommand \BibitemOpen [0]{}%
\providecommand \bibitemStop [0]{}%
\providecommand \bibitemNoStop [0]{.\EOS\space}%
\providecommand \EOS [0]{\spacefactor3000\relax}%
\providecommand \BibitemShut  [1]{\csname bibitem#1\endcsname}%
\let\auto@bib@innerbib\@empty
%</preamble>
\bibitem [{\citenamefont {Amaro-Seoane}\ \emph {et~al.}(2017)\citenamefont {Amaro-Seoane}, \citenamefont {Audley}, \citenamefont {Babak}, \citenamefont {Baker}, \citenamefont {Barausse}, \citenamefont {Bender}, \citenamefont {Berti}, \citenamefont {Binetruy}, \citenamefont {Born}, \citenamefont {Bortoluzzi} \emph {et~al.}}]{amaroseoane2017laser}%
  \BibitemOpen
  \bibfield  {author} {\bibinfo {author} {\bibfnamefont {P.}~\bibnamefont {Amaro-Seoane}}, \bibinfo {author} {\bibfnamefont {H.}~\bibnamefont {Audley}}, \bibinfo {author} {\bibfnamefont {S.}~\bibnamefont {Babak}}, \bibinfo {author} {\bibfnamefont {J.}~\bibnamefont {Baker}}, \bibinfo {author} {\bibfnamefont {E.}~\bibnamefont {Barausse}}, \bibinfo {author} {\bibfnamefont {P.}~\bibnamefont {Bender}}, \bibinfo {author} {\bibfnamefont {E.}~\bibnamefont {Berti}}, \bibinfo {author} {\bibfnamefont {P.}~\bibnamefont {Binetruy}}, \bibinfo {author} {\bibfnamefont {M.}~\bibnamefont {Born}}, \bibinfo {author} {\bibfnamefont {D.}~\bibnamefont {Bortoluzzi}}, \emph {et~al.},\ }\bibfield  {title} {\bibinfo {title} {{Laser Interferometer Space Antenna}},\ }\href@noop {} {\bibfield  {journal} {\bibinfo  {journal} {arXiv e-prints}\ ,\ \bibinfo {eid} {arXiv:1702.00786}} (\bibinfo {year} {02/2017})},\ \Eprint {https://arxiv.org/abs/1702.00786} {arXiv:1702.00786 [astro-ph.IM]} \BibitemShut {NoStop}%
\bibitem [{\citenamefont {Thorpe}\ \emph {et~al.}(2006)\citenamefont {Thorpe}, \citenamefont {Cruz}, \citenamefont {Hartman},\ and\ \citenamefont {Mueller}}]{Thorpe2006}%
  \BibitemOpen
  \bibfield  {author} {\bibinfo {author} {\bibfnamefont {J.}~\bibnamefont {Thorpe}}, \bibinfo {author} {\bibfnamefont {R.~J.}\ \bibnamefont {Cruz}}, \bibinfo {author} {\bibfnamefont {M.}~\bibnamefont {Hartman}},\ and\ \bibinfo {author} {\bibfnamefont {G.}~\bibnamefont {Mueller}},\ }\bibfield  {title} {\bibinfo {title} {{Arm‐Locking in a LISA‐like Hardware Model: A Status Report}},\ }\href {https://doi.org/10.1063/1.2405114} {\ \textbf {\bibinfo {volume} {873}},\ \bibinfo {pages} {661} (\bibinfo {year} {2006})}\BibitemShut {NoStop}%
\bibitem [{\citenamefont {Shaddock}(2008)}]{Shaddock2008}%
  \BibitemOpen
  \bibfield  {author} {\bibinfo {author} {\bibfnamefont {D.}~\bibnamefont {Shaddock}},\ }\bibfield  {title} {\bibinfo {title} {{Space-based gravitational wave detection with LISA}},\ }\href {https://doi.org/10.1088/0264-9381/25/11/114012} {\bibfield  {journal} {\bibinfo  {journal} {Classical and Quantum Gravity}\ }\textbf {\bibinfo {volume} {25}},\ \bibinfo {pages} {114012} (\bibinfo {year} {2008})}\BibitemShut {NoStop}%
\bibitem [{\citenamefont {Dhurandhar}\ \emph {et~al.}(2010)\citenamefont {Dhurandhar}, \citenamefont {Nayak},\ and\ \citenamefont {Vinet}}]{Dhurandhar2010}%
  \BibitemOpen
  \bibfield  {author} {\bibinfo {author} {\bibfnamefont {S.}~\bibnamefont {Dhurandhar}}, \bibinfo {author} {\bibfnamefont {K.~R.}\ \bibnamefont {Nayak}},\ and\ \bibinfo {author} {\bibfnamefont {J.}~\bibnamefont {Vinet}},\ }\bibfield  {title} {\bibinfo {title} {{Time-delay interferometry for LISA with one arm dysfunctional}},\ }\href {https://doi.org/10.1088/0264-9381/27/13/135013} {\bibfield  {journal} {\bibinfo  {journal} {Classical and Quantum Gravity}\ }\textbf {\bibinfo {volume} {27}},\ \bibinfo {pages} {135013} (\bibinfo {year} {2010})}\BibitemShut {NoStop}%
\bibitem [{\citenamefont {Martens}\ and\ \citenamefont {Joffre}(2021)}]{Martens2021}%
  \BibitemOpen
  \bibfield  {author} {\bibinfo {author} {\bibfnamefont {W.}~\bibnamefont {Martens}}\ and\ \bibinfo {author} {\bibfnamefont {E.}~\bibnamefont {Joffre}},\ }\bibfield  {title} {\bibinfo {title} {{Trajectory Design for the {ESA} {LISA} Mission}},\ }\href {https://doi.org/10.1007/s40295-021-00263-2} {\bibfield  {journal} {\bibinfo  {journal} {The Journal of the Astronautical Sciences}\ }\textbf {\bibinfo {volume} {68}},\ \bibinfo {pages} {402} (\bibinfo {year} {2021})}\BibitemShut {NoStop}%
\bibitem [{\citenamefont {Bayle}\ \emph {et~al.}(2018{\natexlab{a}})\citenamefont {Bayle}, \citenamefont {Lilley}, \citenamefont {Petiteau},\ and\ \citenamefont {Halloin}}]{Bayle2018a}%
  \BibitemOpen
  \bibfield  {author} {\bibinfo {author} {\bibfnamefont {J.-B.}\ \bibnamefont {Bayle}}, \bibinfo {author} {\bibfnamefont {M.}~\bibnamefont {Lilley}}, \bibinfo {author} {\bibfnamefont {A.}~\bibnamefont {Petiteau}},\ and\ \bibinfo {author} {\bibfnamefont {H.}~\bibnamefont {Halloin}},\ }\bibfield  {title} {\bibinfo {title} {{Analytic Model and Simulations of Residual Laser Noise after Time-Delay Interferometry in LISA}},\ }\href@noop {} {\bibfield  {journal} {\bibinfo  {journal} {arXiv: Instrumentation and Methods for Astrophysics}\ } (\bibinfo {year} {2018}{\natexlab{a}})}\BibitemShut {NoStop}%
\bibitem [{\citenamefont {Bayle}\ \emph {et~al.}(2018{\natexlab{b}})\citenamefont {Bayle}, \citenamefont {Lilley}, \citenamefont {Petiteau},\ and\ \citenamefont {Halloin}}]{Bayle2018b}%
  \BibitemOpen
  \bibfield  {author} {\bibinfo {author} {\bibfnamefont {J.-B.}\ \bibnamefont {Bayle}}, \bibinfo {author} {\bibfnamefont {M.}~\bibnamefont {Lilley}}, \bibinfo {author} {\bibfnamefont {A.}~\bibnamefont {Petiteau}},\ and\ \bibinfo {author} {\bibfnamefont {H.}~\bibnamefont {Halloin}},\ }\bibfield  {title} {\bibinfo {title} {{Effect of filters on the time-delay interferometry residual laser noise for LISA}},\ }\href {https://doi.org/10.1103/PHYSREVD.99.084023} {\bibfield  {journal} {\bibinfo  {journal} {Physical Review D}\ }\textbf {\bibinfo {volume} {99}},\ \bibinfo {pages} {084023} (\bibinfo {year} {2018}{\natexlab{b}})}\BibitemShut {NoStop}%
\bibitem [{\citenamefont {Tinto}\ \emph {et~al.}(2002)\citenamefont {Tinto}, \citenamefont {Estabrook},\ and\ \citenamefont {Armstrong}}]{Tinto2002TDI1stGen}%
  \BibitemOpen
  \bibfield  {author} {\bibinfo {author} {\bibfnamefont {M.}~\bibnamefont {Tinto}}, \bibinfo {author} {\bibfnamefont {F.~B.}\ \bibnamefont {Estabrook}},\ and\ \bibinfo {author} {\bibfnamefont {J.~W.}\ \bibnamefont {Armstrong}},\ }\bibfield  {title} {\bibinfo {title} {Time-delay interferometry for {LISA}},\ }\bibfield  {journal} {\bibinfo  {journal} {Physical Review D}\ }\textbf {\bibinfo {volume} {65}},\ \href {https://doi.org/10.1103/physrevd.65.082003} {10.1103/physrevd.65.082003} (\bibinfo {year} {2002})\BibitemShut {NoStop}%
\bibitem [{\citenamefont {Tinto}\ and\ \citenamefont {Dhurandhar}(2005)}]{TDITINTO2005}%
  \BibitemOpen
  \bibfield  {author} {\bibinfo {author} {\bibfnamefont {M.}~\bibnamefont {Tinto}}\ and\ \bibinfo {author} {\bibfnamefont {S.~V.}\ \bibnamefont {Dhurandhar}},\ }\bibfield  {title} {\bibinfo {title} {{Time-Delay Interferometry}},\ }\bibfield  {journal} {\bibinfo  {journal} {Living Reviews in Relativity}\ }\textbf {\bibinfo {volume} {8}},\ \href {https://doi.org/10.12942/lrr-2005-4} {10.12942/lrr-2005-4} (\bibinfo {year} {2005})\BibitemShut {NoStop}%
\bibitem [{\citenamefont {Hartig}\ \emph {et~al.}(2023)\citenamefont {Hartig}, \citenamefont {Schuster}, \citenamefont {Heinzel},\ and\ \citenamefont {Wanner}}]{hartig2021nongeometric}%
  \BibitemOpen
  \bibfield  {author} {\bibinfo {author} {\bibfnamefont {M.-S.}\ \bibnamefont {Hartig}}, \bibinfo {author} {\bibfnamefont {S.}~\bibnamefont {Schuster}}, \bibinfo {author} {\bibfnamefont {G.}~\bibnamefont {Heinzel}},\ and\ \bibinfo {author} {\bibfnamefont {G.}~\bibnamefont {Wanner}},\ }\href {https://doi.org/10.1088/2040-8986/acc3ac} {\bibinfo {title} {Non-geometric tilt-to-length coupling in precision interferometry: mechanisms and analytical descriptions}} (\bibinfo {year} {2023})\BibitemShut {NoStop}%
\bibitem [{\citenamefont {Hartig}\ \emph {et~al.}(2022)\citenamefont {Hartig}, \citenamefont {Schuster},\ and\ \citenamefont {Wanner}}]{hartig2022geometric}%
  \BibitemOpen
  \bibfield  {author} {\bibinfo {author} {\bibfnamefont {M.-S.}\ \bibnamefont {Hartig}}, \bibinfo {author} {\bibfnamefont {S.}~\bibnamefont {Schuster}},\ and\ \bibinfo {author} {\bibfnamefont {G.}~\bibnamefont {Wanner}},\ }\href {https://doi.org/10.1088/2040-8986/ac675e} {\bibinfo {title} {Geometric tilt-to-length coupling in precision interferometry: mechanisms and analytical descriptions}} (\bibinfo {year} {2022})\BibitemShut {NoStop}%
\bibitem [{\citenamefont {Armano}\ \emph {et~al.}(2018)\citenamefont {Armano}, \citenamefont {Audley}, \citenamefont {Baird}, \citenamefont {Binetruy}, \citenamefont {Born}, \citenamefont {Bortoluzzi}, \citenamefont {Castelli}, \citenamefont {Cavalleri}, \citenamefont {Cesarini}, \citenamefont {Cruise} \emph {et~al.}}]{Armano2018}%
  \BibitemOpen
  \bibfield  {author} {\bibinfo {author} {\bibfnamefont {M.}~\bibnamefont {Armano}}, \bibinfo {author} {\bibfnamefont {H.}~\bibnamefont {Audley}}, \bibinfo {author} {\bibfnamefont {J.}~\bibnamefont {Baird}}, \bibinfo {author} {\bibfnamefont {P.}~\bibnamefont {Binetruy}}, \bibinfo {author} {\bibfnamefont {M.}~\bibnamefont {Born}}, \bibinfo {author} {\bibfnamefont {D.}~\bibnamefont {Bortoluzzi}}, \bibinfo {author} {\bibfnamefont {E.}~\bibnamefont {Castelli}}, \bibinfo {author} {\bibfnamefont {A.}~\bibnamefont {Cavalleri}}, \bibinfo {author} {\bibfnamefont {A.}~\bibnamefont {Cesarini}}, \bibinfo {author} {\bibfnamefont {A.~M.}\ \bibnamefont {Cruise}}, \emph {et~al.},\ }\bibfield  {title} {\bibinfo {title} {{Beyond the Required LISA Free-Fall Performance: New LISA Pathfinder Results down to $20\text{ }\text{ }\ensuremath{\mu}\mathrm{Hz}$}},\ }\href {https://doi.org/10.1103/PhysRevLett.120.061101} {\bibfield  {journal} {\bibinfo  {journal} {Phys. Rev. Lett.}\ }\textbf {\bibinfo {volume} {120}},\ \bibinfo {pages}
  {061101} (\bibinfo {year} {2018})}\BibitemShut {NoStop}%
\bibitem [{\citenamefont {Robson}\ and\ \citenamefont {Cornish}(2019)}]{Robson2019}%
  \BibitemOpen
  \bibfield  {author} {\bibinfo {author} {\bibfnamefont {T.}~\bibnamefont {Robson}}\ and\ \bibinfo {author} {\bibfnamefont {N.~J.}\ \bibnamefont {Cornish}},\ }\bibfield  {title} {\bibinfo {title} {{Detecting gravitational wave bursts with LISA in the presence of instrumental glitches}},\ }\bibfield  {journal} {\bibinfo  {journal} {Physical Review D}\ }\textbf {\bibinfo {volume} {99}},\ \href {https://doi.org/10.1103/physrevd.99.024019} {10.1103/physrevd.99.024019} (\bibinfo {year} {2019})\BibitemShut {NoStop}%
\bibitem [{\citenamefont {de~Vine}\ \emph {et~al.}(2010)\citenamefont {de~Vine}, \citenamefont {Ware}, \citenamefont {McKenzie}, \citenamefont {Spero}, \citenamefont {Klipstein},\ and\ \citenamefont {Shaddock}}]{PhysRevLett.104.211103}%
  \BibitemOpen
  \bibfield  {author} {\bibinfo {author} {\bibfnamefont {G.}~\bibnamefont {de~Vine}}, \bibinfo {author} {\bibfnamefont {B.}~\bibnamefont {Ware}}, \bibinfo {author} {\bibfnamefont {K.}~\bibnamefont {McKenzie}}, \bibinfo {author} {\bibfnamefont {R.~E.}\ \bibnamefont {Spero}}, \bibinfo {author} {\bibfnamefont {W.~M.}\ \bibnamefont {Klipstein}},\ and\ \bibinfo {author} {\bibfnamefont {D.~A.}\ \bibnamefont {Shaddock}},\ }\bibfield  {title} {\bibinfo {title} {{Experimental Demonstration of Time-Delay Interferometry for the Laser Interferometer Space Antenna}},\ }\href {https://doi.org/10.1103/PhysRevLett.104.211103} {\bibfield  {journal} {\bibinfo  {journal} {Phys. Rev. Lett.}\ }\textbf {\bibinfo {volume} {104}},\ \bibinfo {pages} {211103} (\bibinfo {year} {2010})}\BibitemShut {NoStop}%
\bibitem [{\citenamefont {Mitryk}\ \emph {et~al.}(2010)\citenamefont {Mitryk}, \citenamefont {Wand},\ and\ \citenamefont {Mueller}}]{Mitryk2010}%
  \BibitemOpen
  \bibfield  {author} {\bibinfo {author} {\bibfnamefont {S.}~\bibnamefont {Mitryk}}, \bibinfo {author} {\bibfnamefont {V.}~\bibnamefont {Wand}},\ and\ \bibinfo {author} {\bibfnamefont {G.}~\bibnamefont {Mueller}},\ }\bibfield  {title} {\bibinfo {title} {{Verification of time-delay interferometry techniques using the University of Florida LISA interferometry simulator}},\ }\href {https://doi.org/10.1088/0264-9381/27/8/084012} {\bibfield  {journal} {\bibinfo  {journal} {Classical and Quantum Gravity}\ }\textbf {\bibinfo {volume} {27}},\ \bibinfo {pages} {084012} (\bibinfo {year} {2010})}\BibitemShut {NoStop}%
\bibitem [{\citenamefont {Paczkowski}\ \emph {et~al.}(2022)\citenamefont {Paczkowski}, \citenamefont {Giusteri}, \citenamefont {Hewitson}, \citenamefont {Karnesis}, \citenamefont {Fitzsimons}, \citenamefont {Wanner},\ and\ \citenamefont {Heinzel}}]{PhysRevD.106.042005}%
  \BibitemOpen
  \bibfield  {author} {\bibinfo {author} {\bibfnamefont {S.}~\bibnamefont {Paczkowski}}, \bibinfo {author} {\bibfnamefont {R.}~\bibnamefont {Giusteri}}, \bibinfo {author} {\bibfnamefont {M.}~\bibnamefont {Hewitson}}, \bibinfo {author} {\bibfnamefont {N.}~\bibnamefont {Karnesis}}, \bibinfo {author} {\bibfnamefont {E.~D.}\ \bibnamefont {Fitzsimons}}, \bibinfo {author} {\bibfnamefont {G.}~\bibnamefont {Wanner}},\ and\ \bibinfo {author} {\bibfnamefont {G.}~\bibnamefont {Heinzel}},\ }\bibfield  {title} {\bibinfo {title} {{Postprocessing subtraction of tilt-to-length noise in LISA}},\ }\href {https://doi.org/10.1103/PhysRevD.106.042005} {\bibfield  {journal} {\bibinfo  {journal} {Phys. Rev. D}\ }\textbf {\bibinfo {volume} {106}},\ \bibinfo {pages} {042005} (\bibinfo {year} {2022})}\BibitemShut {NoStop}%
\bibitem [{\citenamefont {Houba}\ \emph {et~al.}(2022{\natexlab{a}})\citenamefont {Houba}, \citenamefont {Delchambre}, \citenamefont {Ziegler}, \citenamefont {Hechenblaikner},\ and\ \citenamefont {Fichter}}]{HoubaPRD}%
  \BibitemOpen
  \bibfield  {author} {\bibinfo {author} {\bibfnamefont {N.}~\bibnamefont {Houba}}, \bibinfo {author} {\bibfnamefont {S.}~\bibnamefont {Delchambre}}, \bibinfo {author} {\bibfnamefont {T.}~\bibnamefont {Ziegler}}, \bibinfo {author} {\bibfnamefont {G.}~\bibnamefont {Hechenblaikner}},\ and\ \bibinfo {author} {\bibfnamefont {W.}~\bibnamefont {Fichter}},\ }\bibfield  {title} {\bibinfo {title} {{LISA spacecraft maneuver design to estimate tilt-to-length noise during gravitational wave events}},\ }\href {https://doi.org/10.1103/PhysRevD.106.022004} {\bibfield  {journal} {\bibinfo  {journal} {Physical review D}\ }\textbf {\bibinfo {volume} {106}} (\bibinfo {year} {2022}{\natexlab{a}})}\BibitemShut {NoStop}%
\bibitem [{\citenamefont {Houba}\ \emph {et~al.}(2023)\citenamefont {Houba}, \citenamefont {Delchambre}, \citenamefont {Hechenblaikner}, \citenamefont {Ziegler},\ and\ \citenamefont {Fichter}}]{houbaJoGEngNote}%
  \BibitemOpen
  \bibfield  {author} {\bibinfo {author} {\bibfnamefont {N.}~\bibnamefont {Houba}}, \bibinfo {author} {\bibfnamefont {S.}~\bibnamefont {Delchambre}}, \bibinfo {author} {\bibfnamefont {G.}~\bibnamefont {Hechenblaikner}}, \bibinfo {author} {\bibfnamefont {T.}~\bibnamefont {Ziegler}},\ and\ \bibinfo {author} {\bibfnamefont {W.}~\bibnamefont {Fichter}},\ }\bibfield  {title} {\bibinfo {title} {Time-delay interferometry infinity for tilt-to-length noise estimation in {LISA}},\ }\href {https://doi.org/10.1088/1361-6382/accbfc} {\bibfield  {journal} {\bibinfo  {journal} {Classical and Quantum Gravity}\ }\textbf {\bibinfo {volume} {40}},\ \bibinfo {pages} {107001} (\bibinfo {year} {2023})}\BibitemShut {NoStop}%
\bibitem [{\citenamefont {Armano}\ \emph {et~al.}(2019)\citenamefont {Armano}, \citenamefont {Audley}, \citenamefont {Baird}, \citenamefont {Binetruy}, \citenamefont {Born}, \citenamefont {Bortoluzzi}, \citenamefont {Castelli}, \citenamefont {Cavalleri}, \citenamefont {Cesarini}, \citenamefont {Cruise} \emph {et~al.}}]{armano2019lisaPF}%
  \BibitemOpen
  \bibfield  {author} {\bibinfo {author} {\bibfnamefont {M.}~\bibnamefont {Armano}}, \bibinfo {author} {\bibfnamefont {H.}~\bibnamefont {Audley}}, \bibinfo {author} {\bibfnamefont {J.}~\bibnamefont {Baird}}, \bibinfo {author} {\bibfnamefont {P.}~\bibnamefont {Binetruy}}, \bibinfo {author} {\bibfnamefont {M.}~\bibnamefont {Born}}, \bibinfo {author} {\bibfnamefont {D.}~\bibnamefont {Bortoluzzi}}, \bibinfo {author} {\bibfnamefont {E.}~\bibnamefont {Castelli}}, \bibinfo {author} {\bibfnamefont {A.}~\bibnamefont {Cavalleri}}, \bibinfo {author} {\bibfnamefont {A.}~\bibnamefont {Cesarini}}, \bibinfo {author} {\bibfnamefont {M.}~\bibnamefont {Cruise}}, \emph {et~al.},\ }\href@noop {} {\bibinfo {title} {{LISA Pathfinder}}} (\bibinfo {year} {2019}),\ \Eprint {https://arxiv.org/abs/1903.08924} {arXiv:1903.08924 [astro-ph.IM]} \BibitemShut {NoStop}%
\bibitem [{\citenamefont {Bortoluzzi}\ \emph {et~al.}(2009)\citenamefont {Bortoluzzi}, \citenamefont {Baglivo}, \citenamefont {Benedetti}, \citenamefont {Biral}, \citenamefont {Bosetti}, \citenamefont {Cavalleri}, \citenamefont {Da~Lio}, \citenamefont {De~Cecco}, \citenamefont {Dolesi}, \citenamefont {Lapolla}, \citenamefont {Weber},\ and\ \citenamefont {Vitale}}]{Bortoluzzi2009}%
  \BibitemOpen
  \bibfield  {author} {\bibinfo {author} {\bibfnamefont {D.}~\bibnamefont {Bortoluzzi}}, \bibinfo {author} {\bibfnamefont {L.}~\bibnamefont {Baglivo}}, \bibinfo {author} {\bibfnamefont {M.}~\bibnamefont {Benedetti}}, \bibinfo {author} {\bibfnamefont {F.}~\bibnamefont {Biral}}, \bibinfo {author} {\bibfnamefont {P.}~\bibnamefont {Bosetti}}, \bibinfo {author} {\bibfnamefont {A.}~\bibnamefont {Cavalleri}}, \bibinfo {author} {\bibfnamefont {M.}~\bibnamefont {Da~Lio}}, \bibinfo {author} {\bibfnamefont {M.}~\bibnamefont {De~Cecco}}, \bibinfo {author} {\bibfnamefont {R.}~\bibnamefont {Dolesi}}, \bibinfo {author} {\bibfnamefont {M.}~\bibnamefont {Lapolla}}, \bibinfo {author} {\bibfnamefont {W.}~\bibnamefont {Weber}},\ and\ \bibinfo {author} {\bibfnamefont {S.}~\bibnamefont {Vitale}},\ }\bibfield  {title} {\bibinfo {title} {{LISA Pathfinder test mass injection in geodesic motion: status of the on-ground testing}},\ }\href {https://doi.org/10.1088/0264-9381/26/9/094011} {\bibfield  {journal} {\bibinfo  {journal}
  {Classical and Quantum Gravity}\ }\textbf {\bibinfo {volume} {26}},\ \bibinfo {pages} {094011} (\bibinfo {year} {2009})}\BibitemShut {NoStop}%
\bibitem [{\citenamefont {Armano}\ \emph {et~al.}(2016)\citenamefont {Armano}, \citenamefont {Audley}, \citenamefont {Auger}, \citenamefont {Baird}, \citenamefont {Bassan}, \citenamefont {Binetruy}, \citenamefont {Born}, \citenamefont {Bortoluzzi}, \citenamefont {Brandt}, \citenamefont {Caleno} \emph {et~al.}}]{PhysRevLett.116.231101}%
  \BibitemOpen
  \bibfield  {author} {\bibinfo {author} {\bibfnamefont {M.}~\bibnamefont {Armano}}, \bibinfo {author} {\bibfnamefont {H.}~\bibnamefont {Audley}}, \bibinfo {author} {\bibfnamefont {G.}~\bibnamefont {Auger}}, \bibinfo {author} {\bibfnamefont {J.~T.}\ \bibnamefont {Baird}}, \bibinfo {author} {\bibfnamefont {M.}~\bibnamefont {Bassan}}, \bibinfo {author} {\bibfnamefont {P.}~\bibnamefont {Binetruy}}, \bibinfo {author} {\bibfnamefont {M.}~\bibnamefont {Born}}, \bibinfo {author} {\bibfnamefont {D.}~\bibnamefont {Bortoluzzi}}, \bibinfo {author} {\bibfnamefont {N.}~\bibnamefont {Brandt}}, \bibinfo {author} {\bibfnamefont {M.}~\bibnamefont {Caleno}}, \emph {et~al.},\ }\bibfield  {title} {\bibinfo {title} {{Sub-Femto-$g$ Free Fall for Space-Based Gravitational Wave Observatories: LISA Pathfinder Results}},\ }\href {https://doi.org/10.1103/PhysRevLett.116.231101} {\bibfield  {journal} {\bibinfo  {journal} {Phys. Rev. Lett.}\ }\textbf {\bibinfo {volume} {116}},\ \bibinfo {pages} {231101} (\bibinfo {year}
  {2016})}\BibitemShut {NoStop}%
\bibitem [{\citenamefont {Wanner}(2019)}]{Wanner2019Space-based}%
  \BibitemOpen
  \bibfield  {author} {\bibinfo {author} {\bibfnamefont {G.}~\bibnamefont {Wanner}},\ }\bibfield  {title} {\bibinfo {title} {{Space-based gravitational wave detection and how LISA Pathfinder successfully paved the way}},\ }\href {https://doi.org/10.1038/S41567-019-0462-3} {\bibfield  {journal} {\bibinfo  {journal} {Nature Physics}\ }\textbf {\bibinfo {volume} {15}},\ \bibinfo {pages} {200} (\bibinfo {year} {2019})}\BibitemShut {NoStop}%
\bibitem [{\citenamefont {Armano}\ \emph {et~al.}(2022)\citenamefont {Armano}, \citenamefont {Audley}, \citenamefont {Baird}, \citenamefont {Binetruy}, \citenamefont {Born}, \citenamefont {Bortoluzzi}, \citenamefont {Castelli}, \citenamefont {Cavalleri}, \citenamefont {Cesarini}, \citenamefont {Chiavegato} \emph {et~al.}}]{PhysRevD.106.062001}%
  \BibitemOpen
  \bibfield  {author} {\bibinfo {author} {\bibfnamefont {M.}~\bibnamefont {Armano}}, \bibinfo {author} {\bibfnamefont {H.}~\bibnamefont {Audley}}, \bibinfo {author} {\bibfnamefont {J.}~\bibnamefont {Baird}}, \bibinfo {author} {\bibfnamefont {P.}~\bibnamefont {Binetruy}}, \bibinfo {author} {\bibfnamefont {M.}~\bibnamefont {Born}}, \bibinfo {author} {\bibfnamefont {D.}~\bibnamefont {Bortoluzzi}}, \bibinfo {author} {\bibfnamefont {E.}~\bibnamefont {Castelli}}, \bibinfo {author} {\bibfnamefont {A.}~\bibnamefont {Cavalleri}}, \bibinfo {author} {\bibfnamefont {A.}~\bibnamefont {Cesarini}}, \bibinfo {author} {\bibfnamefont {V.}~\bibnamefont {Chiavegato}}, \emph {et~al.} (\bibinfo {collaboration} {LISA Pathfinder Collaboration}),\ }\bibfield  {title} {\bibinfo {title} {{Transient acceleration events in LISA Pathfinder data: Properties and possible physical origin}},\ }\href {https://doi.org/10.1103/PhysRevD.106.062001} {\bibfield  {journal} {\bibinfo  {journal} {Phys. Rev. D}\ }\textbf {\bibinfo {volume} {106}},\
  \bibinfo {pages} {062001} (\bibinfo {year} {2022})}\BibitemShut {NoStop}%
\bibitem [{\citenamefont {Baghi}\ \emph {et~al.}(2022)\citenamefont {Baghi}, \citenamefont {Korsakova}, \citenamefont {Slutsky}, \citenamefont {Castelli}, \citenamefont {Karnesis},\ and\ \citenamefont {Bayle}}]{BaghiPhysRevD.105.042002}%
  \BibitemOpen
  \bibfield  {author} {\bibinfo {author} {\bibfnamefont {Q.}~\bibnamefont {Baghi}}, \bibinfo {author} {\bibfnamefont {N.}~\bibnamefont {Korsakova}}, \bibinfo {author} {\bibfnamefont {J.}~\bibnamefont {Slutsky}}, \bibinfo {author} {\bibfnamefont {E.}~\bibnamefont {Castelli}}, \bibinfo {author} {\bibfnamefont {N.}~\bibnamefont {Karnesis}},\ and\ \bibinfo {author} {\bibfnamefont {J.-B.}\ \bibnamefont {Bayle}},\ }\bibfield  {title} {\bibinfo {title} {{Detection and characterization of instrumental transients in LISA Pathfinder and their projection to LISA}},\ }\href {https://doi.org/10.1103/PhysRevD.105.042002} {\bibfield  {journal} {\bibinfo  {journal} {Phys. Rev. D}\ }\textbf {\bibinfo {volume} {105}},\ \bibinfo {pages} {042002} (\bibinfo {year} {2022})}\BibitemShut {NoStop}%
\bibitem [{\citenamefont {Spadaro}\ \emph {et~al.}(2023)\citenamefont {Spadaro}, \citenamefont {Buscicchio}, \citenamefont {Vetrugno}, \citenamefont {Klein}, \citenamefont {Gerosa}, \citenamefont {Vitale}, \citenamefont {Dolesi}, \citenamefont {Weber},\ and\ \citenamefont {Colpi}}]{spadaro2023glitch}%
  \BibitemOpen
  \bibfield  {author} {\bibinfo {author} {\bibfnamefont {A.}~\bibnamefont {Spadaro}}, \bibinfo {author} {\bibfnamefont {R.}~\bibnamefont {Buscicchio}}, \bibinfo {author} {\bibfnamefont {D.}~\bibnamefont {Vetrugno}}, \bibinfo {author} {\bibfnamefont {A.}~\bibnamefont {Klein}}, \bibinfo {author} {\bibfnamefont {D.}~\bibnamefont {Gerosa}}, \bibinfo {author} {\bibfnamefont {S.}~\bibnamefont {Vitale}}, \bibinfo {author} {\bibfnamefont {R.}~\bibnamefont {Dolesi}}, \bibinfo {author} {\bibfnamefont {W.~J.}\ \bibnamefont {Weber}},\ and\ \bibinfo {author} {\bibfnamefont {M.}~\bibnamefont {Colpi}},\ }\href@noop {} {\bibinfo {title} {{Glitch systematics on the observation of massive black-hole binaries with LISA}}} (\bibinfo {year} {2023}),\ \Eprint {https://arxiv.org/abs/2306.03923} {arXiv:2306.03923 [gr-qc]} \BibitemShut {NoStop}%
\bibitem [{\citenamefont {{LDC Working Group}}(2022)}]{SpritzLDF}%
  \BibitemOpen
  \bibfield  {author} {\bibinfo {author} {\bibnamefont {{LDC Working Group}}},\ }\href {https://lisa-ldc.lal.in2p3.fr/static/data/pdf/LDC-manual-Spritz.pdf} {\bibinfo {title} {{LISA Data Challenge: Spritz}}} (\bibinfo {year} {2022}),\ \bibinfo {note} {{LISA-LCST-XXX-TN-001}}\BibitemShut {NoStop}%
\bibitem [{\citenamefont {Hourihane}\ \emph {et~al.}(2022)\citenamefont {Hourihane}, \citenamefont {Chatziioannou}, \citenamefont {Wijngaarden}, \citenamefont {Davis}, \citenamefont {Littenberg},\ and\ \citenamefont {Cornish}}]{PhysRevD.106.042006}%
  \BibitemOpen
  \bibfield  {author} {\bibinfo {author} {\bibfnamefont {S.}~\bibnamefont {Hourihane}}, \bibinfo {author} {\bibfnamefont {K.}~\bibnamefont {Chatziioannou}}, \bibinfo {author} {\bibfnamefont {M.}~\bibnamefont {Wijngaarden}}, \bibinfo {author} {\bibfnamefont {D.}~\bibnamefont {Davis}}, \bibinfo {author} {\bibfnamefont {T.}~\bibnamefont {Littenberg}},\ and\ \bibinfo {author} {\bibfnamefont {N.}~\bibnamefont {Cornish}},\ }\bibfield  {title} {\bibinfo {title} {{Accurate modeling and mitigation of overlapping signals and glitches in gravitational-wave data}},\ }\href {https://doi.org/10.1103/PhysRevD.106.042006} {\bibfield  {journal} {\bibinfo  {journal} {Phys. Rev. D}\ }\textbf {\bibinfo {volume} {106}},\ \bibinfo {pages} {042006} (\bibinfo {year} {2022})}\BibitemShut {NoStop}%
\bibitem [{\citenamefont {Bayle}\ \emph {et~al.}(2021)\citenamefont {Bayle}, \citenamefont {Castelli},\ and\ \citenamefont {Korsakova}}]{lisaglitch21}%
  \BibitemOpen
  \bibfield  {author} {\bibinfo {author} {\bibfnamefont {J.-B.}\ \bibnamefont {Bayle}}, \bibinfo {author} {\bibfnamefont {E.}~\bibnamefont {Castelli}},\ and\ \bibinfo {author} {\bibfnamefont {N.}~\bibnamefont {Korsakova}},\ }\href@noop {} {\bibinfo {title} {{LISA Glitch}}} (\bibinfo {year} {2021})\BibitemShut {NoStop}%
\bibitem [{\citenamefont {Levy}(2019)}]{Levy2019}%
  \BibitemOpen
  \bibfield  {author} {\bibinfo {author} {\bibfnamefont {B.~C.}\ \bibnamefont {Levy}},\ }\bibinfo {title} {{Wiener Process and White Gaussian Noise}},\ in\ \href {https://doi.org/10.1007/978-3-030-22297-0_6} {\emph {\bibinfo {booktitle} {Random Processes with Applications to Circuits and Communications}}}\ (\bibinfo  {publisher} {Springer International Publishing},\ \bibinfo {year} {2019})\ p.\ \bibinfo {pages} {207–234}\BibitemShut {NoStop}%
\bibitem [{\citenamefont {Houba}(2023)}]{Houba2023}%
  \BibitemOpen
  \bibfield  {author} {\bibinfo {author} {\bibfnamefont {N.}~\bibnamefont {Houba}},\ }\href@noop {} {\emph {\bibinfo {title} {Tilt-to-Length Noise Estimation and Reduction Algorithms for Spaceborne Gravitational-Wave Observatories}}},\ \bibinfo {series} {Fortschrittsberichte des Instituts für Flugmechanik und Flugregelung}, Vol.\ \bibinfo {volume} {15, ISBN: 9783844092813}\ (\bibinfo  {publisher} {Shaker Verlag},\ \bibinfo {address} {Düren},\ \bibinfo {year} {2023})\BibitemShut {NoStop}%
\bibitem [{\citenamefont {Bayle}(2019)}]{phdthesisJB}%
  \BibitemOpen
  \bibfield  {author} {\bibinfo {author} {\bibfnamefont {J.-B.}\ \bibnamefont {Bayle}},\ }\emph {\bibinfo {title} {Simulation and Data Analysis for LISA (Instrumental Modeling, Time-Delay Interferometry, Noise-Reduction Performance Study, and Discrimination of Transient Gravitational Signals)}},\ \href@noop {} {Ph.D. thesis} (\bibinfo {year} {2019})\BibitemShut {NoStop}%
\bibitem [{\citenamefont {Hartwig}(2021)}]{https://doi.org/10.15488/11372}%
  \BibitemOpen
  \bibfield  {author} {\bibinfo {author} {\bibfnamefont {O.}~\bibnamefont {Hartwig}},\ }\href {https://doi.org/10.15488/11372} {\bibinfo {title} {{Instrumental modelling and noise reduction algorithms for the Laser Interferometer Space Antenna}}} (\bibinfo {year} {2021})\BibitemShut {NoStop}%
\bibitem [{\citenamefont {Houba}\ \emph {et~al.}(2022{\natexlab{b}})\citenamefont {Houba}, \citenamefont {Delchambre}, \citenamefont {Ziegler},\ and\ \citenamefont {Fichter}}]{HoubaJoG}%
  \BibitemOpen
  \bibfield  {author} {\bibinfo {author} {\bibfnamefont {N.}~\bibnamefont {Houba}}, \bibinfo {author} {\bibfnamefont {S.}~\bibnamefont {Delchambre}}, \bibinfo {author} {\bibfnamefont {T.}~\bibnamefont {Ziegler}},\ and\ \bibinfo {author} {\bibfnamefont {W.}~\bibnamefont {Fichter}},\ }\bibfield  {title} {\bibinfo {title} {Optimal estimation of tilt-to-length noise for spaceborne gravitational-wave observatories},\ }\href {https://doi.org/10.2514/1.G006064} {\bibfield  {journal} {\bibinfo  {journal} {Journal of Guidance, Control, and Dynamics}\ }\textbf {\bibinfo {volume} {45}},\ \bibinfo {pages} {1} (\bibinfo {year} {2022}{\natexlab{b}})}\BibitemShut {NoStop}%
\bibitem [{\citenamefont {Fleddermann}\ \emph {et~al.}(2018)\citenamefont {Fleddermann}, \citenamefont {Diekmann}, \citenamefont {Steier}, \citenamefont {Tr\"{o}bs}, \citenamefont {Heinzel},\ and\ \citenamefont {Danzmann}}]{Fleddermann2018}%
  \BibitemOpen
  \bibfield  {author} {\bibinfo {author} {\bibfnamefont {R.}~\bibnamefont {Fleddermann}}, \bibinfo {author} {\bibfnamefont {C.}~\bibnamefont {Diekmann}}, \bibinfo {author} {\bibfnamefont {F.}~\bibnamefont {Steier}}, \bibinfo {author} {\bibfnamefont {M.}~\bibnamefont {Tr\"{o}bs}}, \bibinfo {author} {\bibfnamefont {G.}~\bibnamefont {Heinzel}},\ and\ \bibinfo {author} {\bibfnamefont {K.}~\bibnamefont {Danzmann}},\ }\bibfield  {title} {\bibinfo {title} {{Sub-pm${{\sqrt{{\rm Hz}}^{-1}}}$ non-reciprocal noise in the LISA backlink fiber}},\ }\href {https://doi.org/10.1088/1361-6382/aaa276} {\bibfield  {journal} {\bibinfo  {journal} {Classical and Quantum Gravity}\ }\textbf {\bibinfo {volume} {35}},\ \bibinfo {pages} {075007} (\bibinfo {year} {2018})}\BibitemShut {NoStop}%
\bibitem [{\citenamefont {Otto}(2015)}]{https://doi.org/10.15488/8545}%
  \BibitemOpen
  \bibfield  {author} {\bibinfo {author} {\bibfnamefont {M.}~\bibnamefont {Otto}},\ }\href {https://doi.org/10.15488/8545} {\bibinfo {title} {Time-delay interferometry simulations for the laser interferometer space antenna}} (\bibinfo {year} {2015})\BibitemShut {NoStop}%
\bibitem [{\citenamefont {Muratore}\ \emph {et~al.}(2020)\citenamefont {Muratore}, \citenamefont {Vetrugno},\ and\ \citenamefont {Vitale}}]{muratore2020revisitation}%
  \BibitemOpen
  \bibfield  {author} {\bibinfo {author} {\bibfnamefont {M.}~\bibnamefont {Muratore}}, \bibinfo {author} {\bibfnamefont {D.}~\bibnamefont {Vetrugno}},\ and\ \bibinfo {author} {\bibfnamefont {S.}~\bibnamefont {Vitale}},\ }\href@noop {} {\bibinfo {title} {{Revisitation of time delay interferometry combinations that suppress laser noise in LISA}}} (\bibinfo {year} {2020}),\ \Eprint {https://arxiv.org/abs/2001.11221} {arXiv:2001.11221 [astro-ph.IM]} \BibitemShut {NoStop}%
\bibitem [{\citenamefont {{LISA Science Study Team}}(2018)}]{SciReqDoc}%
  \BibitemOpen
  \bibfield  {author} {\bibinfo {author} {\bibnamefont {{LISA Science Study Team}}},\ }\href@noop {} {\bibinfo {title} {{LISA Science Requirements document}}},\ \bibinfo {howpublished} {Technical Report, European Space Agency} (\bibinfo {year} {2018}),\ \bibinfo {note} {technical Report ESA-L3-EST-SCI-RS-001}\BibitemShut {NoStop}%
\bibitem [{\citenamefont {Bayle}(2022)}]{lisagwresponse21}%
  \BibitemOpen
  \bibfield  {author} {\bibinfo {author} {\bibfnamefont {J.-B.}\ \bibnamefont {Bayle}},\ }\href@noop {} {\bibinfo {title} {{LISA GW Response}}} (\bibinfo {year} {2022})\BibitemShut {NoStop}%
\bibitem [{\citenamefont {Bayle}\ \emph {et~al.}(2022)\citenamefont {Bayle}, \citenamefont {Hartwig},\ and\ \citenamefont {Staab}}]{lisainstrument22}%
  \BibitemOpen
  \bibfield  {author} {\bibinfo {author} {\bibfnamefont {J.-B.}\ \bibnamefont {Bayle}}, \bibinfo {author} {\bibfnamefont {O.}~\bibnamefont {Hartwig}},\ and\ \bibinfo {author} {\bibfnamefont {M.}~\bibnamefont {Staab}},\ }\href@noop {} {\bibinfo {title} {{LISA Instrument}}} (\bibinfo {year} {2022})\BibitemShut {NoStop}%
\bibitem [{\citenamefont {Bayle}\ and\ \citenamefont {Staab}(2023)}]{pytdi23}%
  \BibitemOpen
  \bibfield  {author} {\bibinfo {author} {\bibfnamefont {J.-B.}\ \bibnamefont {Bayle}}\ and\ \bibinfo {author} {\bibfnamefont {M.}~\bibnamefont {Staab}},\ }\href@noop {} {\bibinfo {title} {{PyTDI}}} (\bibinfo {year} {2023})\BibitemShut {NoStop}%
\bibitem [{\citenamefont {León}(2020)}]{Len2020}%
  \BibitemOpen
  \bibfield  {author} {\bibinfo {author} {\bibfnamefont {C.}~\bibnamefont {León}},\ }\bibfield  {title} {\bibinfo {title} {{Detecting anomalous payments networks: A dimensionality-reduction approach}},\ }\href {https://doi.org/10.1016/j.latcb.2020.100001} {\bibfield  {journal} {\bibinfo  {journal} {Latin American Journal of Central Banking}\ }\textbf {\bibinfo {volume} {1}},\ \bibinfo {pages} {100001} (\bibinfo {year} {2020})}\BibitemShut {NoStop}%
\bibitem [{\citenamefont {Imani}(2018)}]{Imani2018Anomaly}%
  \BibitemOpen
  \bibfield  {author} {\bibinfo {author} {\bibfnamefont {M.}~\bibnamefont {Imani}},\ }\bibfield  {title} {\bibinfo {title} {{Anomaly Detection from Hyperspectral Images Using Clustering Based Feature Reduction}},\ }\href {https://doi.org/10.1007/s12524-018-0784-0} {\bibfield  {journal} {\bibinfo  {journal} {Journal of the Indian Society of Remote Sensing}\ }\textbf {\bibinfo {volume} {46}},\ \bibinfo {pages} {1389} (\bibinfo {year} {2018})}\BibitemShut {NoStop}%
\bibitem [{\citenamefont {Rawat}\ and\ \citenamefont {Wang}(2017)}]{rawat2017deep}%
  \BibitemOpen
  \bibfield  {author} {\bibinfo {author} {\bibfnamefont {W.}~\bibnamefont {Rawat}}\ and\ \bibinfo {author} {\bibfnamefont {Z.}~\bibnamefont {Wang}},\ }\bibfield  {title} {\bibinfo {title} {{Deep Convolutional Neural Networks for Image Classification: A Comprehensive Review}},\ }\href@noop {} {\bibfield  {journal} {\bibinfo  {journal} {Neural Computation}\ }\textbf {\bibinfo {volume} {29}},\ \bibinfo {pages} {2352} (\bibinfo {year} {2017})}\BibitemShut {NoStop}%
\bibitem [{\citenamefont {Alzubaidi}\ \emph {et~al.}(2021)\citenamefont {Alzubaidi} \emph {et~al.}}]{alzubaidi2021review}%
  \BibitemOpen
  \bibfield  {author} {\bibinfo {author} {\bibfnamefont {L.}~\bibnamefont {Alzubaidi}} \emph {et~al.},\ }\bibfield  {title} {\bibinfo {title} {{Review of deep learning: concepts, CNN architectures, challenges, applications, future directions}},\ }\href@noop {} {\bibfield  {journal} {\bibinfo  {journal} {Journal of Big Data}\ }\textbf {\bibinfo {volume} {8}} (\bibinfo {year} {2021})}\BibitemShut {NoStop}%
\bibitem [{\citenamefont {Golinko}\ \emph {et~al.}(2018)\citenamefont {Golinko}, \citenamefont {Sonderman},\ and\ \citenamefont {Zhu}}]{golinko2018learning}%
  \BibitemOpen
  \bibfield  {author} {\bibinfo {author} {\bibfnamefont {E.}~\bibnamefont {Golinko}}, \bibinfo {author} {\bibfnamefont {T.}~\bibnamefont {Sonderman}},\ and\ \bibinfo {author} {\bibfnamefont {X.}~\bibnamefont {Zhu}},\ }\bibfield  {title} {\bibinfo {title} {{Learning Convolutional Neural Networks from Ordered Features of Generic Data}},\ }in\ \href@noop {} {\emph {\bibinfo {booktitle} {2018 17th IEEE International Conference on Machine Learning and Applications (ICMLA)}}}\ (\bibinfo {organization} {IEEE},\ \bibinfo {year} {2018})\ pp.\ \bibinfo {pages} {897--900}\BibitemShut {NoStop}%
\bibitem [{\citenamefont {Roeder}(2023)}]{netron22}%
  \BibitemOpen
  \bibfield  {author} {\bibinfo {author} {\bibfnamefont {L.}~\bibnamefont {Roeder}},\ }\href@noop {} {\bibinfo {title} {{Netron}}},\ \bibinfo {howpublished} {\url{https://github.com/lutzroeder/netron}} (\bibinfo {year} {2023}),\ \bibinfo {note} {{Netron is a viewer for neural network, deep learning and machine learning models.}}\BibitemShut {Stop}%
\bibitem [{\citenamefont {Abadi}\ \emph {et~al.}(2015)\citenamefont {Abadi}, \citenamefont {Agarwal}, \citenamefont {Barham}, \citenamefont {Brevdo}, \citenamefont {Chen}, \citenamefont {Citro}, \citenamefont {Corrado}, \citenamefont {Davis}, \citenamefont {Dean}, \citenamefont {Devin} \emph {et~al.}}]{tensorflow2015whitepaper}%
  \BibitemOpen
  \bibfield  {author} {\bibinfo {author} {\bibfnamefont {M.}~\bibnamefont {Abadi}}, \bibinfo {author} {\bibfnamefont {A.}~\bibnamefont {Agarwal}}, \bibinfo {author} {\bibfnamefont {P.}~\bibnamefont {Barham}}, \bibinfo {author} {\bibfnamefont {E.}~\bibnamefont {Brevdo}}, \bibinfo {author} {\bibfnamefont {Z.}~\bibnamefont {Chen}}, \bibinfo {author} {\bibfnamefont {C.}~\bibnamefont {Citro}}, \bibinfo {author} {\bibfnamefont {G.~S.}\ \bibnamefont {Corrado}}, \bibinfo {author} {\bibfnamefont {A.}~\bibnamefont {Davis}}, \bibinfo {author} {\bibfnamefont {J.}~\bibnamefont {Dean}}, \bibinfo {author} {\bibfnamefont {M.}~\bibnamefont {Devin}}, \emph {et~al.},\ }\href {https://www.tensorflow.org/} {\bibinfo {title} {{TensorFlow}: Large-scale machine learning on heterogeneous systems}} (\bibinfo {year} {2015}),\ \bibinfo {note} {software available from tensorflow.org}\BibitemShut {NoStop}%
\bibitem [{\citenamefont {Galanis}\ \emph {et~al.}(2022)\citenamefont {Galanis}, \citenamefont {Vafiadis}, \citenamefont {Mirzaev},\ and\ \citenamefont {Papakostas}}]{Galanis2022Convolutional}%
  \BibitemOpen
  \bibfield  {author} {\bibinfo {author} {\bibfnamefont {N.}~\bibnamefont {Galanis}}, \bibinfo {author} {\bibfnamefont {P.}~\bibnamefont {Vafiadis}}, \bibinfo {author} {\bibfnamefont {K.-G.}\ \bibnamefont {Mirzaev}},\ and\ \bibinfo {author} {\bibfnamefont {G.}~\bibnamefont {Papakostas}},\ }\bibfield  {title} {\bibinfo {title} {{Convolutional Neural Networks: A Roundup and Benchmark of Their Pooling Layer Variants}},\ }\href {https://doi.org/10.3390/a15110391} {\bibfield  {journal} {\bibinfo  {journal} {Algorithms}\ }\textbf {\bibinfo {volume} {15}},\ \bibinfo {pages} {391} (\bibinfo {year} {2022})}\BibitemShut {NoStop}%
\bibitem [{\citenamefont {Mahmoudi}\ \emph {et~al.}(2020)\citenamefont {Mahmoudi}, \citenamefont {Chetouani}, \citenamefont {Boufera},\ and\ \citenamefont {Tabia}}]{Mahmoudi2020Learnable}%
  \BibitemOpen
  \bibfield  {author} {\bibinfo {author} {\bibfnamefont {M.~A.}\ \bibnamefont {Mahmoudi}}, \bibinfo {author} {\bibfnamefont {A.}~\bibnamefont {Chetouani}}, \bibinfo {author} {\bibfnamefont {F.}~\bibnamefont {Boufera}},\ and\ \bibinfo {author} {\bibfnamefont {H.}~\bibnamefont {Tabia}},\ }\bibfield  {title} {\bibinfo {title} {{Learnable pooling weights for facial expression recognition}},\ }\href {https://doi.org/10.1016/j.patrec.2020.09.001} {\bibfield  {journal} {\bibinfo  {journal} {Pattern Recognit. Lett.}\ }\textbf {\bibinfo {volume} {138}},\ \bibinfo {pages} {644} (\bibinfo {year} {2020})}\BibitemShut {NoStop}%
\bibitem [{\citenamefont {Prince}\ \emph {et~al.}(2002)\citenamefont {Prince}, \citenamefont {Tinto},\ and\ \citenamefont {Larson}}]{LISAOptSens}%
  \BibitemOpen
  \bibfield  {author} {\bibinfo {author} {\bibfnamefont {T.}~\bibnamefont {Prince}}, \bibinfo {author} {\bibfnamefont {M.}~\bibnamefont {Tinto}},\ and\ \bibinfo {author} {\bibfnamefont {S.}~\bibnamefont {Larson}},\ }\bibfield  {title} {\bibinfo {title} {{LISA optimal sensitivity}},\ }\href {https://doi.org/10.1103/PhysRevD.66.122002} {\bibfield  {journal} {\bibinfo  {journal} {Phys. Rev. D}\ }\textbf {\bibinfo {volume} {66}} (\bibinfo {year} {2002})}\BibitemShut {NoStop}%
\bibitem [{\citenamefont {Sampson}\ \emph {et~al.}(2013)\citenamefont {Sampson}, \citenamefont {Cornish},\ and\ \citenamefont {Yunes}}]{Sampson2013Mismodeling}%
  \BibitemOpen
  \bibfield  {author} {\bibinfo {author} {\bibfnamefont {L.}~\bibnamefont {Sampson}}, \bibinfo {author} {\bibfnamefont {N.}~\bibnamefont {Cornish}},\ and\ \bibinfo {author} {\bibfnamefont {N.}~\bibnamefont {Yunes}},\ }\bibfield  {title} {\bibinfo {title} {{Mismodeling in gravitational-wave astronomy: The trouble with templates}},\ }\href {https://doi.org/10.1103/PhysRevD.89.064037} {\bibfield  {journal} {\bibinfo  {journal} {Physical Review D}\ }\textbf {\bibinfo {volume} {89}},\ \bibinfo {pages} {064037} (\bibinfo {year} {2013})}\BibitemShut {NoStop}%
\bibitem [{\citenamefont {Ajith}\ \emph {et~al.}(2008)\citenamefont {Ajith}, \citenamefont {Babak}, \citenamefont {Chen}, \citenamefont {Hewitson}, \citenamefont {Krishnan}, \citenamefont {Sintes}, \citenamefont {Whelan}, \citenamefont {Br\"ugmann}, \citenamefont {Diener}, \citenamefont {Dorband}, \citenamefont {Gonzalez}, \citenamefont {Hannam}, \citenamefont {Husa}, \citenamefont {Pollney}, \citenamefont {Rezzolla}, \citenamefont {Santamar\'{\i}a}, \citenamefont {Sperhake},\ and\ \citenamefont {Thornburg}}]{PhysRevD.77.104017}%
  \BibitemOpen
  \bibfield  {author} {\bibinfo {author} {\bibfnamefont {P.}~\bibnamefont {Ajith}}, \bibinfo {author} {\bibfnamefont {S.}~\bibnamefont {Babak}}, \bibinfo {author} {\bibfnamefont {Y.}~\bibnamefont {Chen}}, \bibinfo {author} {\bibfnamefont {M.}~\bibnamefont {Hewitson}}, \bibinfo {author} {\bibfnamefont {B.}~\bibnamefont {Krishnan}}, \bibinfo {author} {\bibfnamefont {A.~M.}\ \bibnamefont {Sintes}}, \bibinfo {author} {\bibfnamefont {J.~T.}\ \bibnamefont {Whelan}}, \bibinfo {author} {\bibfnamefont {B.}~\bibnamefont {Br\"ugmann}}, \bibinfo {author} {\bibfnamefont {P.}~\bibnamefont {Diener}}, \bibinfo {author} {\bibfnamefont {N.}~\bibnamefont {Dorband}}, \bibinfo {author} {\bibfnamefont {J.}~\bibnamefont {Gonzalez}}, \bibinfo {author} {\bibfnamefont {M.}~\bibnamefont {Hannam}}, \bibinfo {author} {\bibfnamefont {S.}~\bibnamefont {Husa}}, \bibinfo {author} {\bibfnamefont {D.}~\bibnamefont {Pollney}}, \bibinfo {author} {\bibfnamefont {L.}~\bibnamefont {Rezzolla}}, \bibinfo {author} {\bibfnamefont {L.}~\bibnamefont
  {Santamar\'{\i}a}}, \bibinfo {author} {\bibfnamefont {U.}~\bibnamefont {Sperhake}},\ and\ \bibinfo {author} {\bibfnamefont {J.}~\bibnamefont {Thornburg}},\ }\bibfield  {title} {\bibinfo {title} {{Template bank for gravitational waveforms from coalescing binary black holes: Nonspinning binaries}},\ }\href {https://doi.org/10.1103/PhysRevD.77.104017} {\bibfield  {journal} {\bibinfo  {journal} {Phys. Rev. D}\ }\textbf {\bibinfo {volume} {77}},\ \bibinfo {pages} {104017} (\bibinfo {year} {2008})}\BibitemShut {NoStop}%
\bibitem [{\citenamefont {Strub}\ \emph {et~al.}(2023)\citenamefont {Strub}, \citenamefont {Ferraioli}, \citenamefont {Schmelzbach}, \citenamefont {St\"ahler},\ and\ \citenamefont {Giardini}}]{StrubGPU}%
  \BibitemOpen
  \bibfield  {author} {\bibinfo {author} {\bibfnamefont {S.~H.}\ \bibnamefont {Strub}}, \bibinfo {author} {\bibfnamefont {L.}~\bibnamefont {Ferraioli}}, \bibinfo {author} {\bibfnamefont {C.}~\bibnamefont {Schmelzbach}}, \bibinfo {author} {\bibfnamefont {S.~C.}\ \bibnamefont {St\"ahler}},\ and\ \bibinfo {author} {\bibfnamefont {D.}~\bibnamefont {Giardini}},\ }\bibfield  {title} {\bibinfo {title} {{Accelerating global parameter estimation of gravitational waves from Galactic binaries using a genetic algorithm and GPUs}},\ }\href {https://doi.org/10.1103/PhysRevD.108.103018} {\bibfield  {journal} {\bibinfo  {journal} {Phys. Rev. D}\ }\textbf {\bibinfo {volume} {108}},\ \bibinfo {pages} {103018} (\bibinfo {year} {2023})}\BibitemShut {NoStop}%
\bibitem [{\citenamefont {Strub}\ \emph {et~al.}(2022)\citenamefont {Strub}, \citenamefont {Ferraioli}, \citenamefont {Schmelzbach}, \citenamefont {St\"ahler},\ and\ \citenamefont {Giardini}}]{PhysRevD.106.062003}%
  \BibitemOpen
  \bibfield  {author} {\bibinfo {author} {\bibfnamefont {S.~H.}\ \bibnamefont {Strub}}, \bibinfo {author} {\bibfnamefont {L.}~\bibnamefont {Ferraioli}}, \bibinfo {author} {\bibfnamefont {C.}~\bibnamefont {Schmelzbach}}, \bibinfo {author} {\bibfnamefont {S.~C.}\ \bibnamefont {St\"ahler}},\ and\ \bibinfo {author} {\bibfnamefont {D.}~\bibnamefont {Giardini}},\ }\bibfield  {title} {\bibinfo {title} {{Bayesian parameter estimation of Galactic binaries in LISA data with Gaussian process regression}},\ }\href {https://doi.org/10.1103/PhysRevD.106.062003} {\bibfield  {journal} {\bibinfo  {journal} {Phys. Rev. D}\ }\textbf {\bibinfo {volume} {106}},\ \bibinfo {pages} {062003} (\bibinfo {year} {2022})}\BibitemShut {NoStop}%
\bibitem [{\citenamefont {Cheng}\ \emph {et~al.}(2016)\citenamefont {Cheng}, \citenamefont {Koc}, \citenamefont {Harmsen}, \citenamefont {Shaked}, \citenamefont {Chandra}, \citenamefont {Aradhye}, \citenamefont {Anderson}, \citenamefont {Corrado}, \citenamefont {Chai}, \citenamefont {Ispir} \emph {et~al.}}]{10.1145/2988450.2988454}%
  \BibitemOpen
  \bibfield  {author} {\bibinfo {author} {\bibfnamefont {H.-T.}\ \bibnamefont {Cheng}}, \bibinfo {author} {\bibfnamefont {L.}~\bibnamefont {Koc}}, \bibinfo {author} {\bibfnamefont {J.}~\bibnamefont {Harmsen}}, \bibinfo {author} {\bibfnamefont {T.}~\bibnamefont {Shaked}}, \bibinfo {author} {\bibfnamefont {T.}~\bibnamefont {Chandra}}, \bibinfo {author} {\bibfnamefont {H.}~\bibnamefont {Aradhye}}, \bibinfo {author} {\bibfnamefont {G.}~\bibnamefont {Anderson}}, \bibinfo {author} {\bibfnamefont {G.}~\bibnamefont {Corrado}}, \bibinfo {author} {\bibfnamefont {W.}~\bibnamefont {Chai}}, \bibinfo {author} {\bibfnamefont {M.}~\bibnamefont {Ispir}}, \emph {et~al.},\ }\bibfield  {title} {\bibinfo {title} {{Wide \& Deep Learning for Recommender Systems}}\ }(\bibinfo  {publisher} {Association for Computing Machinery},\ \bibinfo {address} {New York, NY, USA},\ \bibinfo {year} {2016})\ p.\ \bibinfo {pages} {7–10}\BibitemShut {NoStop}%
\bibitem [{\citenamefont {Goodfellow}\ \emph {et~al.}(2016)\citenamefont {Goodfellow}, \citenamefont {Bengio},\ and\ \citenamefont {Courville}}]{Goodfellow-et-al-2016}%
  \BibitemOpen
  \bibfield  {author} {\bibinfo {author} {\bibfnamefont {I.}~\bibnamefont {Goodfellow}}, \bibinfo {author} {\bibfnamefont {Y.}~\bibnamefont {Bengio}},\ and\ \bibinfo {author} {\bibfnamefont {A.}~\bibnamefont {Courville}},\ }\href@noop {} {\emph {\bibinfo {title} {{Deep Learning}}}}\ (\bibinfo  {publisher} {MIT Press},\ \bibinfo {year} {2016})\ \bibinfo {note} {\url{http://www.deeplearningbook.org}}\BibitemShut {NoStop}%
\bibitem [{\citenamefont {Dubey}\ \emph {et~al.}(2022)\citenamefont {Dubey}, \citenamefont {Singh},\ and\ \citenamefont {Chaudhuri}}]{DUBEY202292}%
  \BibitemOpen
  \bibfield  {author} {\bibinfo {author} {\bibfnamefont {S.~R.}\ \bibnamefont {Dubey}}, \bibinfo {author} {\bibfnamefont {S.~K.}\ \bibnamefont {Singh}},\ and\ \bibinfo {author} {\bibfnamefont {B.~B.}\ \bibnamefont {Chaudhuri}},\ }\bibfield  {title} {\bibinfo {title} {{Activation functions in deep learning: A comprehensive survey and benchmark}},\ }\href {https://doi.org/https://doi.org/10.1016/j.neucom.2022.06.111} {\bibfield  {journal} {\bibinfo  {journal} {Neurocomputing}\ }\textbf {\bibinfo {volume} {503}},\ \bibinfo {pages} {92} (\bibinfo {year} {2022})}\BibitemShut {NoStop}%
\bibitem [{\citenamefont {Gimeno}\ \emph {et~al.}(2021)\citenamefont {Gimeno}, \citenamefont {Mingote}, \citenamefont {Ortega}, \citenamefont {Miguel},\ and\ \citenamefont {Lleida}}]{gimeno:hal-03447745}%
  \BibitemOpen
  \bibfield  {author} {\bibinfo {author} {\bibfnamefont {P.}~\bibnamefont {Gimeno}}, \bibinfo {author} {\bibfnamefont {V.}~\bibnamefont {Mingote}}, \bibinfo {author} {\bibfnamefont {A.}~\bibnamefont {Ortega}}, \bibinfo {author} {\bibfnamefont {A.}~\bibnamefont {Miguel}},\ and\ \bibinfo {author} {\bibfnamefont {E.}~\bibnamefont {Lleida}},\ }\bibfield  {title} {\bibinfo {title} {{Generalising AUC Optimisation to Multiclass Classification for Audio Segmentation with Limited Training Data}},\ }\href {https://doi.org/10.1109/LSP.2021.3084501} {\bibfield  {journal} {\bibinfo  {journal} {{IEEE Signal Processing Letters}}\ }\textbf {\bibinfo {volume} {28}},\ \bibinfo {pages} {1135} (\bibinfo {year} {2021})}\BibitemShut {NoStop}%
\bibitem [{\citenamefont {Hochreiter}\ and\ \citenamefont {Schmidhuber}(1997)}]{Hochreiter1997}%
  \BibitemOpen
  \bibfield  {author} {\bibinfo {author} {\bibfnamefont {S.}~\bibnamefont {Hochreiter}}\ and\ \bibinfo {author} {\bibfnamefont {J.}~\bibnamefont {Schmidhuber}},\ }\bibfield  {title} {\bibinfo {title} {{Long Short-Term Memory}},\ }\href {https://doi.org/10.1162/neco.1997.9.8.1735} {\bibfield  {journal} {\bibinfo  {journal} {Neural Computation}\ }\textbf {\bibinfo {volume} {9}},\ \bibinfo {pages} {1735} (\bibinfo {year} {1997})}\BibitemShut {NoStop}%
\bibitem [{\citenamefont {Yu}\ \emph {et~al.}(2019)\citenamefont {Yu}, \citenamefont {Si}, \citenamefont {Hu},\ and\ \citenamefont {xun Zhang}}]{Yu2019}%
  \BibitemOpen
  \bibfield  {author} {\bibinfo {author} {\bibfnamefont {Y.}~\bibnamefont {Yu}}, \bibinfo {author} {\bibfnamefont {X.}~\bibnamefont {Si}}, \bibinfo {author} {\bibfnamefont {C.}~\bibnamefont {Hu}},\ and\ \bibinfo {author} {\bibfnamefont {J.}~\bibnamefont {xun Zhang}},\ }\bibfield  {title} {\bibinfo {title} {{A Review of Recurrent Neural Networks: LSTM Cells and Network Architectures}},\ }\href {https://doi.org/10.1162/neco_a_01199} {\bibfield  {journal} {\bibinfo  {journal} {Neural Computation}\ }\textbf {\bibinfo {volume} {31}},\ \bibinfo {pages} {1235} (\bibinfo {year} {2019})}\BibitemShut {NoStop}%
\bibitem [{\citenamefont {Gers}\ \emph {et~al.}(2000)\citenamefont {Gers}, \citenamefont {Schmidhuber},\ and\ \citenamefont {Cummins}}]{Gers2000}%
  \BibitemOpen
  \bibfield  {author} {\bibinfo {author} {\bibfnamefont {F.~A.}\ \bibnamefont {Gers}}, \bibinfo {author} {\bibfnamefont {J.}~\bibnamefont {Schmidhuber}},\ and\ \bibinfo {author} {\bibfnamefont {F.}~\bibnamefont {Cummins}},\ }\bibfield  {title} {\bibinfo {title} {{Learning to Forget: Continual Prediction with LSTM}},\ }\href {https://doi.org/10.1162/089976600300015015} {\bibfield  {journal} {\bibinfo  {journal} {Neural Computation}\ }\textbf {\bibinfo {volume} {12}},\ \bibinfo {pages} {2451} (\bibinfo {year} {2000})}\BibitemShut {NoStop}%
\bibitem [{\citenamefont {Virtanen}\ \emph {et~al.}(2020)\citenamefont {Virtanen}, \citenamefont {Gommers}, \citenamefont {Oliphant}, \citenamefont {Haberland}, \citenamefont {Reddy}, \citenamefont {Cournapeau}, \citenamefont {Burovski}, \citenamefont {Peterson}, \citenamefont {Weckesser}, \citenamefont {Bright} \emph {et~al.}}]{2020SciPy-NMeth}%
  \BibitemOpen
  \bibfield  {author} {\bibinfo {author} {\bibfnamefont {P.}~\bibnamefont {Virtanen}}, \bibinfo {author} {\bibfnamefont {R.}~\bibnamefont {Gommers}}, \bibinfo {author} {\bibfnamefont {T.~E.}\ \bibnamefont {Oliphant}}, \bibinfo {author} {\bibfnamefont {M.}~\bibnamefont {Haberland}}, \bibinfo {author} {\bibfnamefont {T.}~\bibnamefont {Reddy}}, \bibinfo {author} {\bibfnamefont {D.}~\bibnamefont {Cournapeau}}, \bibinfo {author} {\bibfnamefont {E.}~\bibnamefont {Burovski}}, \bibinfo {author} {\bibfnamefont {P.}~\bibnamefont {Peterson}}, \bibinfo {author} {\bibfnamefont {W.}~\bibnamefont {Weckesser}}, \bibinfo {author} {\bibfnamefont {J.}~\bibnamefont {Bright}}, \emph {et~al.},\ }\bibfield  {title} {\bibinfo {title} {{{SciPy} 1.0: Fundamental Algorithms for Scientific Computing in Python}},\ }\href {https://doi.org/10.1038/s41592-019-0686-2} {\bibfield  {journal} {\bibinfo  {journal} {Nature Methods}\ }\textbf {\bibinfo {volume} {17}},\ \bibinfo {pages} {261} (\bibinfo {year} {2020})}\BibitemShut {NoStop}%
\bibitem [{\citenamefont {Fabian}\ \emph {et~al.}(2019)\citenamefont {Fabian}, \citenamefont {Guainazzi}, \citenamefont {Mcnamara}, \citenamefont {Piro},\ and\ \citenamefont {Tanvir}}]{Fabian2019AthenaLISAS}%
  \BibitemOpen
  \bibfield  {author} {\bibinfo {author} {\bibfnamefont {A.~C.}\ \bibnamefont {Fabian}}, \bibinfo {author} {\bibfnamefont {M.}~\bibnamefont {Guainazzi}}, \bibinfo {author} {\bibfnamefont {P.}~\bibnamefont {Mcnamara}}, \bibinfo {author} {\bibfnamefont {L.}~\bibnamefont {Piro}},\ and\ \bibinfo {author} {\bibfnamefont {N.~R.}\ \bibnamefont {Tanvir}},\ }\bibfield  {title} {\bibinfo {title} {{Athena-LISA Synergies}}\ }(\bibinfo {year} {2019})\ \bibinfo {note} {\url{https://api.semanticscholar.org/CorpusID:218551158}}\BibitemShut {NoStop}%
\end{thebibliography}%

\end{document}